\documentclass[AER]{AEA}

\usepackage{natbib}

\usepackage{epstopdf}
\usepackage[caption=false]{subfig}

\usepackage{rotating} 

\usepackage[usenames, dvipsnames]{color} 

\usepackage{hyperref}
\hypersetup{colorlinks=true,linkcolor=black,anchorcolor=black,citecolor=black,filecolor=black,menucolor=black,runcolor=black,urlcolor=black, breaklinks=true}

\usepackage[british]{babel} 

\usepackage[export]{adjustbox} 
\usepackage[section]{placeins} 

\newcommand{\source}[1]{\small {\hfill Source: {#1}} } 

\usepackage{subfig} 

\usepackage[disable]{todonotes} 

\newcommand{\comment}[1]{}  

\usepackage{longtable}

\usepackage{soul} 

\usepackage{amsmath}

\usepackage{array}

\draftSpacing{1.5}

\begin{document}

\title{The Economic Complexity of US Metropolitan Areas}
\shortTitle{The Economic Complexity of US Metropolitan Areas}
\author{Benedikt S. L. Fritz and Robert A. Manduca\thanks{Fritz: Zeppelin University, Am Seemooser Horn 20, 88045 Friedrichshafen, Germany, b.fritz@zeppelin-university.net. 
		Manduca: Harvard University, Department of Sociology, 33 Kirkland Street, Cambridge, MA  02138,  rmanduca@g.harvard.edu. Acknowledgements: We thank Dominik Hartmann, Jarko Fidrmuc, Marcel Tyrell, Akira Sasahara, and participants of the North American Regional Sciences Conference in Vancouver for helpful comments and ideas. Benedikt Fritz is furthermore indebted to Cusanuswerk for funding.}}
\date{\today}
\pubMonth{January}
\pubYear{2019}
\pubVolume{}
\pubIssue{}
\JEL{}
\Keywords{}

\begin{abstract}
We calculate measures of economic complexity for US metropolitan areas for the years 2007-2015 based on industry employment data. We show that the concept of economic complexity translates well from the cross-country to the regional setting, and is able to incorporate local as well as traded industries. The largest cities and the Northeast of the US have the highest average complexity, while traded industries are more complex than local-serving ones on average, but with some exceptions. On average, regions with higher complexity have a higher income per capita, but those regions also were more affected by the financial crisis. Finally, economic complexity is a significant predictor of within-decreases in income per capita and population. Our findings highlight the importance of subnational regions, and particularly metropolitan areas, as units of economic geography.
\end{abstract}

\maketitle

\section{Introduction}
Over the past decade there has been great deal of progress in characterizing and analyzing the productive structures of countries and the paths national economies take as they develop. New measures of economic complexity show great promise in inferring information about the productive capabilities of countries from the products they export \citep{Hidalgo2009, Tacchella2012}. These metrics are more effective at predicting national economic growth than traditional predictors like education levels, institutions, and current GDP per capita \citep{Hausmann2014, Cristelli2013}, and they have become a widely used indicator of national economic performance \citep{OECD2017}.

Given the effectiveness of complexity measures at the national level, a natural extension is to apply them to subnational and metropolitan regions, which are the fundamental unit of economic geography \citep{Jacobs1969, Storper1997}. Attempts have been made to apply these metrics in China, Australia, and the United States \citep{Gao2018, Reynolds2017, Sbardella2017}. However, key questions remain about the applicability of methods designed for countries to the subnational setting. Additionally, there are questions as to whether these methods are extendable to service sectors, which dominate the economies of most developed countries \citep{Buera2012}, or if the methods are limited to the physical goods that they were developed for. A related question is whether to include only traded industries or local industries as well \citep{Porter2003a}.

In this paper, we produce economic complexity measures for metropolitan areas in the United States between 2007 and 2015. Using employment data from the US Census County Business Patterns, we are able to include almost all types of economic activity in the country. We show that the spatial distribution of economic activity across US metro areas meets the criteria for economic complexity analysis. Major cities with developed economies tend to have industries of all types, common as well as specialized: Los Angeles produces hand bags as well as guided missiles. But less developed metros tend to have only common industries.  This pattern, well documented across countries, is core to American economic geography. As this finding holds for local as well as traded industries, we argue that both should be included in the analysis. 

Furthermore, we show that the differences between competing economic complexity algorithms \citep{Albeaik2017b, Pietronero2017a} are relatively small in the context of the United States. More consequential is the method of determining industry presence in a region. We compare five different ways of designing the input matrix, and adopt two measures combining a relative and an absolute employment threshold. This approach is better able to capture the myriad capabilities of large and diverse cities.

We find that the highest complexity areas are major cities, especially Los Angeles, New York, and Chicago, while traded industries tend to rate higher on complexity than local serving ones. The Northeast of the US emerges as the most complex region on average, though the stronger distinction is between metropolitan and rural areas. In cross-sectional regressions, cities with higher complexity have significantly higher populations and higher per capita incomes, controlling for a number of social, economic, and institutional characteristics. However, in panel regressions increases in economic complexity are a significant negative predictor of decreases in population and possibly income per capita. Regressions predicting future growth based on current complexity show that higher complexity regions suffered more in the Great Recession.  

The rest of the paper is organized as follows: The next section gives a short review of previous research on economic complexity and regional economic development. The following sections introduce the data and the complexity methodology and discuss several theoretical and methodological challenges associated with constructing a subnational version of complexity. After analyzing the general transferability of economic complexity into the regional context, we conduct our regression analyses. The paper concludes with a discussion of implications for future research.

\section{Literature Review}
\subsection{Studying Productive Structures: The Capabilities Approach}
Economic complexity measures draw on what researchers have termed the ``capabilities approach'' to understanding economic development \citep{Hausmann2014}. In this approach, an economy is seen as a system of knowledge accumulation, and its prosperity depends upon whether it can make ever more information grow \citep{Hidalgo2015}. Productive knowledge is thought to exist in discrete units, or capabilities, such as the capability to weld metal or the capability to spin thread. These capabilities are combined to produce the prime economic outputs of the system, its products and services \citep{Hidalgo2007, Hidalgo2009}. The key claim of complexity researchers is that while it is typically impossible to directly observe capabilities, they can be inferred from the presence of industries: If an economy is able to competitively produce handbags, say, then it must have whatever capabilities go into handbag production. By studying the collection of products that are produced,  the portfolio of capabilities that a given region must have can be determined. Implications for development policy are that countries can most easily achieve growth by diversifying into closely related industries/products and that it is hard to expand into less similar (perhaps more complex) industries/products \citep{Hidalgo2018}. The same is observed for regions \citep{Kogler2013}.
 
The Economic Complexity Index ($ECI$) is designed to measure the overall complexity level of an economy's capability portfolio and to predict its future development \citep{Hausmann2014}. There has been some debate about the formulation of the algorithm \citep{Albeaik2017b, Pietronero2017a} and the Fitness Index ($FI$) has been established as an alternative measure for economic complexity \citep{Tacchella2012}. 

\subsection{Studying Regional Economies}
Most studies of economic complexity have been undertaken at the national level. However, there are compelling theoretical and empirical reasons to use metropolitan regions as the base unit of economic analysis. From a theoretical perspective, regions are the geographic entity at which agglomeration begins \citep{Glaeser1992a, Marshall1890a}, and at which new industries are developed \citep{Jacobs1969}. They can be functionally defined as the geographies in which face to face meetings can be undertaken without friction \citep{Storper2004}.  

Empirically, the variation in economic performance among regions within a country is often comparable to the difference between countries, even countries at very different levels of development. For instance, the GDP per capita of the San Fransisco MSA in 2017 was \$90,000, more than four times that of the McAllen Texas MSA \citep{BEA2017}. This is roughly the ratio between the GDP per capita of the United States as a whole and that of Peru \citep{IMF2017}. 

For these reasons, regional scientists have long sought to understand the economic performance of subnational regions and have used a wide swath of metrics and concepts to characterize regional economies \citep{Isard1966, Storper2011}.  Many of these, like the capabilities approach, are focused on the productive structure of regions, particularly on their export industries. Examples include the identification of a region's  ``economic base'' \citep{Andrews1953, Heilburn1981}, ``growth poles'' \citep{Parr1973, Perroux1955}, competitive clusters of linked industries \citep{Porter1998a}, and the characterization of the extent and type of industrial variety \citep{Frenken2007}. 

In the case of the United States, recent years have provided a both normative and a positive case for focusing on regional development \citep{Storper2018}. Since 1980, the long-term trend of regional convergence in income levels has stalled and begun to reverse \citep{Ganong2017, Manduca2018a, Amos2014}. A small set of fortunate regions has gotten richer, while incomes in the rest of the country have stagnated, to the point that they may have stronger connections to their peer cities across the globe than to their own hinterlands  (Sassen 1991). This divergence has led to a great deal of interest in regional economic performance in the popular media, and to calls for policymakers to treat regional economic development as a core national policy concern \citep{Badger2016, Block2019, Leonhardt2018, Longmann2015}.

\subsection{Previous Research on Regional Complexity}
There have been several previous attempts to focus the lens of economic complexity on regional or subnational productive structures. These fall into two major groups. Some studies have used economic complexity measures computed from international trade data, as in the original cross-national studies, and then calculated the complexity of regions by taking (dollar-)weighted averages of their industry portfolios. \cite{Reynolds2017} describe the productive structure of Australia's states and found that minor differences within inter-state as well as rest of world exports matter greatly to relative complexity. Other projects in this vein have placed specific countries in the context of the global product space, such as Turkey \citep{Erkan2015}, the Netherlands \citep{Zaccaria2016}  and Sub-Saharan Africa \citep{Abdon2011}.
 
The second group consists of studies which, like the present paper, calculate complexity from scratch using the national economic network. \cite{Gao2018} calculate the $ECI$ and $FI$ for Chinese provinces based on the number of publicly traded firms. They corroborate cross-country findings of a positive relationship between complexity and growth. Instead of using the number of firms per sector, \cite{Chavez2017} use employment across sectors to compute the $ECI$ of Mexico's states and provide evidence that the higher a state's $ECI$ the higher its GDP per capita. Another notable approach includes \cite{Cicerone}, who try to account for the position of a region within the product space. Combining a measure of the centrality of a province's exports within the export network with the absolute values of the revealed comparative advantage (RCA, see below), they show that better positioned provinces in Italy boast stronger regional development. 

Surprisingly, the US has received relatively little attention so far. One attempt to calculate complexity metrics for subnational regions in the US comes from \cite{Sbardella2017}. They investigate the relationship between economic complexity, GDP per capita, and wage inequality.  Mixing monetary/aggregate development data with fitness in the ``Complex Relative Rank Development Index,'' they cover two different geographical scales based on wage/labor data: countries on the one hand and the US counties on the other. However, for the analysis of counties they have only a very coarse industrial classification, with just 89 3-digit NAICS categories. This may not give enough resolution to distinguish between common and uncommon capabilities (see Appendix \ref{appendix:DigitsFigure}  for an investigation of how the stylized facts underpinning the complexity analysis break down at aggregation levels below 4-digit NAICS). Additionally, using wage data to explain wage inequality and average wages may strengthen general concerns of reversed causality. 

\section{Data}
We use data from the US Census County Business Patterns (CBP).  This production oriented data provides employment counts by NAICS industry classification and county for the entire United States. It is constructed from the Business Register, a Census database of all known single and multi-establishment companies in the United States, and created through a combination of surveys and administrative records \citep{CensusBureau2017}. Data are released annually with estimates of employment during the week of March 12 of each year.  The CBP data cover most economic activity in the United States, with the exceptions of crop and animal production, rail transportation, certain types of financial funds, self-employed individuals, and most government activity.

The CBP data provides counts of total employment by county and NAICS industry code. Observations are suppressed if there are too few employers within a given county-industry cell to maintain anonymity. These observations are assigned to one of twelve categories based on the total employment size. We replace these observations with the midpoint of the identified size range. In total, for 2015 we directly observe employment counts for approximately 99 million employees, and impute employment based on size code for 26 million more. This gives us a total dataset covering approximately 125 million employees, approximately 84\% of the Civilian Employment Level for March 2015 \citep{USBureauofLaborStatistics2017}. We use CBP data for the years 2007-2015 for our measures of economic complexity. 

An advantage of the CBP data is that because it is based on employment rather than exports, it includes service as well as goods producing industries. If most (export/product based) measures of economic complexity are inferring the presence of advanced service industries because they are necessary to manufacture complex products, we are directly observing these industries. Furthermore, because we use employment rather than output we are even more agnostic about the underlying capabilities. The importance of an industry is not merely determined by its productive output as its sheer existence allows for cooperation and technological exchange which may not be captured by the present-day productive output. We refrain from using skill data because they may exclude the role of institutions and positive externalities in contributing to productive capacity. 

\subsection{Industry Classification Scheme}

An important decision when analyzing productive structure is which industry classification scheme to use and at what level of aggregation. This can strongly affect results \citep{Arcaute2015a, Youn2016} because the application of the principles of relatedness \citep{Hidalgo2018} fundamentally depend on the underlying classifications.  

The CBP use the NAICS as a classification scheme for employment, ranging from the two-digit level (sectoral level, 19 categories) to the six-digit level (national industry level, 977 categories). Determining which digit level to use is far from trivial as data structure is shown to heavily influence results \citep{Wixe2017}. However, following the majority of papers on economic complexity, we exploit the most detailed level (e.g. six-digit) as the basis for our measures of complexity. This allows us to exploit as much information as possible. But we also report and discuss results for other digit levels in Appendix \ref{appendix:DifferentDigits}. 

Further accounting for the possible sensitivity of our results toward the data structure, we also employ the Business Cluster Definitions (BCD) designed by Porter and colleagues \citep{Delgado2016}. In contrast to the merely production oriented NAICS, the BCD combines NAICS industries into clusters based on ``inter-industry linkages based on co-location patterns, input-output links, and similarities in labor occupations'' \citep{Delgado2016}. The BCD provide the user with 67 clusters and 316 sub-clusters. We report results for sub-cluster because of the finer resolution.

\subsection{Geographic Aggregation}

Having decided upon the classification for productive structures, we are left with the question of which regional classification we should employ--a question of similar importance for our results \citep{Arcaute2015a}. Of the various possibilities, including the creation of our own regional clustering approach (ibid., \citeyear{Arcaute2015a}), we opt to use Core Based Statistical Areas (CBSAs). CBSAs consist of Metropolitan (MSA) and Micropolitan Statistical ($\mu$SA) Areas, and have at least one urbanized area at their core ($>$50,000 people for MSA, 10,000-50,000 for $\mu$SA), to which neighboring counties are tied based on their ``degree of social and economic integration with the central county or counties as measured by commuting ties'' \citep{U.S.CensusBureau2012}. For the purposes of calculating economic complexity measures, we include non-metropolitan counties as individual units, since these counties are deemed to be economically independent. Because of data limitations on control variables, our regression analysis is limited primarily to MSAs, and covers roughly 65\% of the US population (210 million inhabitants). 

\subsection{Control and Dependent Variables}

For the purposes of predicting economic growth based on complexity, we use a number of control variables – most of them from the American Community Survey. Our control variables can be clustered into economic controls, sociodemographic controls and institutional factors \citep{Breau2014, Florida2016}. Our economic controls consist of the unemployment rate, the share of people working in the manufacturing sector, and the number of patents field per year. Our sociodemographic controls include the number of people living in a region, the median age of the population, the share of people older than 18 with more than a high school degree, the share of the population being of African American origin, and the share of foreign-born people. The three institutional controls we employ are the share of workers under union coverage, the minimum wage level (of the state to which a region belongs), and the share of the eligible voting population which turned out in November elections. Economic outcome variables that we consider are per capita income and population, both sourced from the Bureau of Economic Analysis Regional Economic Accounts \citep{BEA2017}. The full list of sources and further explanation for all variables employed in our analysis can be found in the Appendix \ref{appendix:Data Sources}. We use one-year survey data to avoid the problem of moving averages in panel regressions.

\section{Methodology}

\subsection{The Index and the Input Matrices}

\subsubsection{The logic behind economic complexity}
The concept of economic complexity stems directly from the capabilities approach and the cross-industry knowledge spillovers described by \citep{Jacobs1969}. The idea is to infer the presence of capabilities, which are not directly observable, from the basket of goods an economy is able to produce \citep{Hausmann2014}. 

At the international level, the ability to produce a given product has typically been measured using Balassa's concept of revealed comparative advantage (RCA, \citealp{Balassa1965}). Empirically, and in contrast to standard international trade theories, the most advanced economies tend to have RCA in common as well as specialized products, whereas less diverse entities have RCAs only in products exported by many countries. Ordering the country-product RCA matrix by diversity (e.g. the number of products in which a country has an RCA) and ubiquity (e.g. the number of countries with an RCA in the specific product), results in a triangular shape \citep{Tacchella2012}. This pattern has been observed at both the cross country \citep{Cristelli2013} and the subnational level \citep{Sbardella2017}, based on a relatively linear transition from low diversity to high diversity regions.

To produce the final indicator of economic complexity, ubiquity and diversity are clustered together in a bipartite network (product space) and then used to correct for each other in a set of theoretically infinite iterative linear equations (the ``method of reflections,'' \citealp{Hidalgo2009}). The purpose of this correction is to avoid identifying industries as complex that are infrequent solely because they use primary inputs only found in a few areas, as opposed to requiring a large number of capabilities \citep{Hidalgo2009}. Thereby a refined indicator is produced with information on the average ubiquity of the industries present in a region, the average diversification of regions with an RCA in a specific industry, the average diversification of regions with a similar distribution of RCAs across industries, and so on.

The most common indicator constructed through this method is the Economic Complexity Index ($ECI$, \citealp{Hidalgo2009,Hausmann2014}). More recently, a second indicator, the Fitness Index ($FI$, \citealp{Tacchella2012}), has been proposed which differs from the $ECI$ inasmuch as it tries to avoid penalizing regional units with RCAs in both high and low value industries (which it claims the $ECI$ method does). Both indicators are explained in detail in Appendix \ref{appendix:Formulae}. 

A core assumption of the $ECI$, and the entire capabilities approach, is that the capabilities that go into a given product are similar across places - and in particular, that the difference in methods by which two regions produce a given type of product is smaller on average than the difference in methods by which one region produces two different products \citep{Hartmann2017}. This is a very difficult assumption for us to test, though it may be that because regions of the US share a common culture, educational system, and intellectual property regime there is less variation in methods of production across CBSAs than there is in the international context. 
 
A further concern when constructing economic complexity measures for subnational regions specifically is that the total lack of trade barriers between regions of a country may induce regional specialization that is due to economies of scale rather than capability limitations. There may be products that are relatively simple to produce that are nonetheless only made in few regions, just because a few places are enough to meet the demand of the whole country. Whether this is problematic is debatable: Even if other regions could easily develop the capabilities to produce simple products, they don't currently have them. This phenomenon may occur for more sophisticated industries as well. For instance, the 12 regions chosen as headquarters of Federal Reserve branches might not be the only ones endowed with the necessary capabilities for central banking, but nevertheless they suffice under the present system in supplying the whole nation.

Ultimately, both concerns are empirical questions. We aim to answer the second one by inspecting the most and least complex industries identified in our analysis, and tentatively trust that the first one has a positive answer if we are able to detect the triangular structure of ubiquity and diversity. The first concern especially is to some degree beyond the scope of this paper, and we join the large body of literature on economic complexity that more or less implicitly assumes this critical feature as given.

\subsubsection{How to measure industry presence }
Economic complexity measures are constructed from a matrix of industry presence by geography. A key question is how to measure the presence of an industry. In cross-national studies, the typical approach has been to identify countries as producing products if they have a RCA in a product greater than 1 - that is, if the product occupies a greater portion of their export basket than that of the average country. Results are then binarized to 1 and 0 respectively (labelled as input matrix $BM_{r,i}$). In our context, the equivalent of the RCA is the Location Quotient ($LQ$), which measures an industry's employment share in a particular region as a fraction of its employment share in the nation as a whole. Let $X_{r,i}$ denote the number of employees in region $r$ who are employed in industry $i$. Then  〖$BM_{r,i}$ corresponds to:

\begin{equation}
\begin{split}
LQ_{r,i} & = \frac{X_{r,i}}{\sum_{i}X_{r,i}} / \frac{\sum_{r}X_{r,i}}{\sum_{r,i}X_{r,i}} \\
BM_{r,i} & = 1 \quad if \quad LQ_{r,i} \geq 1 \\
BM_{r,i} & = 0 \quad if \quad LQ_{r,i} < 1
\end{split}
\end{equation}

This is a reasonable starting point, but it has some weaknesses. The most notable of these is that it tends to underreport the production of common goods in diversified regions.  Consider a country with two regions, one that produces just apples and one that produces equal quantities of apples and computers. The latter is clearly the more complex economy because it has both the capabilities that go into making apples and those that go into making computers. However, it would have a location quotient in apples of less than 1, because apples form a smaller portion of its economy than they do for the nation as a whole. Arithmetically, regions with more diverse economies will tend to have location quotients less than 1 for ubiquitous products simply because they are also producing uncommon products.  

This issue is of particular concern given the inclusion of non-traded sectors and services in our analysis. Previous research has documented that most local services scale sublinearly with city size \citep{Youn2016}. This means that ubiquitous industries are likely to be underrepresented in terms of $LQ$ in highly diverse regions. As an example, consider the case of gas stations (NAICS code 447110) in New York City. In 2015, the New York MSA had 13,972 employees at gas stations, clearly an indication of the capability to provide cars with fuel.  Yet, as a share of total city employment, gas stations were substantially underrepresented in New York, with a location quotient of just 0.27. Using the standard approach of $LQ > 1$, New York would be identified as lacking the capabilities involved in running gas stations. 

A simple solution would be to keep the raw LQ values (denoted as 〖$RLQ_{r,i}$) - so that a region having an $LQ$ close to 1 still receives some ``merit'' for this. Instead of only depicting whether a comparative advantage exits or not, RLQ reveals by which factor a region's industry exceeds/undermatches the employment we would expect given the region's overall employment relative to national employment. We thereby see different foci of development as a heavy specialization in a certain industry (a high $LQ$) signals a higher competitive advantage than a moderate one (an $LQ$ just above 1). However, 〖$RLQ_{r,i}$ to some degree overexaggerates the idea of $BM_{r,i}$ - an $LQ$ of 8 does not imply twice the specialization of a $LQ$ value of 4. For completeness,  $RLQ_{r,i}$ can be written as:

\begin{equation}
RLQ_{r,i} = LQ_{r,i}
\end{equation}

Alternatively, we follow Cristelli et al (2013) and build a weighted matrix $WM_{r,i}$, which measures a region's employment in an industry as a share of total national employment in that industry. $WM_{r,i}$ reveals information about the relative importance of a region in industrial employment and allows us to better detect the most influential regions in the national dynamics of different industries, too. But thereby, $WM_{r,i}$ heavily favors larger regions by design.\footnote{e.g. New York has 1011 employees in NAICS 512240 ``Sound Recording Studios'', 0.01\% of its total employment. The same industry makes up about 0.05\% of employment, five times as much, in Lubbock TX and Santa Maria CA--but in each case comprises only about 60 employees. Thereby, New York's share of nation-wide employment in this particular industry is more than 15 times that of the two smaller regions.} $WM_{r,i}$ is defined as
\begin{equation}
WM_{r,i} = \frac{X_{r,i}}{\sum_{r}X_{r,i}}
\end{equation}

Facing all of these problems, we consider alternative two matrices. The first measure, $Presence_{r,i}$, is based on the presence of any employment in an industry whatsoever. E.g. as soon as an industry is represented at all in a region, we assign to it a value of 1 and only give it a value of 0 if not a single worker is present in the respective industry:

\begin{equation}
\begin{split}
Presence_{r,i} & = 1 \quad if \quad X_{r,i} \geq 1 \\
Presence_{r,i} & = 0 \quad if \quad X_{r,i} = 0
\end{split}
\end{equation}

The second alternative, the cut-off matrix ($CM_{r,i}$), mimics $BM_{r,i}$, but modifies it by also counting an industry as ``present'' in a region if it employs more than 50 people there, regardless of the location quotient. Thereby, $CM_{r,i}$ corrects for artificially low $LQs$ in large or diverse regions. In the gas station example above, New York would have $CM_{r,i}=1$ because its total employment at gas stations is well over 50, even though the $LQ < 1$. At the same time, small regions will still be identified as specializing in an industry if its location quotient is greater than one, even if the absolute number of employees is small. Mathematically, 〖$CM_{r,i}$ can be written as follows:

\begin{equation}
CM_{r,i} = 1  \quad if \quad LQ_{r,i} \geq 1 \quad or \quad X_{r,i} > 50
\end{equation}

Both designs circumvent most of the disadvantages of the three input matrices described above. Common industries in diverse areas are not underreported - $Presence_{r,i}$ and $CM_{r,i}$ are immune to sublinear scaling laws because (nearly) all employment counts towards their complexity score. Similarly, RCAs are not overexaggerated because once the basic employment threshold has been met all other observations across industries per region are treated equally.  Lastly, large regions are not favored ex ante as we restrain from using national employment shares.

However, one might argue that the central weakness both of $Presence_{r,i}$ and of $CM_{r,i}$ is that both blur the differences between the most complex and medium complexity regions. It is possible that the latter has presence in a similarly high number of industries which would make the two types indistinguishable from each other. Both matrices also have individual weaknesses: The cut-off value of $CM_{r,i}$ is to some degree arbitrary. Why not use a cut-off of 100 workers or 200 or 1000? Surely cut-offs could differ across regions with different population sizes. Therefore, $CM_{r,i}$ might neglect the importance of smaller region's specializations. $Presence_{r,i}$, on the other hand, may be too sensitive to industries with small levels of employment in a region. The extent of a region's capabilities is perhaps questionable when less then a dozen of people work in a specific industry. Moreover, in some cases there might be actually no employment at all in an industry that is recorded as having fewer than 19 employees due to the way we impute employment from size codes.

Ultimately, we consider the question of which input matrix is most suitable for measuring subnational regional structures an empirical one and will analyze it according to the criterion of triangularity to then make a theoretically and empirically informed decision. 

\subsection{Exploring the Indices and the Input Matrices}
\subsubsection{Triangularity}
\begin{figure}[!h]
	\centering
	\subfloat[$CM_{r,i}$]{{\includegraphics[width=6cm]{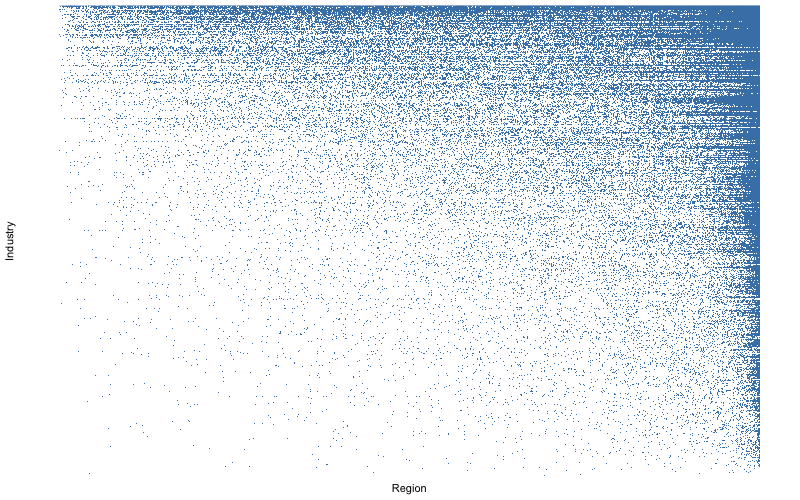}}}%
	\qquad
	\subfloat[$Presence_{r,i}$]{{\includegraphics[width=6cm]{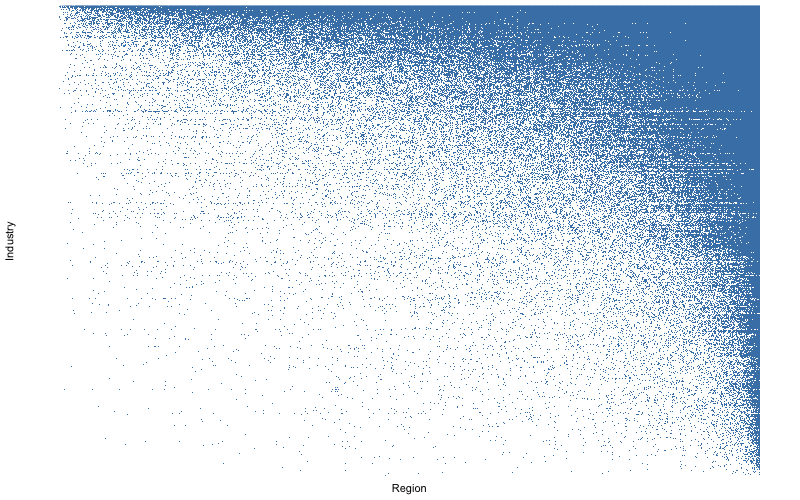}}}%
	\\
	\subfloat[$BM_{r,i}$]{{\includegraphics[width=6cm]{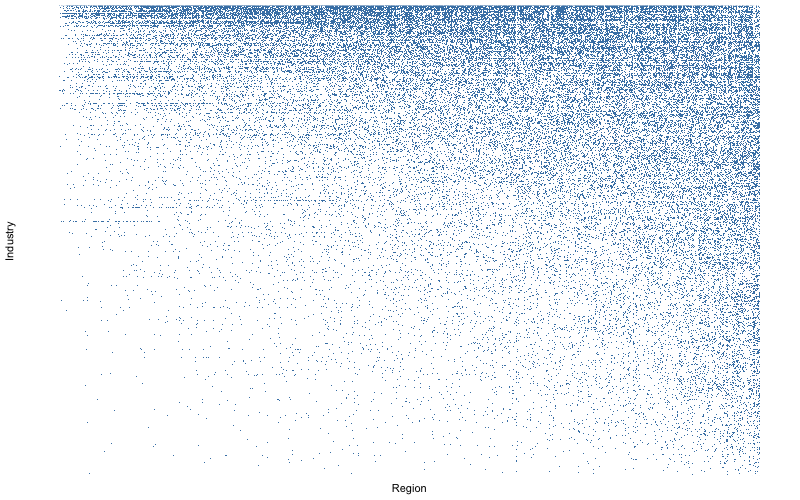}}}%
	\qquad
	\subfloat[$RLQ_{r,i}$]{{\includegraphics[width=6cm]{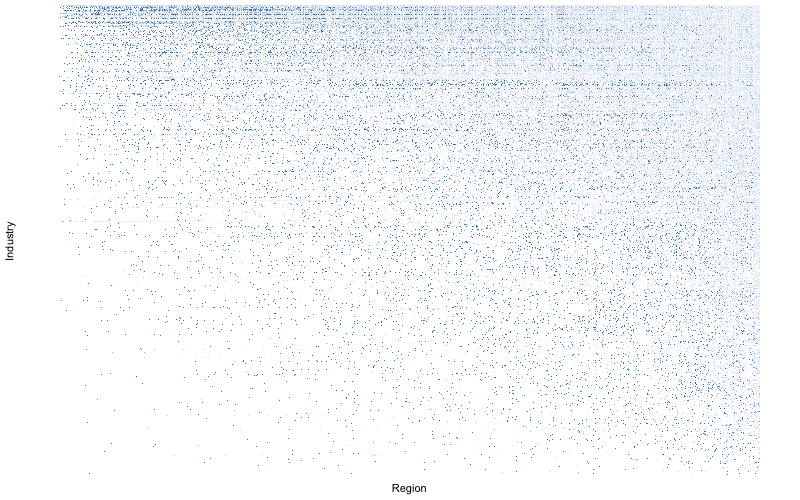}}}%
	\\
	\subfloat[$WM_{r,i}$]{{\includegraphics[width=6cm]{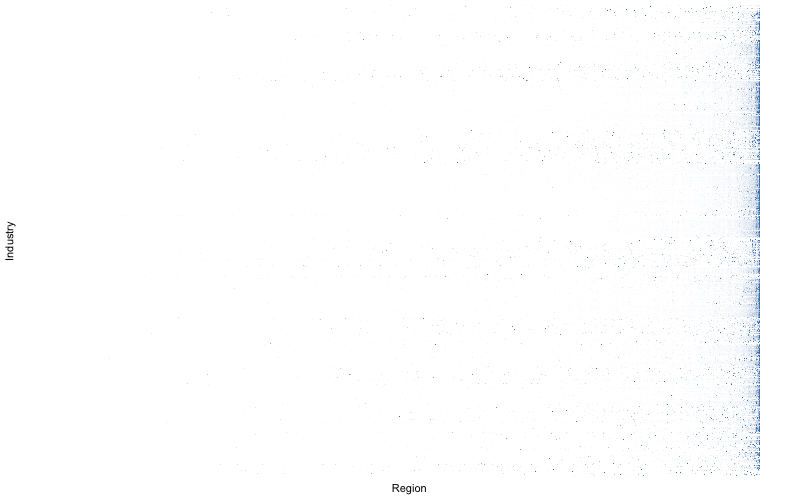}}}%
	\caption{Input matrices: Ubiquity vs. Diversity (NAICS, 2015)}%
	\label{Fig1}%
	\source{Own calculations and CBP.}
\end{figure}

In Fig \ref{Fig1}, we show the graphical representation of industry location across regions (using CBSAs) and industries (using NAICS), where regions are ordered from left to right by diversity and industries are sorted from bottom to top by ubiquity. The five different graphs correspond to the five different input matrices discussed above. 

The plots of $BM_{r,i}$, $CM_{r,i}$, and $Presence_{r,i}$ are relatively comparable and reveal a triangular structure. The triangle is most strongly marked in the cases of $CM_{r,i}$ and $Presence_{r,i}$.  The adjacent leg and the opposite leg of those two triangles are more marked than the other ones - e.g. the most ubiquitous industries are really present in all regions and the most diverse regions have more (if not all) of the industries present. Regions with the lowest level of diversity only have a presence in high ubiquity industries. 

The $RLQ_{r,i}$ triangle is less visible--the upper-right corner is only lightly filled and the area around the upper left-hand corner is a bit more strongly shaded. The implication of this finding is that the most diverse regions tend to have extremely high $LQs$ in the least ubiquitous industries and lower $LQs$ in the most ubiquitous ones. Note that for the purposes of visualization we top-code location quotients at 10. In the case of $WM_{r,i}$, we see no triangular structure at all - in contrast to the results by \cite{Cristelli2013} on country exports. A reason for this non-triangularity might be that the regions with the overall highest share in employment across all industries are simply not the ones which boast the number of industries that attract most of the employment - and vice versa.  

We cannot make a clear distinction based on the visualizations of $BM_{r,i}$, $CM_{r,i}$, $RLQ_{r,i}$, and $Presence_{r,i}$. Comparing the four indices, we see that the correlation coefficient of the $ECIs$ based on the respective input matrices is very high: All of the coefficients lie between 0.966 and 0.990. Interestingly, the correlations for $FI$ are lower (between 0.841 and 0.941).  There overall similarity leaves us some room for decision. Because of the theoretical reasoning outlined above, we opt for $CM_{r,i}$ and $Presence_{r,i}$ and stick to the former as our main input matrix. Results for the other input matrices can be found in the Appendix \ref{appendix:InpMatr:all}. 

\subsubsection{The role of local industries}
A second question when constructing the complexity measure for regions is whether to include local industries or just traded ones. Local industries are those that meet the needs of people living in their region, while traded industries are primarily aimed at producing exports to other regions. Because most previous work on economic complexity is based on international trade data, it is limited to traded industries by necessity. We do not face this data constraint, but the question remains as to whether local industries logically fit with the complexity analysis. 

The concern about including local industries in the complexity score is that they play a fundamentally different role in the regional economy from that of traded industries. Many theorists have noted the importance of exports in bringing money into a region from the outside \citep{Parr1973, Andrews1953}. The money earned by these export industries in turn supports local industries, which form the bulk of employment \citep{Porter2003a, Moretti2010}. For the purposes of measuring regional economic performance, it may thus make sense to focus only on the export industries believed to drive economic growth. 

However, there are a number of counterarguments to this reasoning.  First of all, local services can vary in complexity and ubiquity just like anything else. Consider the case of restaurants - perhaps the quintessential local service. Although restaurants exist within every region of the country, a big city like New York has a much wider variety of restaurants than a small town. This diversity of restaurant options has even been proposed as a source of competitive advantage for cities \citep{Clark2002, Florida2002}. 

An example at the industry level is leather goods and luggage stores (NAICS 448320). These are local retail establishments serving local areas, but they are present in only 136 MSAs, compared to a median of 160 for traded industries and 656 for other local ones. People who live in places with leather goods stores are atypical and have a better portfolio of consumer choices available than those who do not. It is also possible that the capabilities that support leather stores could be applied in spin-off industries, such as the development of new ``smart luggage''.  

\begin{figure}[!h]
	\centering
	\subfloat[Traded Industries]{{\includegraphics[width=6cm]{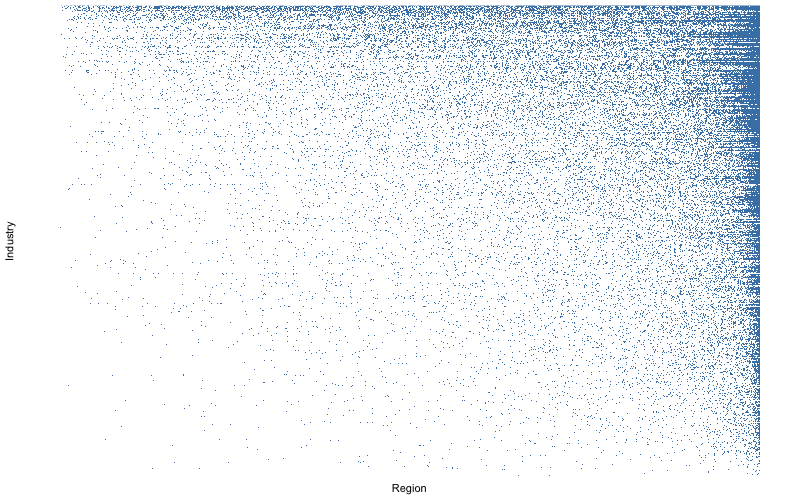}}}%
	\qquad
	\subfloat[Local Industries]{{\includegraphics[width=6cm]{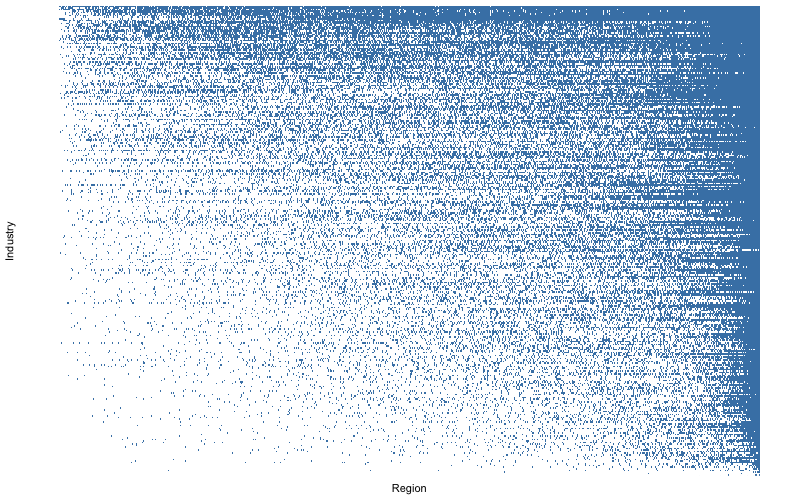}}}%
	\caption{Input matrices: Ubiquity vs. Diversity ($CM_{r,i}$, local vs. traded industries, 2015)}%
	\label{Fig2}%
	\source{Own calculations, \citep{Delgado2016}, and CBP.}
\end{figure}

Looking at local industries collectively, it turns out that they have a similar spatial distribution to traded ones. Figure \ref{Fig2} reproduces the triangular graphs based on $CM_{r,i}$ separately for traded and local industries, as defined by \cite{Delgado2016}. We can see that the overall triangular structure that motivates the complexity measure is present in both, although the triangle is much more filled and has a somewhat convex hypotenuse in the case of local industries - vice versa in the case of traded industries. Given the importance of local industries and their comparable geographic distributions, we opt to use both traded and local industries. 

\subsubsection{ECI vs FI}
\begin{figure}[!h]
	\centering
	\subfloat[$ECI$ vs. $FI$]{{\includegraphics[width=6cm]{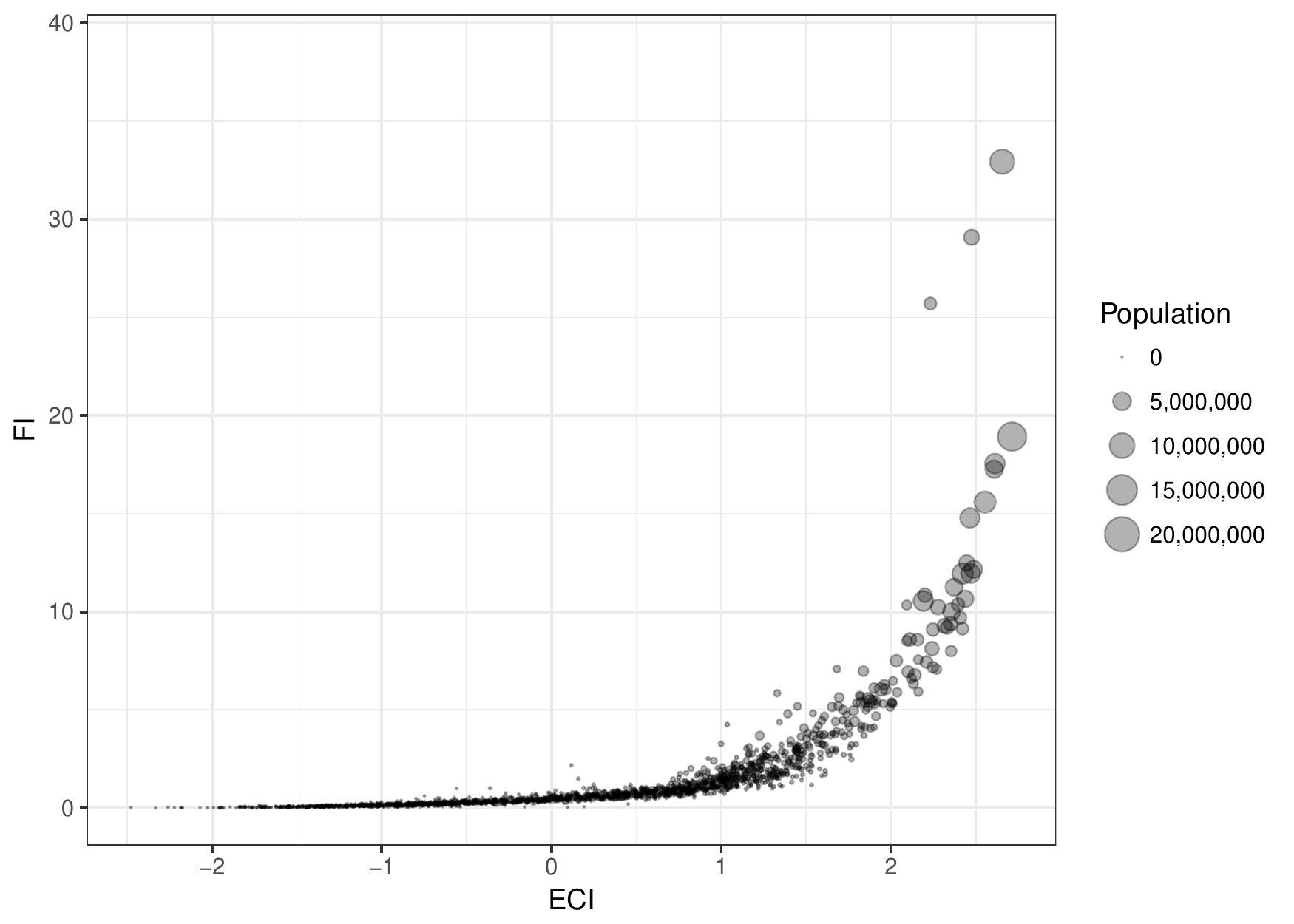}}}%
	\qquad
	\subfloat[$ECI$ vs. logged $FI$]{{\includegraphics[width=6cm]{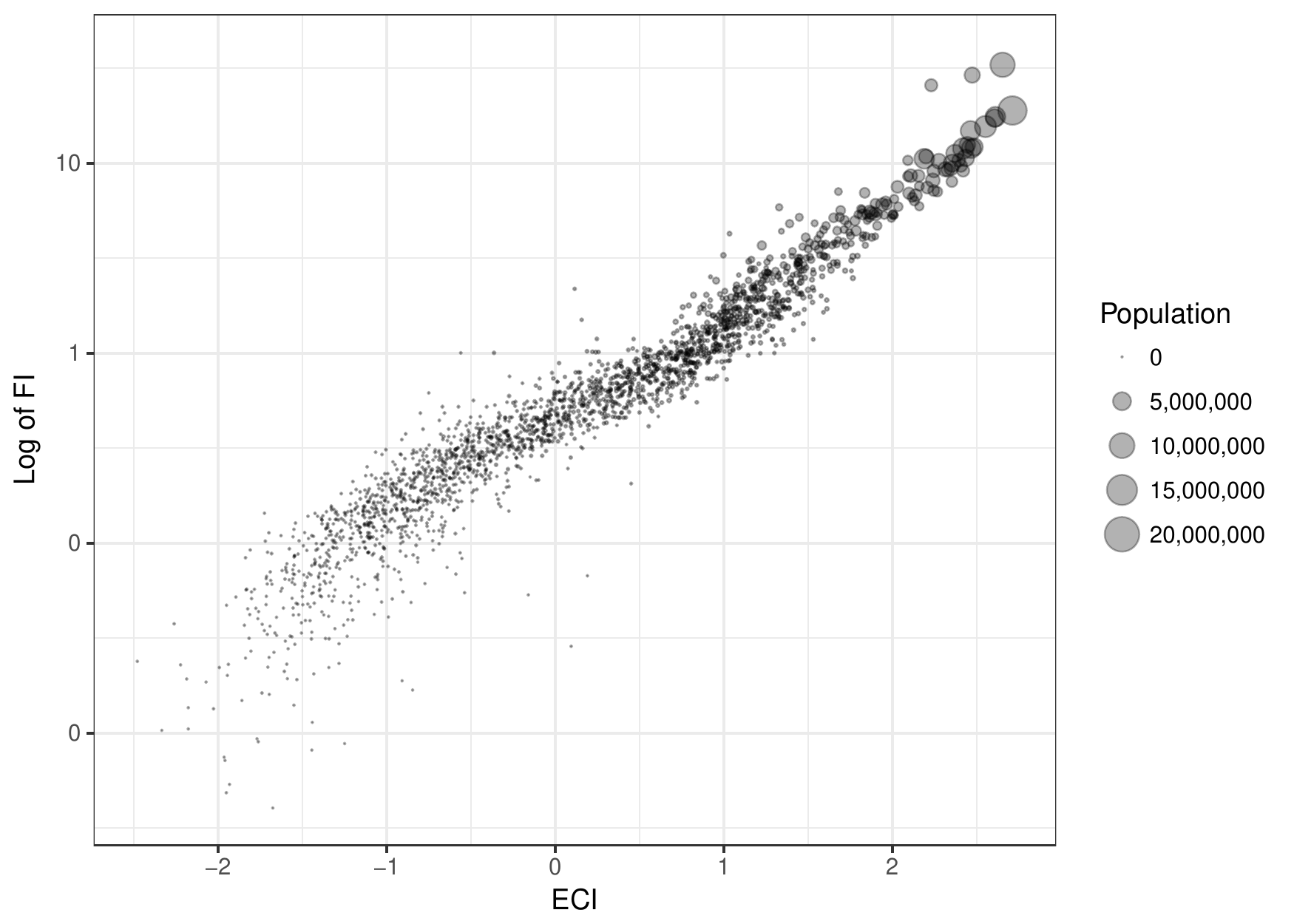}}}%
	\caption{Scatterplots $ECI$ vs. $FI$ ($CM_{r,i}$, NAICS, 2015)}%
	\label{Fig3}%
	\source{Own calculations and CBP.}
\end{figure}

There has been a great deal of discussion in recent literature about the relative merits of the $ECI$ and the $FI$ \citep{Albeaik2017b, Albeaik2017, Gabrielli2017, Pietronero2017a}. However, for regions of the United States they are extremely highly correlated. Building the correlation of $ECI$ and $FI$ based on $CM_{r,i}$, we see that they have a correlation of 0.60 (see Fig \ref{Fig3}a). The correlation is even stronger if we take the log of $FI$, rising to 0.95 (Fig \ref{Fig3}b). Thus, it appears that the primary difference between the $ECI$ and the $FI$, at least as computed on these data, is one of scaling rather than a fundamental difference in concept.  Because it produces a less skewed distribution of values, we choose to focus on $ECI$ here, though we reiterate that both measures appear to be capturing the same underlying construct with remarkable consistency. In the remainder of the paper we report results using the $ECI$. Results using the $FI$ are shown in the Appendix \ref{appendix:Fitness:all}. 

\section{Results}

\subsection{Descriptive Results}
\subsubsection{Zooming into industries}

In Appendix \ref{TabPresenceCM}, we present a list of the 20 most and 20 least complex industries according to the ECI algorithm and the $CM_{r,i}$ input matrix. We expect them not to be driven by exogenous outliers (e.g. NAICS code 211111 Crude Petroleum and Natural Gas Extraction should not fare as a top complexity industry as it is driven by natural resource distribution rather than by capabilities). Furthermore, despite our considerations from above we would expect traded industries to be overrepresented among the top industries given that they have a higher probability of being the source of meaningful capabilities for regional development.

The expectation regarding traded and local industries appears to hold. On average, local industries have a complexity value of -0.639 whereas traded ones have an average value of 0.287 (note that complexity values are normalized to have mean 0 and variance 1). Of the 20 highest complexity industries, only two are local ones. Those two are both subsumed under the industry group 4851 ``Urban Transit Systems.'' This points to the underrepresentation of (public) transit systems among many of the smaller regional units included in our analysis. The two most complex industries are both finance related. One could argue about the merit of having Central Banking as the second most complex industry - after all the headquarters of the 12 Federal Reserve Districts are determined by the Federal Reserve System and not allocated by an economic process of regional selection. However, although those 12 regions are pre-selected, having a regional Fed bank undoubtedly leads to the allocation of highly specialized capabilities within a region. The remainder of the 20 most complex industries are largely filled with advanced manufacturing. Turning to the bottom of the table, the 20 least complex industries are dominated by local industries and resource extraction, in line with our expectations from above.

However, there are examples of complex local industries, speaking more to the idea of local industries being built on complex capabilities. Coming back to the example of 448320 ``Leather Goods and Luggage Stores,'' not only are these stores uncommon, they are also only present in regions with a special set of capabilities. Hence their placement in the 100 most complex industries - among six other local industries on that list.

\subsubsection{Mapping the most complex regions}
\begin{figure}[!h]
	\centering
	\includegraphics[width=15cm]{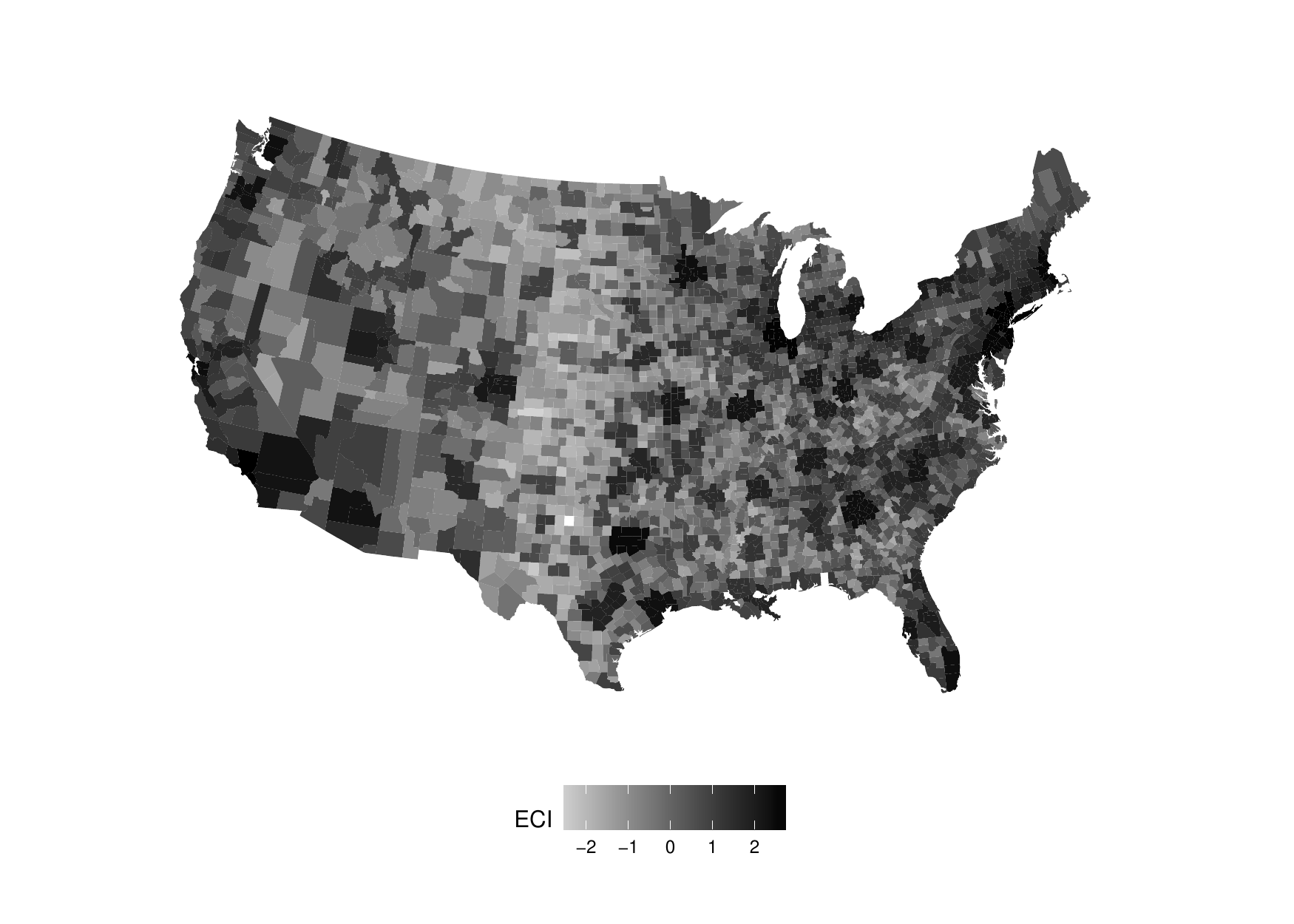}
	\caption{Map: ECI ($CM_{r,i}$, NAICS, 2015)}%
	\label{Fig4}%
	\source{Own calculations and CBP.}
\end{figure}

Fig \ref{Fig4} maps regions by economic complexity. The most complex regions are the metropolitan regions of Los Angeles, New York, Chicago, Philadelphia, and Boston. These are large cities with very diverse economies, producing all manner of goods and services. The least complex regions tend to be rural counties, most notably those in the Great Plains. These areas' employment is centered on agriculture. But those industries are also found in many more diverse regions, so they are determined to have low complexity.\footnote{It may also be the case that complexity estimates for the Great Plains are artificially low because agricultural employment is underreported in the CBP data. However, agriculture is found in all parts of the country, so we expect that if the full data were available it would show similar patterns.} The dominance of very large cities and the strong correlation between complexity and population stands in some contrast to the results at the country level, where many relatively small countries rank quite high. This may be because there is less overall variation in the level of development within the US than there is across countries, which means that the diversity of products a region can make is constrained more by population than by technology. Alternately, the free movement of labor and capital within the country may be responsible.\footnote{Both indicators of economic complexity have been shown to be biased towards more developed countries on the cross-country level \citep{Morrison2017}. The dominance of larger cities might therefore also be explained by the bias of the $ECI$ and the $FI$ towards regions with higher levels of development.}

The level of variation is relatively high, with regions contrasted starkly against each other and the overall distribution being normally distributed with short tails (e.g. no regions  1.5 times above/below the interquartile range). In terms of broader regions, the Midwest has the lowest (unweighted) average complexity (-0.140), followed in increasing order by the South, the West and the Northeast (1.04). But the larger difference is by population size, between Metropolitan Statistical Areas (average $ECI$ = 1.390), Micropolitan Statistical Areas (0.560), and non-metropolitan counties (-0.654). 

\subsection{Regression Results}
The final part of our research is testing for the power of complexity in explaining variation and changes regional economic outcomes. We use two outcome variables: population and per capita income. We consider the bivariate relationships between $ECI$ and these outcome measures as well as the relationship after controlling for a range of economic, social, and institutional variables.  

We begin by testing whether the $ECI$ can explain cross-sectional variation in regional economic outcomes. Results for 2007 and 2015 are presented in Appendix \ref{appendix:MainReg:CrossReg}. 

As shown in the four tables, higher complexity regions have a significantly higher per capita income as well as a higher population. This pattern is stable across years and after including controls.\footnote{Only in 2015 when including all controls in the regression of income per capita, ECI looses significance.} For example, a region with a one-unit higher $ECI$ in 2007 had a per capita income 2,275 USD higher and  1,300,000 more inhabitants than a comparable region of lower complexity. This is in line with the standard literature on economic complexity \citep{Hidalgo2009, Tacchella2012} on the cross-country level. Cross-regional as well as cross-country differences in economic output can be explained by differences in economic complexity. 

Turning to regressions of changes over a period, we use $ECI$ (and our other controls) at the start of a given period as regressor for changes in population and per capita income during that period. We consider two periods: 2007-2009 (as the period of the crisis) and 2010-2015 (the after-crisis period). 

In Table \ref{Tab2}, we see that regions with a higher $ECI$ underwent a decrease in per capita income during the Great Recession, implying that higher complexity places got hit harder by the effects of the Financial Crisis. This finding is significant independent of whether and which controls we include (columns 2 to 5). Apparently, more complex regions with their more refined and interlinked web of capabilities have been more susceptible to this shock. While there is some indication that higher complexity regions had a rebound in income (see Appendix \ref{appendix:MainReg:PeriodRegressions}), this finding is not significant across all controls (there is some susceptibility to our vector of sociodemographic controls). In the same vein, we see that the regions with higher complexity were also the ones that saw a bigger growth in population during both periods. But again, results are not stable.

\begin{table}[!htbp] \centering 
  \caption{Period Regression: Income per Capita (in Pct-Change) between 2007 and 2009 (NAICS, $CM_{r,i}$, ECI)} 
  \label{Tab2} 
\footnotesize 
\begin{tabular}{@{\extracolsep{5pt}}lccccc} 
\\[-1.8ex]\hline 
\hline \\[-1.8ex] 
 & \multicolumn{5}{c}{\textit{Dependent variable:}} \\ 
\cline{2-6} 
\\[-1.8ex] & \multicolumn{5}{c}{Income per Capita (in Pct-Change) Change between 2007 and 2009} \\ 
\\[-1.8ex] & (1) & (2) & (3) & (4) & (5)\\ 
\hline \\[-1.8ex] 
 ECI & $-$0.010$^{***}$ & $-$0.010$^{***}$ & $-$0.008$^{**}$ & $-$0.012$^{***}$ & $-$0.008$^{**}$ \\ 
  & (0.003) & (0.003) & (0.003) & (0.003) & (0.003) \\ 
  & & & & & \\ 
 Income per Capita in 2007 & $-$0.00000$^{***}$ & $-$0.00000$^{***}$ & $-$0.00000$^{**}$ & $-$0.00000$^{***}$ & $-$0.00000$^{*}$ \\ 
  & (0.00000) & (0.00000) & (0.00000) & (0.00000) & (0.00000) \\ 
  & & & & & \\ 
 Unemployment &  & 0.001 &  &  & 0.001 \\ 
  &  & (0.001) &  &  & (0.001) \\ 
  & & & & & \\ 
 ManuShare &  & 0.0001 &  &  & $-$0.0002 \\ 
  &  & (0.0002) &  &  & (0.0002) \\ 
  & & & & & \\ 
 Patent &  & $-$0.00000 &  &  & 0.00000 \\ 
  &  & (0.00000) &  &  & (0.00000) \\ 
  & & & & & \\ 
 Population &  &  & 0.000 &  & $-$0.000 \\ 
  &  &  & (0.000) &  & (0.000) \\ 
  & & & & & \\ 
 MedianAge &  &  & $-$0.001$^{***}$ &  & $-$0.001$^{***}$ \\ 
  &  &  & (0.0003) &  & (0.0004) \\ 
  & & & & & \\ 
 Education &  &  & $-$0.0002 &  & $-$0.0003 \\ 
  &  &  & (0.0002) &  & (0.0002) \\ 
  & & & & & \\ 
 BlackShare &  &  & $-$0.0002 &  & $-$0.00002 \\ 
  &  &  & (0.0001) &  & (0.0002) \\ 
  & & & & & \\ 
 Foreign &  &  & $-$0.001$^{***}$ &  & $-$0.001$^{***}$ \\ 
  &  &  & (0.0002) &  & (0.0002) \\ 
  & & & & & \\ 
 UnionCoverage &  &  &  & 0.002$^{***}$ & 0.001$^{***}$ \\ 
  &  &  &  & (0.0003) & (0.0003) \\ 
  & & & & & \\ 
 MinimumWage &  &  &  & $-$0.005$^{***}$ & $-$0.002 \\ 
  &  &  &  & (0.002) & (0.002) \\ 
  & & & & & \\ 
 Vote &  &  &  & $-$0.00005 & $-$0.0002 \\ 
  &  &  &  & (0.0002) & (0.0002) \\ 
  & & & & & \\ 
 Constant & 0.032$^{***}$ & 0.024$^{***}$ & 0.067$^{***}$ & 0.048$^{***}$ & 0.073$^{***}$ \\ 
  & (0.005) & (0.009) & (0.013) & (0.011) & (0.016) \\ 
  & & & & & \\ 
\hline \\[-1.8ex] 
Observations & 292 & 292 & 292 & 292 & 292 \\ 
R$^{2}$ & 0.216 & 0.220 & 0.272 & 0.302 & 0.350 \\ 
Adjusted R$^{2}$ & 0.210 & 0.206 & 0.254 & 0.290 & 0.319 \\ 
Residual Std. Error & 0.021 & 0.021 & 0.020 & 0.020 & 0.019 \\ 
F Statistic & 39.708$^{***}$ & 16.122$^{***}$ & 15.172$^{***}$ & 24.784$^{***}$ & 11.498$^{***}$ \\ 
\hline 
\hline \\[-1.8ex] 
\textit{Note:}  & \multicolumn{5}{r}{$^{*}$p$<$0.1; $^{**}$p$<$0.05; $^{***}$p$<$0.01} \\ 
\end{tabular} 
\end{table}

A parallel to the result of the effect of $ECI$ on per capita income can be found in research on resilience and system tightness. Using data on US urban employment during the Great Recession, \cite{Shutters2015a} found that regions with lower tightness (degree of connection of labor in the industry network) were more resilient during the crisis. Lower tightness corresponds to a region not having its capabilities concentrated within single sectors but with nodes spread-out throughout the industrial network and with only light connections between its industries, thereby creating an ever-thicker web. In contrast, complex capabilities are thought to be primarily developing when regions branch into industries that are technologically related to the preexisting industries \citep{Hidalgo2018}. Based on this logic, higher complexity regions may have a tighter network and thus be less resilient. 

Finally, we consider how change in $ECI$ compares to changes in economic outcomes using a panel regression. Here, we lag the $ECI$ to achieve quasi exogeneity. All our panel regressions use time fixed effects, dummies for every region and robust standard errors.

\begin{table}[!htbp] \centering 
  \caption{Panel Regression: Income per Capita (in 1,000 USD) 2015 (NAICS, $CM_{r,i}$, ECI)} 
  \label{Tab3} 
\footnotesize 
\begin{tabular}{@{\extracolsep{5pt}}lccccc} 
\\[-1.8ex]\hline 
\hline \\[-1.8ex] 
 & \multicolumn{5}{c}{\textit{Dependent variable:}} \\ 
\cline{2-6} 
\\[-1.8ex] & \multicolumn{5}{c}{Income per Capita (in 1,000 USD) 2015} \\ 
\\[-1.8ex] & (1) & (2) & (3) & (4) & (5)\\ 
\hline \\[-1.8ex] 
 l.ECI & $-$2.019$^{***}$ & $-$1.561$^{**}$ & $-$2.125$^{***}$ & $-$1.904$^{***}$ & $-$1.772$^{***}$ \\ 
  & (0.696) & (0.697) & (0.523) & (0.700) & (0.529) \\ 
  & & & & & \\ 
 Unemployment &  & $-$0.147$^{***}$ &  &  & $-$0.140$^{***}$ \\ 
  &  & (0.022) &  &  & (0.020) \\ 
  & & & & & \\ 
 ManuShare &  & 0.042 &  &  & 0.044 \\ 
  &  & (0.037) &  &  & (0.037) \\ 
  & & & & & \\ 
 Patent &  & 0.001$^{***}$ &  &  & 0.001$^{***}$ \\ 
  &  & (0.0003) &  &  & (0.0003) \\ 
  & & & & & \\ 
 Population &  &  & $-$0.00000 &  & $-$0.00000$^{**}$ \\ 
  &  &  & (0.00000) &  & (0.00000) \\ 
  & & & & & \\ 
 MedianAge &  &  & $-$0.035 &  & $-$0.007 \\ 
  &  &  & (0.122) &  & (0.119) \\ 
  & & & & & \\ 
 Education &  &  & 0.030 &  & 0.027 \\ 
  &  &  & (0.025) &  & (0.025) \\ 
  & & & & & \\ 
 BlackShare &  &  & $-$0.249$^{***}$ &  & $-$0.214$^{***}$ \\ 
  &  &  & (0.080) &  & (0.075) \\ 
  & & & & & \\ 
 Foreign &  &  & 0.262$^{**}$ &  & 0.251$^{*}$ \\ 
  &  &  & (0.133) &  & (0.133) \\ 
  & & & & & \\ 
 UnionCoverage &  &  &  & 0.042 & 0.072 \\ 
  &  &  &  & (0.059) & (0.059) \\ 
  & & & & & \\ 
 MinimumWage &  &  &  & 0.037 & 0.005 \\ 
  &  &  &  & (0.170) & (0.159) \\ 
  & & & & & \\ 
 Vote &  &  &  & $-$0.053$^{***}$ & $-$0.045$^{***}$ \\ 
  &  &  &  & (0.011) & (0.010) \\ 
  & & & & & \\ 
\hline \\[-1.8ex] 
Observations & 2,715 & 2,715 & 2,715 & 2,715 & 2,715 \\ 
R$^{2}$ & 0.007 & 0.041 & 0.028 & 0.016 & 0.071 \\ 
Adjusted R$^{2}$ & $-$0.126 & $-$0.089 & $-$0.104 & $-$0.117 & $-$0.058 \\ 
F Statistic & 16.560$^{***}$ & 25.243$^{***}$ & 11.578$^{***}$ & 9.623$^{***}$ & 15.162$^{***}$ \\ 
\hline 
\hline \\[-1.8ex] 
\textit{Note:}  & \multicolumn{5}{r}{$^{*}$p$<$0.1; $^{**}$p$<$0.05; $^{***}$p$<$0.01} \\ 
\end{tabular} 
\end{table}

\begin{table}[!htbp] \centering 
  \caption{Panel Regression: Population (in 1,000) 2015 (NAICS, $CM_{r,i}$, ECI)} 
  \label{Tab4} 
\footnotesize 
\begin{tabular}{@{\extracolsep{5pt}}lccccc} 
\\[-1.8ex]\hline 
\hline \\[-1.8ex] 
 & \multicolumn{5}{c}{\textit{Dependent variable:}} \\ 
\cline{2-6} 
\\[-1.8ex] & \multicolumn{5}{c}{Population (in 1,000) 2015} \\ 
\\[-1.8ex] & (1) & (2) & (3) & (4) & (5)\\ 
\hline \\[-1.8ex] 
 l.ECI & $-$71.131$^{***}$ & $-$36.561$^{**}$ & $-$63.342$^{***}$ & $-$72.122$^{***}$ & $-$36.077$^{**}$ \\ 
  & (25.007) & (16.838) & (24.275) & (24.579) & (16.459) \\ 
  & & & & & \\ 
 Unemployment &  & 0.231 &  &  & 0.102 \\ 
  &  & (0.262) &  &  & (0.256) \\ 
  & & & & & \\ 
 ManuShare &  & 0.351 &  &  & 0.456 \\ 
  &  & (0.609) &  &  & (0.617) \\ 
  & & & & & \\ 
 Patent &  & 0.097$^{***}$ &  &  & 0.097$^{***}$ \\ 
  &  & (0.031) &  &  & (0.031) \\ 
  & & & & & \\ 
 Income per Capita &  & $-$1.707 &  &  & $-$1.690$^{*}$ \\ 
  &  & (1.064) &  &  & (0.976) \\ 
  & & & & & \\ 
 MedianAge &  &  & $-$3.647$^{**}$ &  & $-$0.734 \\ 
  &  &  & (1.619) &  & (0.904) \\ 
  & & & & & \\ 
 Education &  &  & 1.040$^{**}$ &  & 0.413 \\ 
  &  &  & (0.453) &  & (0.288) \\ 
  & & & & & \\ 
 BlackShare &  &  & 0.965 &  & 2.474 \\ 
  &  &  & (3.164) &  & (2.376) \\ 
  & & & & & \\ 
 Foreign &  &  & 2.332 &  & 0.720 \\ 
  &  &  & (2.011) &  & (1.473) \\ 
  & & & & & \\ 
 UnionCoverage &  &  &  & 6.081$^{***}$ & 5.071$^{***}$ \\ 
  &  &  &  & (1.889) & (1.471) \\ 
  & & & & & \\ 
 MinimumWage &  &  &  & 7.073$^{*}$ & 7.834$^{**}$ \\ 
  &  &  &  & (4.153) & (3.526) \\ 
  & & & & & \\ 
 Vote &  &  &  & $-$0.068 & 0.498$^{*}$ \\ 
  &  &  &  & (0.412) & (0.274) \\ 
  & & & & & \\ 
\hline \\[-1.8ex] 
Observations & 2,715 & 2,715 & 2,715 & 2,715 & 2,715 \\ 
R$^{2}$ & 0.011 & 0.337 & 0.019 & 0.027 & 0.352 \\ 
Adjusted R$^{2}$ & $-$0.121 & 0.247 & $-$0.114 & $-$0.104 & 0.262 \\ 
F Statistic & 27.787$^{***}$ & 242.795$^{***}$ & 9.457$^{***}$ & 16.711$^{***}$ & 107.954$^{***}$ \\ 
\hline 
\hline \\[-1.8ex] 
\textit{Note:}  & \multicolumn{5}{r}{$^{*}$p$<$0.1; $^{**}$p$<$0.05; $^{***}$p$<$0.01} \\ 
\end{tabular} 
\end{table}

Interestingly, both in the regression of income per capita (see Table \ref{Tab3}) as well as in the regression of population (see Table \ref{Tab4}), lagged ECI emerges as a significant negative variable across all regressions. For example, a one-unit deviation above the mean is associated with a decrease of 36,000 inhabitants and 1,772 USD below the mean of the average region when we employ all of our controls. Employing simply $l.ECI$ as a regressor, this number rises to 71,000 and 2,000 USD respectively. 

However, looking at the results based on $Presence_{r,i}$ in Appendix \ref{appendix:MainReg:PeriodRegressions}, we find a different result. Whereas results for our period and cross-sectional regressions and for the panel regression of population are similar to those using $CM_{r,i}$, lagged ECI computed based on $Presence_{r,i}$ is not a stable significant regressor for income per capita.  

\section{Discussion}
Using industry employment data, this paper has applied the ideas and methods of economic complexity to the context of US regions between 2007 and 2015. After modifying part of the underlying methodology and adding scrutiny to the underlying assumptions of complexity in a regional context, we show that economic complexity is well suited to analyzing the productive structure of US regions. Specifically, the general triangular pattern of the region-industry matrix that has been observed on the country level also applies to to the US regions. This highlights the primacy of regions as a focus of economic geography - although they cannot control their borders, they exhibit some of the same patterns as countries. 

We have also argued and shown tentatively that the inclusion of local sectors is valid and strengthens the analytic power of the complexity indicators. Furthermore, we introduced two new input matrices, $Presence_{r,i}$ and $CM_{r,i}$, which circumvent many of the standard problems of an industry-product matrix, such as sublinear scaling with city size, by taking into account (almost) every industry that is present in a region. According to our results, $Presence_{r,i}$ and $CM_{r,i}$ are well suited to function as input matrices for indicators of economic complexity. In terms of indicators, we argue that differences between $ECI$ and $FI$ are less dramatic than they seem, at least in this context. 

Our empirical analysis revealed that the largest cities have the highest complexity and that metropolitan areas and the Northeast of the US have a higher complexity than non-metropolitan areas and other US regions respectively. Traded industries dominate the list of most complex industries, but local ones can also have high complexity values. In general, higher $ECI$ regions have a significantly higher population and income per capita. Economic complexity is  predictive of other economic outcomes - especially decreases in population. When computing ECI based on $CM_{r,i}$, it is also a significant negative predictor of within changes in income per capita. This result can however not be found when ECI is recomputed based on $Presence_{r,i}$. 

An explanation for the negative effect of ECI in the income per capita panel regression might stem from our finding that higher complexity in regions in 2007 predicted a bigger decrease in income per capita during the financial crisis and that this contraction was not significantly recovered in high complexity areas afterwards. This finding can be linked to literature on resilience and we thus suspect that higher complexity regions may be more susceptible to economic shocks in terms of their income. But this should be corroborated through future research.

One possible cause of the instable results (e.g. dependence on the input matrix) of the income per capita panel regression in the US context - which differ from findings at the international level - might be the lack of barriers to migration, which in turn creates a safety valve by which migration equilibrates wage differentials across regions. 

Based on our study, we believe that complexity measures at the subnational level form a good proxy for economic performance or capacity. We are more cautious of complexity as a predictor of future growth, and more research is needed on to deal with the insecurities named above.
 
Future research should work on these and related issues more closely. Despite our extensive list of controls, much remains to be learned about regional economic complexity and its correlates. We cannot exclude omitted variable bias/endogeneity at this moment. In a similar vein, it might be useful to construct an employment network according to more refined (i.e. independently developed) terms instead of relying on the NAICS (and the BCD) as manna from heaven. This would not only strengthen the credibility of our results but also make clearer what actually constitutes relatedness - in the spirit of \cite{Wixe2017} - in the US/regional setting. Finally, economic complexity should be compared to other measures of productive structure, such as related and unrelated variety and other indicators of entropy and diversity, to get a more holistic impression of regional economies and their trajectories.

\newpage

\bibliographystyle{aea}


\newpage

\appendix

\section{Formulae of Economic Complexity}
\label{appendix:Formulae}

The $ECI$ was the original indicator and is a combination of two different pieces of information: The ubiquity of an industry across different regional units and the diversity of a regional unit. The metric corrects both values in a set of theoretically infinite iterative linear equations for each other. A higher regional $ECI$ value assigns a higher complexity to its productive structure.

Thereby, the diversity of a regional unit is defined as the number of industries for which it has an RCA:

\begin{equation}
Diversity = k_{r,0} = \sum_{i=1}^{n}M_{r,i}
\end{equation}

with $n$ being the number of industries. And accordingly, ubiquity of an industry is defined as the number of regional units it has an RCA in:

\begin{equation}
Ubiquity = k_{i,0} = \sum_{r=1}^{m}M_{r,i}
\end{equation}

with $m$ being the number of regions. To correct a regional unit's diversity by ubiquity and by average diversity and so forth, we calculate the average ubiquity of its RCA industries, the average diversity of the regional units also having an RCA in the same industries and so forth. This can be expressed as:

\begin{equation}
k_{r,N} = \sum_{r'}k_{r',N-2} \sum\frac{M_{r,i}M_{r',i}}{k_{r,0}k_{i,0}} = \sum_{r'}\tilde{M}_{rr'}k_{r',N-2} \: | \: with M_{rr'} = \sum_{r}\frac{M_{r,i}M_{r',i}}{k_{r,0}k_{i,0}}
\end{equation}

The $ECI$ is thereby defined as:

\begin{equation}
ECI = \frac{ K_{r}-\langle K_{r} \rangle }{std(K_{r})}
\end{equation}

where $K$ is the second largest eigenvector (the largest eigenvector is just a vector of 1s) and $〈K_{r}$〉 is the mean of the respective eigenvector.

The $FI$ is closely related to the $ECI$ and differs from the latter only in few albeit relevant aspects. The first step, constructing the RCAs, thereby determining diversity and ubiquity, is exactly the same one as for $ECI$. In the second step, again the ubiquity of industries and diversity of regions are iteratively used to correct each other, such that an industry is labeled more complex if it tends to be found in more complex regions, and vice versa. As opposed to the $ECI$, the $FI$ binds the complexity of industries by the fitness of the least competitive regional units which export them in order to avoid penalizing regional units with high $LQs$ in both high and low value industries (which is what the $ECI$ method does). This results in the following set of non-linear equations

\begin{equation}
\left\{
\begin{array}{ll}
\tilde{F}_{r}^{(n)}= \sum_{i} M_{r,i}Q_{i}^{n-1} \\
\tilde{Q}_{i}^{(n)} = \frac{ 1 }{ \sum_{r}M_{r,i} \frac{1}{F_{C}^{n-1} }}\\
\end{array}
\right.
\rightarrow
\left\{
\begin{array}{ll}
F_{r}^{(n)} = \frac{ \tilde{F}_{r}^{(n)} }{ \langle\tilde{F}_{r}^{(n)} \rangle_{r} } \\
Q_{i}^{(n)} = \frac{ \tilde{Q}_{i}^{(n)} }{ \langle \tilde{Q}_{i}^{(n)} \rangle_{i} }
\end{array}
\right.
\end{equation}

with $F_{r}^{(n)}$ the fitness of a regional unit (proportional to the sum of industries weighted by their complexity $Q_{i}^{(n)}$) and $Q_{i}^{(n)}$ the industry complexity (the inverse proportional to the number of regional units possessing it corrected/weighted by the fitness of the regional units; the higher the fitness the smaller the weight) being normalized in each new step.

\section{Stylized Facts on Economic Complexity on Different Digit Levels}   
\begin{figure}[!h]
	\centering
	\includegraphics[width=10cm]{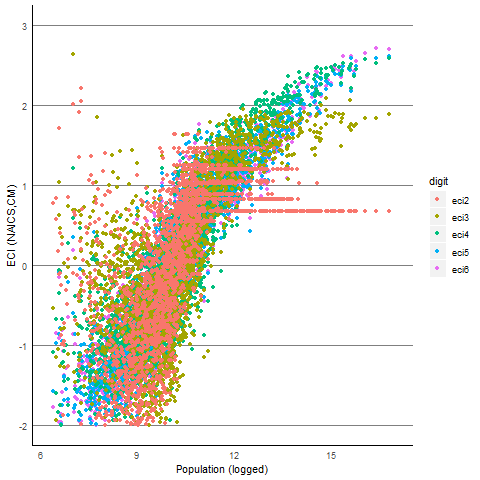}%
	\caption{Scatterplots Different Digit ECI ($CM_{r,i}$, NAICS, 2015)}%
	\source{Own calculations and CBP.}
	\label{appendix:DigitsFigure}
\end{figure}

The correlation plot in Figure \ref{appendix:DigitsFigure} shows the relative strong consistency of ECI values across the 4- to 6-digit level. However, below the 4-digit level, the general correlation pattern falters - especially the 2- and the 3-digit level showing a very different correlation between Population and ECI. 

The correlation coefficients between ECI on the 6- and the equivalent on the 5- and 6-digit level is 0.99 and 0.94 respectively, whereas correlation between the 6-digit and the 3-digit ECI is only 0.79. Thus, we consider the three most disaggregated digit levels roughly equivalent and stick to the highest resolution (6-digit) in order to make use of as much information as possible.  
 
\label{appendix:DifferentDigits}

\FloatBarrier
\newpage

\section{Data Sources}
\label{appendix:Data Sources}
\begin{small}
\begin{longtable}{|p{2cm}|p{4.5cm}|p{3cm}|p{3cm}|}

	KILLED & LINE!!!! \kill
	\caption{Data and Data Sources}\\
	\hline
	\textbf{Variable Names} & \textbf{Description} & \textbf{Source} & \textbf{Additional \newline Information}\\
	\hline
	\endhead
	\multicolumn{4}{|c|}%
	{\textit{Dependent Variables}}\\
	\hline
	\textsc{Income per Capita} & Income per Capita & BEA Regional Economic Accounts &  XX 
	\\
	\hline
	\textsc{Population} & Number of people living in a region & BEA Regional Economic Accounts &  
	\\
	\hline
	\multicolumn{4}{|c|}%
	{\textit{Measures of Productive Structures}}\\
	\hline
	\textsc{ECI} & Economic Complexity Index calculated either based on NAICS or BCD. For both on 6digit level & County Business Patterns &  Variables NAICS: \newline NAICS.ECI \newline Variables BCD: \newline BCD.BCD
	\\
	\hline
	\textsc{FI} & Fitness Index calculated either based on NAICS or BCD. For both on 6digit level & Community Business Patterns &  Variables NAICS: \newline NAICS.FI \newline Variables BCD: \newline BCD.FI
	\\
	\hline
	\multicolumn{4}{|c|}%
	{\textit{Economic Controls}}\\
	\hline
	\textsc{Unemploy-ment} & Unemployment as a percentage of the total working population & American Community Survey &  
	\\
	\hline
	\textsc{ManuShare} & Share of the employed population working in manufacturing sector & County Business Patterns &  
	\\
	\hline
	\textsc{Patent} & Number of patents filed per year & U.S. Patent and Trademark Office &  
	\\
	\hline
	\multicolumn{4}{|c|}%
	{\textit{Sociodemographic Controls}}\\
	\hline
	\textsc{MedianAge} & Median age of the population living in the region & American Community Survey &  
	\\
	\hline
	\textsc{Education} & Share of people older than 18 with more than a high school degree & American Community Survey &  
	\\
	\hline
	\textsc{Black Share} & Share of the population identifying as being of African American origin (alone or in combination with other races) & American Community Survey &  
	\\
	\hline
	\textsc{Foreign} & Share of foreign born people in the total population & American Community Survey &  
	\\
	\hline
	\multicolumn{4}{|c|}%
	{\textit{Institutional Controls}}\\
	\hline
	\textsc{Union Coverage} & Union coverage at the state level in which the region is mainly situated  & unionstats.com & Only available at the State level.  
	\\
	\hline
	\textsc{Minimum Wage} & Minimum Wage at the state level in which the region is situated. & Department of Labor, State Minimum Wage Rate (via FRED) & On the State level as only in recent years few cities have introduced their own minimum wages  
	\\
	\hline
	\textsc{Vote} & Eligible voting population voter turnout on November elections in even years & electproject.org & The votes from even years are duplicated for the odd years (no November elections) | Always the votes for the highest office (in presidential election years the presidential vote, in other years either governor, senator or congressional vote) \\
	\hline
	
\end{longtable}
\end{small}

\section{Addenda to main text}
\subsection{Cross-Sectional Regressions}
\label{appendix:MainReg:CrossReg}

\begin{table}[!htbp] \centering 
  \caption{Cross-Section Regression: Income per Capita (in 1,000 USD) 2007 (NAICS, $CM_{r,i}$, ECI)} 
  \label{} 
\footnotesize 
\begin{tabular}{@{\extracolsep{5pt}}lccccc} 
\\[-1.8ex]\hline 
\hline \\[-1.8ex] 
 & \multicolumn{5}{c}{\textit{Dependent variable:}} \\ 
\cline{2-6} 
\\[-1.8ex] & \multicolumn{5}{c}{Income per Capita (in 1,000 USD) 2007} \\ 
\\[-1.8ex] & (1) & (2) & (3) & (4) & (5)\\ 
\hline \\[-1.8ex] 
 ECI & 9.462$^{***}$ & 6.818$^{***}$ & 2.081$^{**}$ & 9.287$^{***}$ & 2.275$^{**}$ \\ 
  & (0.930) & (1.018) & (1.044) & (0.928) & (1.104) \\ 
  & & & & & \\ 
 Unemployment &  & $-$1.009$^{***}$ &  &  & $-$0.670$^{***}$ \\ 
  &  & (0.208) &  &  & (0.200) \\ 
  & & & & & \\ 
 ManuShare &  & $-$0.262$^{***}$ &  &  & $-$0.069 \\ 
  &  & (0.054) &  &  & (0.054) \\ 
  & & & & & \\ 
 Patent &  & 0.003$^{***}$ &  &  & 0.002$^{**}$ \\ 
  &  & (0.001) &  &  & (0.001) \\ 
  & & & & & \\ 
 Population &  &  & 0.00000 &  & $-$0.00000 \\ 
  &  &  & (0.00000) &  & (0.00000) \\ 
  & & & & & \\ 
 MedianAge &  &  & 0.848$^{***}$ &  & 0.813$^{***}$ \\ 
  &  &  & (0.096) &  & (0.102) \\ 
  & & & & & \\ 
 Education &  &  & 0.600$^{***}$ &  & 0.467$^{***}$ \\ 
  &  &  & (0.052) &  & (0.063) \\ 
  & & & & & \\ 
 BlackShare &  &  & 0.083$^{*}$ &  & 0.131$^{***}$ \\ 
  &  &  & (0.044) &  & (0.050) \\ 
  & & & & & \\ 
 Foreign &  &  & 0.406$^{***}$ &  & 0.415$^{***}$ \\ 
  &  &  & (0.061) &  & (0.074) \\ 
  & & & & & \\ 
 UnionCoverage &  &  &  & $-$0.172 & 0.055 \\ 
  &  &  &  & (0.109) & (0.090) \\ 
  & & & & & \\ 
 MinimumWage &  &  &  & 2.251$^{***}$ & $-$0.361 \\ 
  &  &  &  & (0.744) & (0.664) \\ 
  & & & & & \\ 
 Vote &  &  &  & 0.038 & 0.108 \\ 
  &  &  &  & (0.080) & (0.070) \\ 
  & & & & & \\ 
 Constant & 28.361$^{***}$ & 41.770$^{***}$ & $-$16.282$^{***}$ & 14.615$^{***}$ & $-$9.046$^{*}$ \\ 
  & (1.303) & (2.077) & (4.091) & (4.593) & (5.148) \\ 
  & & & & & \\ 
\hline \\[-1.8ex] 
Observations & 292 & 292 & 292 & 292 & 292 \\ 
R$^{2}$ & 0.263 & 0.411 & 0.582 & 0.290 & 0.609 \\ 
Adjusted R$^{2}$ & 0.261 & 0.402 & 0.573 & 0.280 & 0.592 \\ 
Residual Std. Error & 8.410 & 7.561 & 6.392 & 8.299 & 6.246 \\ 
F Statistic & 103.520$^{***}$ & 49.978$^{***}$ & 66.024$^{***}$ & 29.277$^{***}$ & 36.201$^{***}$ \\ 
\hline 
\hline \\[-1.8ex] 
\textit{Note:}  & \multicolumn{5}{r}{$^{*}$p$<$0.1; $^{**}$p$<$0.05; $^{***}$p$<$0.01} \\ 
\end{tabular} 
\end{table}

\begin{table}[!htbp] \centering 
  \caption{Cross-Section Regression: Income per Capita (in 1,000 USD) 2015 (NAICS, $CM_{r,i}$, ECI)} 
  \label{} 
\footnotesize 
\begin{tabular}{@{\extracolsep{5pt}}lccccc} 
\\[-1.8ex]\hline 
\hline \\[-1.8ex] 
 & \multicolumn{5}{c}{\textit{Dependent variable:}} \\ 
\cline{2-6} 
\\[-1.8ex] & \multicolumn{5}{c}{Income per Capita (in 1,000 USD) 2015} \\ 
\\[-1.8ex] & (1) & (2) & (3) & (4) & (5)\\ 
\hline \\[-1.8ex] 
 ECI & 9.759$^{***}$ & 5.526$^{***}$ & 2.350$^{*}$ & 9.254$^{***}$ & 1.432 \\ 
  & (1.038) & (1.133) & (1.340) & (1.040) & (1.390) \\ 
  & & & & & \\ 
 Unemployment &  & $-$1.053$^{***}$ &  &  & $-$0.809$^{***}$ \\ 
  &  & (0.205) &  &  & (0.221) \\ 
  & & & & & \\ 
 ManuShare &  & $-$0.263$^{***}$ &  &  & $-$0.063 \\ 
  &  & (0.058) &  &  & (0.064) \\ 
  & & & & & \\ 
 Patent &  & 0.002$^{***}$ &  &  & 0.002$^{***}$ \\ 
  &  & (0.0004) &  &  & (0.0005) \\ 
  & & & & & \\ 
 Population &  &  & 0.00000 &  & $-$0.00000$^{*}$ \\ 
  &  &  & (0.00000) &  & (0.00000) \\ 
  & & & & & \\ 
 MedianAge &  &  & 0.394$^{***}$ &  & 0.374$^{***}$ \\ 
  &  &  & (0.086) &  & (0.086) \\ 
  & & & & & \\ 
 Education &  &  & 0.512$^{***}$ &  & 0.374$^{***}$ \\ 
  &  &  & (0.057) &  & (0.069) \\ 
  & & & & & \\ 
 BlackShare &  &  & $-$0.009 &  & 0.083 \\ 
  &  &  & (0.049) &  & (0.054) \\ 
  & & & & & \\ 
 Foreign &  &  & 0.378$^{***}$ &  & 0.379$^{***}$ \\ 
  &  &  & (0.071) &  & (0.082) \\ 
  & & & & & \\ 
 UnionCoverage &  &  &  & 0.072 & 0.186 \\ 
  &  &  &  & (0.125) & (0.114) \\ 
  & & & & & \\ 
 MinimumWage &  &  &  & 1.638$^{**}$ & $-$0.197 \\ 
  &  &  &  & (0.800) & (0.776) \\ 
  & & & & & \\ 
 Vote &  &  &  & 0.064 & 0.035 \\ 
  &  &  &  & (0.063) & (0.059) \\ 
  & & & & & \\ 
 Constant & 29.473$^{***}$ & 44.822$^{***}$ & 3.002 & 13.771$^{**}$ & 12.978$^{**}$ \\ 
  & (1.450) & (2.418) & (4.067) & (5.863) & (6.211) \\ 
  & & & & & \\ 
\hline \\[-1.8ex] 
Observations & 312 & 312 & 312 & 312 & 312 \\ 
R$^{2}$ & 0.222 & 0.367 & 0.443 & 0.252 & 0.494 \\ 
Adjusted R$^{2}$ & 0.219 & 0.359 & 0.432 & 0.242 & 0.473 \\ 
Residual Std. Error & 8.917 & 8.081 & 7.608 & 8.788 & 7.324 \\ 
F Statistic & 88.426$^{***}$ & 44.539$^{***}$ & 40.371$^{***}$ & 25.789$^{***}$ & 24.290$^{***}$ \\ 
\hline 
\hline \\[-1.8ex] 
\textit{Note:}  & \multicolumn{5}{r}{$^{*}$p$<$0.1; $^{**}$p$<$0.05; $^{***}$p$<$0.01} \\ 
\end{tabular} 
\end{table}

\begin{table}[!htbp] \centering 
  \caption{Cross-Section Regression: Population (in 1,000) 2007 (NAICS, $CM_{r,i}$, ECI)} 
  \label{} 
\footnotesize 
\begin{tabular}{@{\extracolsep{5pt}}lccccc} 
\\[-1.8ex]\hline 
\hline \\[-1.8ex] 
 & \multicolumn{5}{c}{\textit{Dependent variable:}} \\ 
\cline{2-6} 
\\[-1.8ex] & \multicolumn{5}{c}{Population (in 1,000) 2007} \\ 
\\[-1.8ex] & (1) & (2) & (3) & (4) & (5)\\ 
\hline \\[-1.8ex] 
 ECI & 2,004.241$^{***}$ & 1,237.834$^{***}$ & 1,787.039$^{***}$ & 2,004.861$^{***}$ & 1,318.535$^{***}$ \\ 
  & (146.113) & (152.481) & (173.645) & (145.690) & (163.392) \\ 
  & & & & & \\ 
 Unemployment &  & 15.318 &  &  & $-$32.822 \\ 
  &  & (30.176) &  &  & (33.231) \\ 
  & & & & & \\ 
 ManuShare &  & $-$27.896$^{***}$ &  &  & $-$34.011$^{***}$ \\ 
  &  & (7.810) &  &  & (8.541) \\ 
  & & & & & \\ 
 Patent &  & 1.341$^{***}$ &  &  & 1.273$^{***}$ \\ 
  &  & (0.110) &  &  & (0.112) \\ 
  & & & & & \\ 
 Income per Capita &  & $-$24.940$^{***}$ &  &  & $-$13.909 \\ 
  &  & (8.219) &  &  & (9.735) \\ 
  & & & & & \\ 
 MedianAge &  &  & 7.929 &  & 5.140 \\ 
  &  &  & (18.699) &  & (18.455) \\ 
  & & & & & \\ 
 Education &  &  & $-$18.363$^{*}$ &  & $-$39.357$^{***}$ \\ 
  &  &  & (9.971) &  & (10.931) \\ 
  & & & & & \\ 
 BlackShare &  &  & 21.353$^{**}$ &  & 17.409$^{**}$ \\ 
  &  &  & (8.383) &  & (8.173) \\ 
  & & & & & \\ 
 Foreign &  &  & 62.366$^{***}$ &  & 20.094 \\ 
  &  &  & (11.239) &  & (12.627) \\ 
  & & & & & \\ 
 UnionCoverage &  &  &  & 6.059 & 26.529$^{*}$ \\ 
  &  &  &  & (17.142) & (14.555) \\ 
  & & & & & \\ 
 MinimumWage &  &  &  & 149.029 & $-$120.618 \\ 
  &  &  &  & (116.777) & (108.140) \\ 
  & & & & & \\ 
 Vote &  &  &  & $-$40.743$^{***}$ & 2.152 \\ 
  &  &  &  & (12.500) & (11.464) \\ 
  & & & & & \\ 
 Constant & $-$1,979.643$^{***}$ & 52.571 & $-$2,097.640$^{***}$ & $-$1,322.604$^{*}$ & 983.515 \\ 
  & (204.731) & (448.863) & (786.835) & (721.322) & (842.635) \\ 
  & & & & & \\ 
\hline \\[-1.8ex] 
Observations & 292 & 292 & 292 & 292 & 292 \\ 
R$^{2}$ & 0.394 & 0.620 & 0.469 & 0.416 & 0.653 \\ 
Adjusted R$^{2}$ & 0.391 & 0.614 & 0.460 & 0.408 & 0.638 \\ 
Residual Std. Error & 1,321.269 & 1,052.737 & 1,244.636 & 1,303.354 & 1,019.329 \\ 
F Statistic & 188.158$^{***}$ & 93.441$^{***}$ & 50.570$^{***}$ & 51.098$^{***}$ & 43.699$^{***}$ \\ 
\hline 
\hline \\[-1.8ex] 
\textit{Note:}  & \multicolumn{5}{r}{$^{*}$p$<$0.1; $^{**}$p$<$0.05; $^{***}$p$<$0.01} \\ 
\end{tabular} 
\end{table}

\FloatBarrier

\subsection{Period Regressions}

\begin{table}[!htbp] \centering 
  \caption{Period Regression: Income per Capita (in Pct-Change) between 2010 and 2015 (NAICS, $CM_{r,i}$, ECI)} 
  \label{} 
\footnotesize 
\begin{tabular}{@{\extracolsep{5pt}}lccccc} 
\\[-1.8ex]\hline 
\hline \\[-1.8ex] 
 & \multicolumn{5}{c}{\textit{Dependent variable:}} \\ 
\cline{2-6} 
\\[-1.8ex] & \multicolumn{5}{c}{Income per Capita (in Pct-Change) Change between 2010 and 2015} \\ 
\\[-1.8ex] & (1) & (2) & (3) & (4) & (5)\\ 
\hline \\[-1.8ex] 
 ECI & 0.003$^{***}$ & 0.001 & 0.003$^{*}$ & 0.003$^{***}$ & 0.001 \\ 
  & (0.001) & (0.001) & (0.002) & (0.001) & (0.002) \\ 
  & & & & & \\ 
 Income per Capita in 2010 & 0.000 & 0.000 & $-$0.00000 & 0.000 & $-$0.00000 \\ 
  & (0.00000) & (0.00000) & (0.00000) & (0.00000) & (0.00000) \\ 
  & & & & & \\ 
 Unemployment &  & 0.00002 &  &  & 0.0003 \\ 
  &  & (0.0002) &  &  & (0.0002) \\ 
  & & & & & \\ 
 ManuShare &  & 0.0001$^{*}$ &  &  & 0.0002$^{***}$ \\ 
  &  & (0.0001) &  &  & (0.0001) \\ 
  & & & & & \\ 
 Patent &  & 0.00000$^{***}$ &  &  & 0.00000$^{**}$ \\ 
  &  & (0.00000) &  &  & (0.00000) \\ 
  & & & & & \\ 
 Population &  &  & 0.000 &  & $-$0.000 \\ 
  &  &  & (0.000) &  & (0.000) \\ 
  & & & & & \\ 
 MedianAge &  &  & 0.00002 &  & $-$0.00001 \\ 
  &  &  & (0.0001) &  & (0.0001) \\ 
  & & & & & \\ 
 Education &  &  & 0.00004 &  & 0.0001 \\ 
  &  &  & (0.0001) &  & (0.0001) \\ 
  & & & & & \\ 
 BlackShare &  &  & $-$0.0002$^{***}$ &  & $-$0.0003$^{***}$ \\ 
  &  &  & (0.0001) &  & (0.0001) \\ 
  & & & & & \\ 
 Foreign &  &  & 0.0002$^{**}$ &  & 0.0002 \\ 
  &  &  & (0.0001) &  & (0.0001) \\ 
  & & & & & \\ 
 UnionCoverage &  &  &  & 0.00002 & $-$0.0002 \\ 
  &  &  &  & (0.0001) & (0.0001) \\ 
  & & & & & \\ 
 MinimumWage &  &  &  & 0.002 & 0.001 \\ 
  &  &  &  & (0.001) & (0.001) \\ 
  & & & & & \\ 
 Vote &  &  &  & $-$0.0001 & $-$0.0001 \\ 
  &  &  &  & (0.0001) & (0.0001) \\ 
  & & & & & \\ 
 Constant & 0.007$^{***}$ & 0.008$^{*}$ & 0.009 & $-$0.003 & 0.004 \\ 
  & (0.003) & (0.004) & (0.006) & (0.009) & (0.010) \\ 
  & & & & & \\ 
\hline \\[-1.8ex] 
Observations & 298 & 298 & 298 & 298 & 298 \\ 
R$^{2}$ & 0.033 & 0.075 & 0.100 & 0.041 & 0.158 \\ 
Adjusted R$^{2}$ & 0.027 & 0.060 & 0.078 & 0.025 & 0.119 \\ 
Residual Std. Error & 0.010 & 0.010 & 0.010 & 0.010 & 0.009 \\ 
F Statistic & 5.085$^{***}$ & 4.762$^{***}$ & 4.610$^{***}$ & 2.513$^{**}$ & 4.096$^{***}$ \\ 
\hline 
\hline \\[-1.8ex] 
\textit{Note:}  & \multicolumn{5}{r}{$^{*}$p$<$0.1; $^{**}$p$<$0.05; $^{***}$p$<$0.01} \\ 
\end{tabular} 
\end{table} 

\label{appendix:MainReg:PeriodRegressions}

\begin{table}[!htbp] \centering 
  \caption{Period Regression: Population (in Pct-Change) between 2007 and 2009 (NAICS, $CM_{r,i}$, ECI)} 
  \label{} 
\footnotesize 
\begin{tabular}{@{\extracolsep{5pt}}lccccc} 
\\[-1.8ex]\hline 
\hline \\[-1.8ex] 
 & \multicolumn{5}{c}{\textit{Dependent variable:}} \\ 
\cline{2-6} 
\\[-1.8ex] & \multicolumn{5}{c}{Population (in Pct-Change) Change between 2007 and 2009} \\ 
\\[-1.8ex] & (1) & (2) & (3) & (4) & (5)\\ 
\hline \\[-1.8ex] 
 ECI & $-$0.001 & 0.0003 & $-$0.002$^{**}$ & 0.001 & 0.0001 \\ 
  & (0.001) & (0.001) & (0.001) & (0.001) & (0.001) \\ 
  & & & & & \\ 
 Population in 2007 & 0.000 & $-$0.000 & $-$0.000 & 0.000 & $-$0.000 \\ 
  & (0.000) & (0.000) & (0.000) & (0.000) & (0.000) \\ 
  & & & & & \\ 
 Unemployment &  & $-$0.001$^{***}$ &  &  & $-$0.001$^{***}$ \\ 
  &  & (0.0002) &  &  & (0.0002) \\ 
  & & & & & \\ 
 ManuShare &  & $-$0.0003$^{***}$ &  &  & $-$0.0001$^{***}$ \\ 
  &  & (0.0001) &  &  & (0.0001) \\ 
  & & & & & \\ 
 Patent &  & 0.00000$^{*}$ &  &  & 0.00000 \\ 
  &  & (0.00000) &  &  & (0.00000) \\ 
  & & & & & \\ 
 Income per Capita &  & $-$0.00000$^{***}$ &  &  & $-$0.000 \\ 
  &  & (0.00000) &  &  & (0.00000) \\ 
  & & & & & \\ 
 MedianAge &  &  & $-$0.001$^{***}$ &  & $-$0.001$^{***}$ \\ 
  &  &  & (0.0001) &  & (0.0001) \\ 
  & & & & & \\ 
 Education &  &  & 0.0002$^{**}$ &  & 0.00000 \\ 
  &  &  & (0.0001) &  & (0.0001) \\ 
  & & & & & \\ 
 BlackShare &  &  & 0.00005 &  & $-$0.0001$^{*}$ \\ 
  &  &  & (0.0001) &  & (0.0001) \\ 
  & & & & & \\ 
 Foreign &  &  & 0.0003$^{***}$ &  & 0.0001$^{*}$ \\ 
  &  &  & (0.0001) &  & (0.0001) \\ 
  & & & & & \\ 
 UnionCoverage &  &  &  & $-$0.001$^{***}$ & $-$0.001$^{***}$ \\ 
  &  &  &  & (0.0001) & (0.0001) \\ 
  & & & & & \\ 
 MinimumWage &  &  &  & 0.001 & 0.002$^{***}$ \\ 
  &  &  &  & (0.001) & (0.001) \\ 
  & & & & & \\ 
 Vote &  &  &  & $-$0.0002$^{***}$ & $-$0.0001 \\ 
  &  &  &  & (0.0001) & (0.0001) \\ 
  & & & & & \\ 
 Constant & 0.008$^{***}$ & 0.029$^{***}$ & 0.036$^{***}$ & 0.017$^{***}$ & 0.055$^{***}$ \\ 
  & (0.002) & (0.004) & (0.005) & (0.005) & (0.005) \\ 
  & & & & & \\ 
\hline \\[-1.8ex] 
Observations & 292 & 292 & 292 & 292 & 292 \\ 
R$^{2}$ & 0.006 & 0.199 & 0.332 & 0.172 & 0.545 \\ 
Adjusted R$^{2}$ & $-$0.001 & 0.182 & 0.318 & 0.158 & 0.524 \\ 
Residual Std. Error & 0.009 & 0.008 & 0.008 & 0.008 & 0.006 \\ 
F Statistic & 0.909 & 11.810$^{***}$ & 23.602$^{***}$ & 11.894$^{***}$ & 25.631$^{***}$ \\ 
\hline 
\hline \\[-1.8ex] 
\textit{Note:}  & \multicolumn{5}{r}{$^{*}$p$<$0.1; $^{**}$p$<$0.05; $^{***}$p$<$0.01} \\ 
\end{tabular} 
\end{table}

\begin{table}[!htbp] \centering 
  \caption{Period Regression: Population (in Pct-Change) between 2010 and 2015 (NAICS, $CM_{r,i}$, ECI)} 
  \label{} 
\footnotesize 
\begin{tabular}{@{\extracolsep{5pt}}lccccc} 
\\[-1.8ex]\hline 
\hline \\[-1.8ex] 
 & \multicolumn{5}{c}{\textit{Dependent variable:}} \\ 
\cline{2-6} 
\\[-1.8ex] & \multicolumn{5}{c}{Population (in Pct-Change) Change between 2010 and 2015} \\ 
\\[-1.8ex] & (1) & (2) & (3) & (4) & (5)\\ 
\hline \\[-1.8ex] 
 ECI & 0.002$^{*}$ & 0.002$^{**}$ & $-$0.001 & 0.003$^{***}$ & 0.001 \\ 
  & (0.001) & (0.001) & (0.001) & (0.001) & (0.001) \\ 
  & & & & & \\ 
 Population in 2010 & 0.000 & $-$0.000 & $-$0.000 & 0.000 & $-$0.000 \\ 
  & (0.000) & (0.000) & (0.000) & (0.000) & (0.000) \\ 
  & & & & & \\ 
 Unemployment &  & $-$0.0002 &  &  & 0.00002 \\ 
  &  & (0.0001) &  &  & (0.0002) \\ 
  & & & & & \\ 
 ManuShare &  & $-$0.0004$^{***}$ &  &  & $-$0.0002$^{***}$ \\ 
  &  & (0.0001) &  &  & (0.0001) \\ 
  & & & & & \\ 
 Patent &  & 0.00000 &  &  & $-$0.00000 \\ 
  &  & (0.00000) &  &  & (0.00000) \\ 
  & & & & & \\ 
 Income per Capita &  & $-$0.00000 &  &  & 0.000 \\ 
  &  & (0.00000) &  &  & (0.00000) \\ 
  & & & & & \\ 
 MedianAge &  &  & $-$0.0001 &  & $-$0.0002$^{**}$ \\ 
  &  &  & (0.0001) &  & (0.0001) \\ 
  & & & & & \\ 
 Education &  &  & 0.0003$^{***}$ &  & 0.0002$^{***}$ \\ 
  &  &  & (0.0001) &  & (0.0001) \\ 
  & & & & & \\ 
 BlackShare &  &  & 0.00004 &  & $-$0.0001$^{**}$ \\ 
  &  &  & (0.00005) &  & (0.00005) \\ 
  & & & & & \\ 
 Foreign &  &  & 0.0004$^{***}$ &  & 0.0003$^{***}$ \\ 
  &  &  & (0.0001) &  & (0.0001) \\ 
  & & & & & \\ 
 UnionCoverage &  &  &  & $-$0.001$^{***}$ & $-$0.001$^{***}$ \\ 
  &  &  &  & (0.0001) & (0.0001) \\ 
  & & & & & \\ 
 MinimumWage &  &  &  & 0.003$^{***}$ & 0.0003 \\ 
  &  &  &  & (0.001) & (0.001) \\ 
  & & & & & \\ 
 Vote &  &  &  & $-$0.00002 & 0.00003 \\ 
  &  &  &  & (0.0001) & (0.0001) \\ 
  & & & & & \\ 
 Constant & 0.002 & 0.011$^{***}$ & 0.00002 & $-$0.010 & 0.010 \\ 
  & (0.001) & (0.003) & (0.004) & (0.007) & (0.007) \\ 
  & & & & & \\ 
\hline \\[-1.8ex] 
Observations & 298 & 298 & 298 & 298 & 298 \\ 
R$^{2}$ & 0.038 & 0.193 & 0.224 & 0.189 & 0.430 \\ 
Adjusted R$^{2}$ & 0.032 & 0.176 & 0.208 & 0.175 & 0.404 \\ 
Residual Std. Error & 0.008 & 0.007 & 0.007 & 0.007 & 0.006 \\ 
F Statistic & 5.890$^{***}$ & 11.568$^{***}$ & 14.019$^{***}$ & 13.570$^{***}$ & 16.506$^{***}$ \\ 
\hline 
\hline \\[-1.8ex] 
\textit{Note:}  & \multicolumn{5}{r}{$^{*}$p$<$0.1; $^{**}$p$<$0.05; $^{***}$p$<$0.01} \\ 
\end{tabular} 
\end{table}

\FloatBarrier

\subsection{Top/Bottom Complexity Industries}
\begin{table}[ht]
\centering
\caption{ICI: Highest and Lowest Complexity Industries ($ECI$, $CM_{r,i}$, NAICS, 2015)} 
\begingroup\footnotesize
\begin{tabular}{l|l|p{10cm}|l|r}
  \hline
 & NAICS & Name & ECI & Type \\ 
  \hline
1 & 523210 & Securities and Commodity Exchanges & 3.70 & Traded \\ 
  2 & 521110 & Monetary Authorities - Central Bank & 2.99 & Traded \\ 
  3 & 485119 & Other Urban Transit Systems & 2.64 & Local \\ 
  4 & 485112 & Commuter Rail Systems & 2.64 & Local \\ 
  5 & 522190 & Other Depository Credit Intermediation & 2.50 & Traded \\ 
  6 & 334517 & Irradiation Apparatus Manufacturing & 2.29 & Traded \\ 
  7 & 336415 & Guided Missile and Space Vehicle Propulsion Unit and Propulsion Unit Parts Manufacturing & 2.21 & Traded \\ 
  8 & 336414 & Guided Missile and Space Vehicle Manufacturing & 2.17 & Traded \\ 
  9 & 524130 & Reinsurance Carriers & 2.15 & Traded \\ 
  10 & 334112 & Computer Storage Device Manufacturing & 2.11 & Traded \\ 
  11 & 334613 & Magnetic and Optical Recording Media Manufacturing & 2.09 & Traded \\ 
  12 & 522294 & Secondary Market Financing & 2.08 & Traded \\ 
  13 & 333242 & Semiconductor Machinery Manufacturing  & 2.04 & Traded \\ 
  14 & 325413 & In-Vitro Diagnostic Substance Manufacturing & 2.00 & Traded \\ 
  15 & 311930 & Flavoring Syrup and Concentrate Manufacturing & 2.00 & Traded \\ 
  16 & 334417 & Electronic Connector Manufacturing & 1.96 & Traded \\ 
  17 & 335921 & Fiber Optic Cable Manufacturing & 1.94 & Traded \\ 
  18 & 522210 & Credit Card Issuing & 1.89 & Traded \\ 
  19 & 335912 & Primary Battery Manufacturing & 1.88 & Traded \\ 
  20 & 483111 & Deep Sea Freight Transportation & 1.85 & Traded \\ 
  21 &  & ... &  & ... \\ 
  959 & 541191 & Title Abstract and Settlement Offices & -2.07 & Local \\ 
  960 & 212291 & Uranium-Radium-Vanadium Ore Mining & -2.09 & Traded \\ 
  961 & 812210 & Funeral Homes and Funeral Services & -2.12 & Local \\ 
  962 & 486110 & Pipeline Transportation of Crude Oil & -2.13 & Traded \\ 
  963 & 541690 & Other Scientific and Technical Consulting Services & -2.16 & Traded \\ 
  964 & 311611 & Animal (except Poultry) Slaughtering & -2.17 & Traded \\ 
  965 & 447110 & Gasoline Stations with Convenience Stores & -2.17 & Local \\ 
  966 & 484230 & Specialized Freight (except Used Goods) Trucking, Long-Distance & -2.19 & Traded \\ 
  967 & 811310 & Commercial and Industrial Machinery and Equipment (except Automotive and Electronic) Repair and Maintenance & -2.20 & Local \\ 
  968 & 211112 & Natural Gas Liquid Extraction & -2.21 & Traded \\ 
  969 & 115111 & Cotton Ginning & -2.22 & Traded \\ 
  970 & 447190 & Other Gasoline Stations & -2.22 & Local \\ 
  971 & 424910 & Farm Supplies Merchant Wholesalers & -2.22 & Traded \\ 
  972 & 115112 & Soil Preparation, Planting, and Cultivating & -2.31 & Traded \\ 
  973 & 484220 & Specialized Freight (except Used Goods) Trucking, Local & -2.32 & Local \\ 
  974 & 213111 & Drilling Oil and Gas Wells & -2.47 & Traded \\ 
  975 & 813910 & Business Associations & -2.52 & Local \\ 
  976 & 424510 & Grain and Field Bean Merchant Wholesalers & -2.53 & Traded \\ 
  977 & 213112 & Support Activities for Oil and Gas Operations & -2.78 & Traded \\ 
  978 & 211111 & Crude Petroleum and Natural Gas Extraction & -2.79 & Traded \\ 
   \hline
\end{tabular}
\endgroup
\end{table}

\label{TabPresenceCM}

\FloatBarrier

\section{Different Input Matrices}
\label{appendix:InpMatr:all}

\subsection{Scatter-plots ECI VS. FI}
\label{appendix:InpMat:FigECIvsFI}

\begin{figure}[!h]
	\centering
	\subfloat[$ECI$ vs. $FI$]{{\includegraphics[width=6cm]{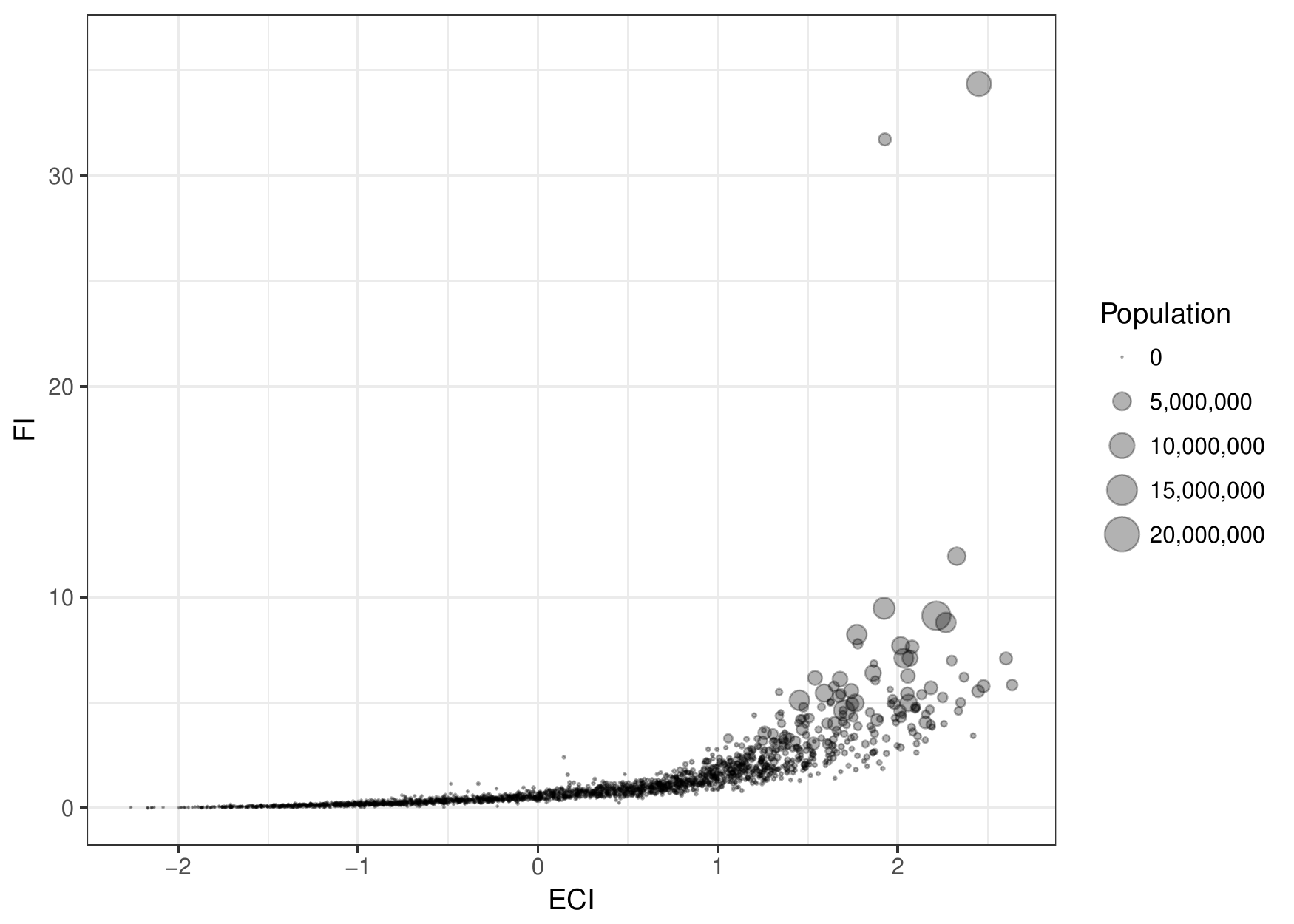}}}%
	\qquad
	\subfloat[$ECI$ vs. logged $FI$]{{\includegraphics[width=6cm]{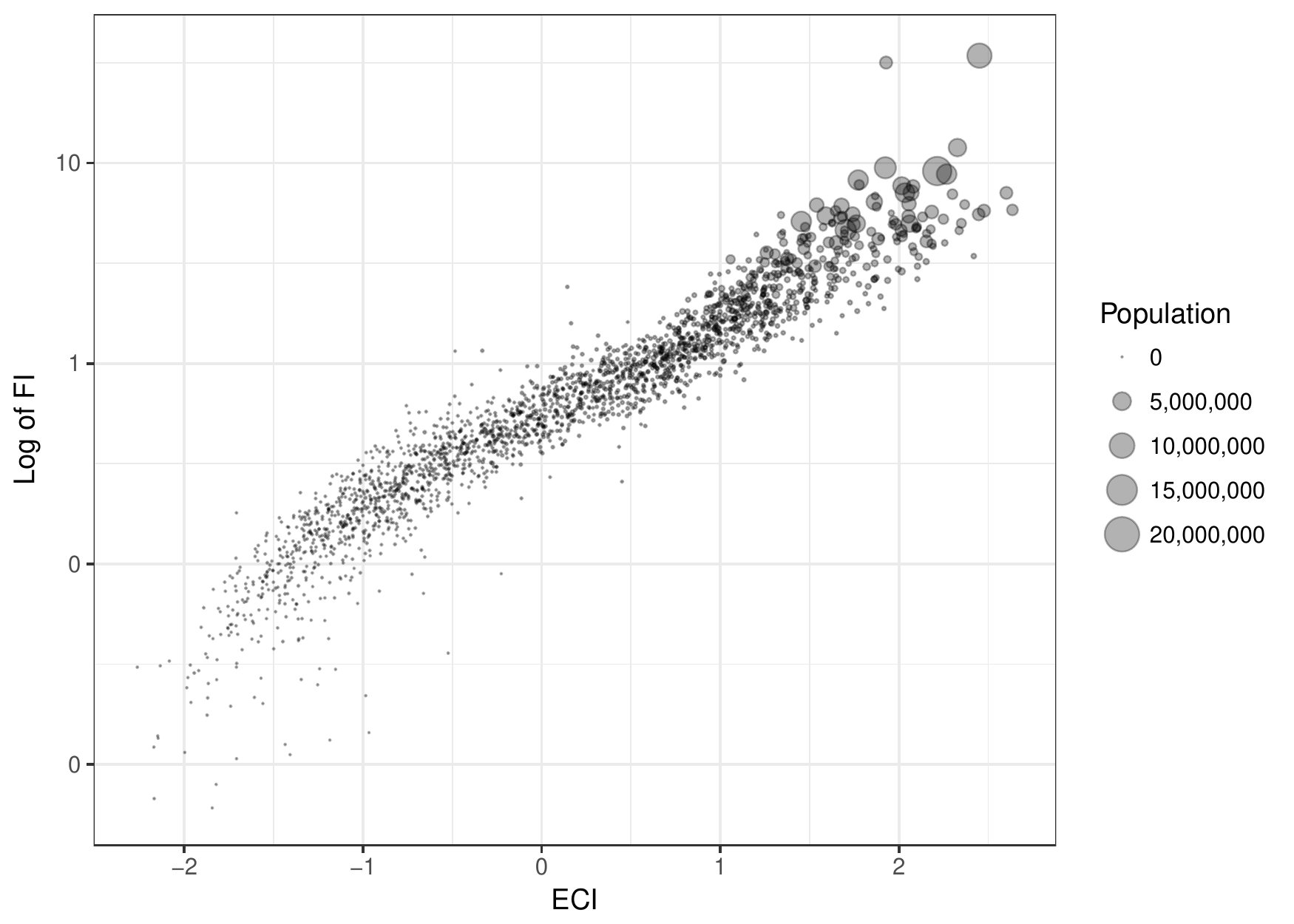}}}	
	\caption{Scatterplots $ECI$ vs. $FI$ ($BM_{r,i}$, NAICS, 2015)}%
	\source{Own calculations and CBP.}
\end{figure}

\begin{figure}[!h]
	\centering
	\subfloat[$ECI$ vs. $FI$]{{\includegraphics[width=6cm]{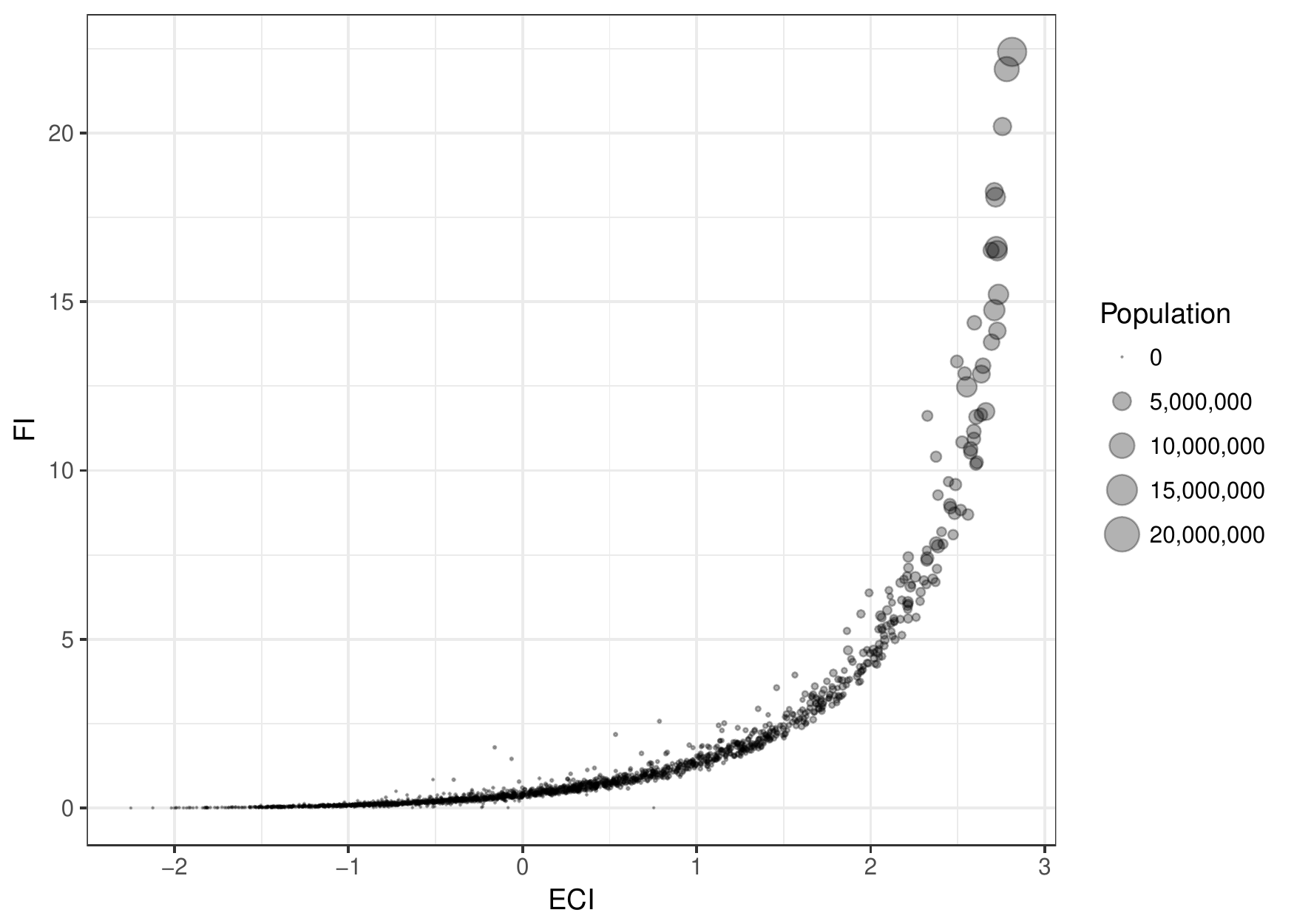}}}%
	\qquad
	\subfloat[$ECI$ vs. logged $FI$]{{\includegraphics[width=6cm]{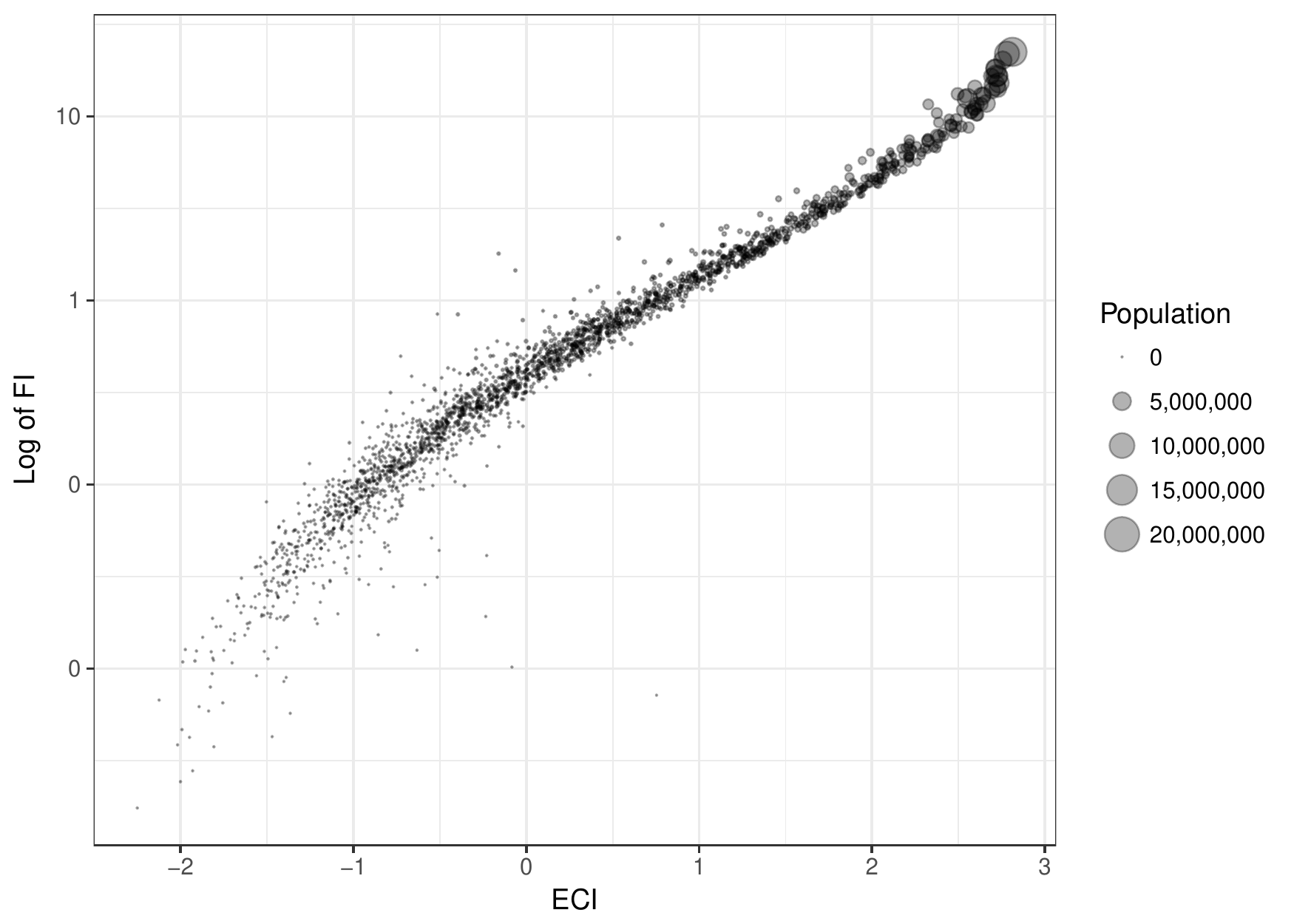}}}	
	\caption{Scatterplots $ECI$ vs. $FI$ ($Presence_{r,i}$, NAICS, 2015)}%
	\source{Own calculations and CBP.}
\end{figure}

\begin{figure}[!h]
	\centering
	\subfloat[$ECI$ vs. $FI$]{{\includegraphics[width=6cm]{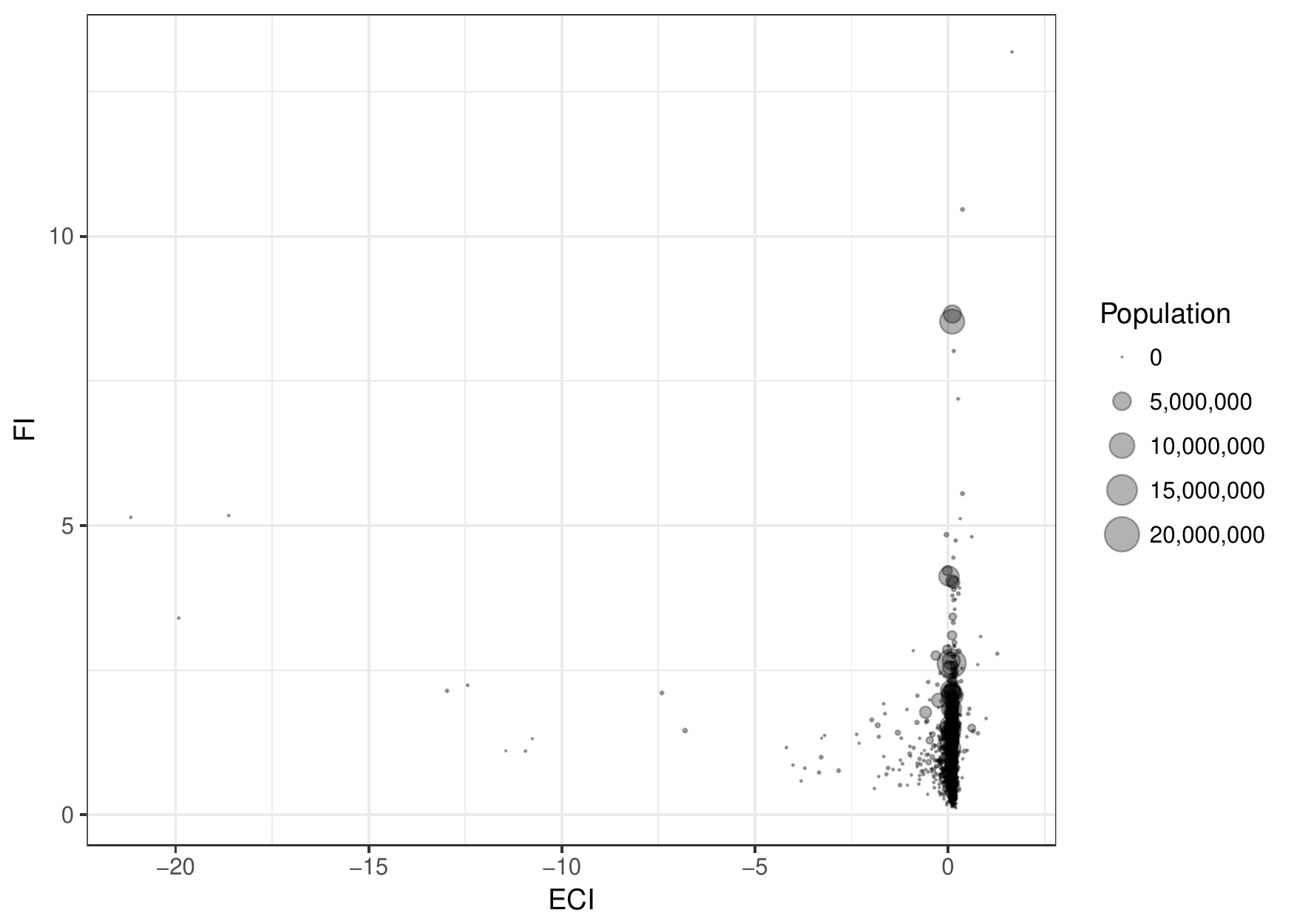}}}%
	\qquad
	\subfloat[$ECI$ vs. logged $FI$]{{\includegraphics[width=6cm]{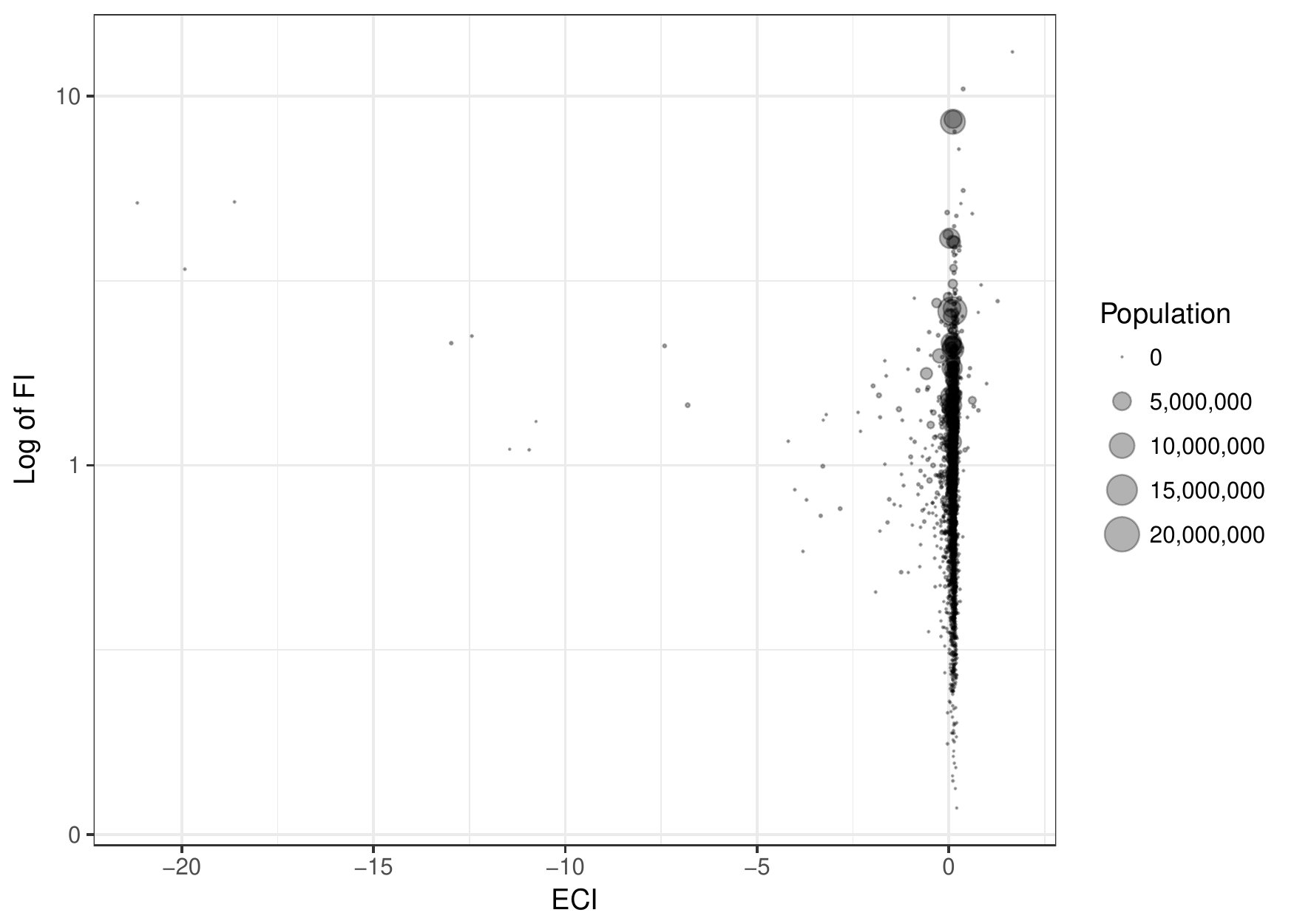}}}	
	\caption{Scatterplots $ECI$ vs. $FI$ ($RLQ_{r,i}$, NAICS, 2015)}%
	\source{Own calculations and CBP.}
\end{figure}

\begin{figure}[!h]
	\centering
	\subfloat[$ECI$ vs. $FI$]{{\includegraphics[width=6cm]{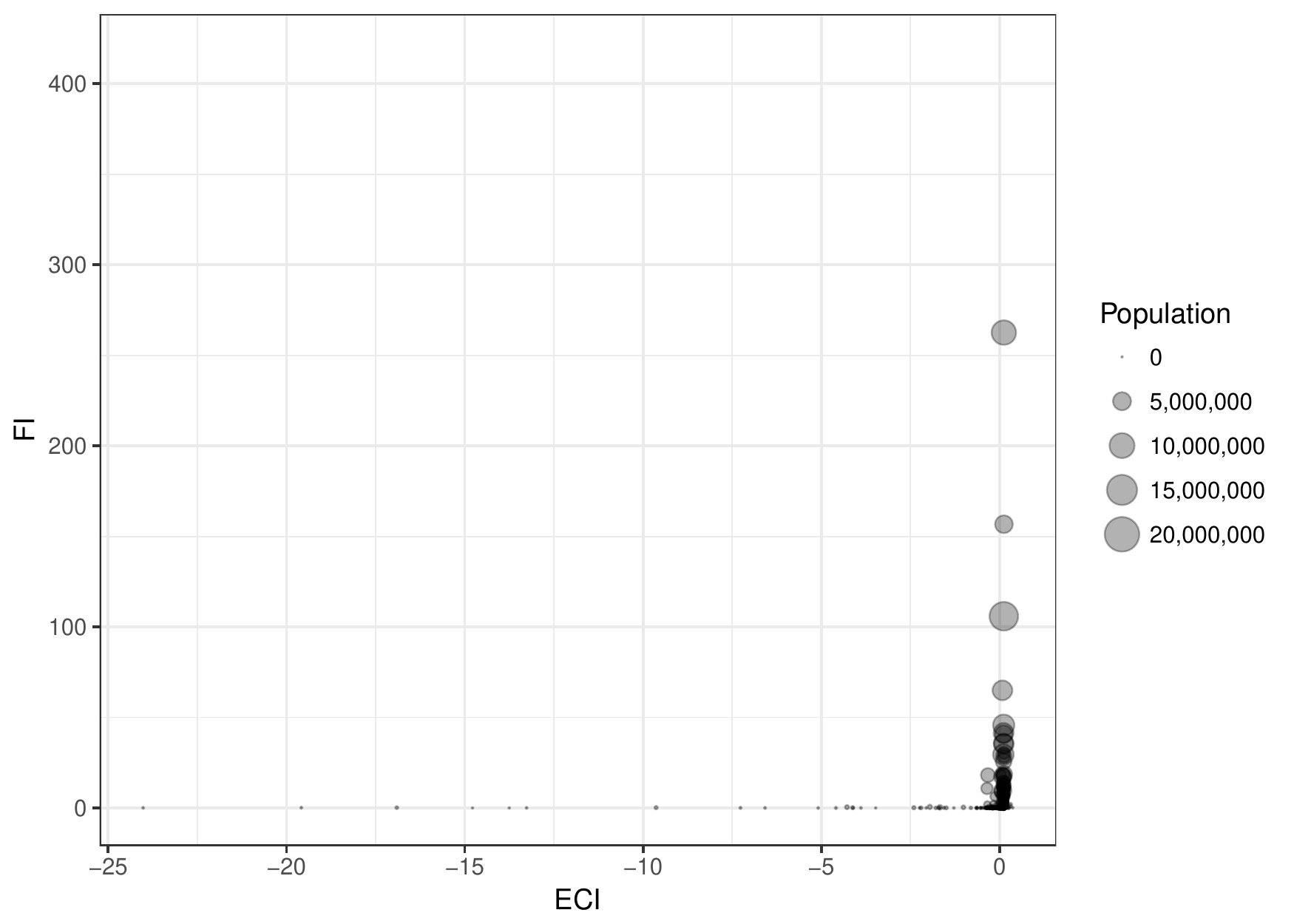}}}%
	\qquad
	\subfloat[$ECI$ vs. logged $FI$]{{\includegraphics[width=6cm]{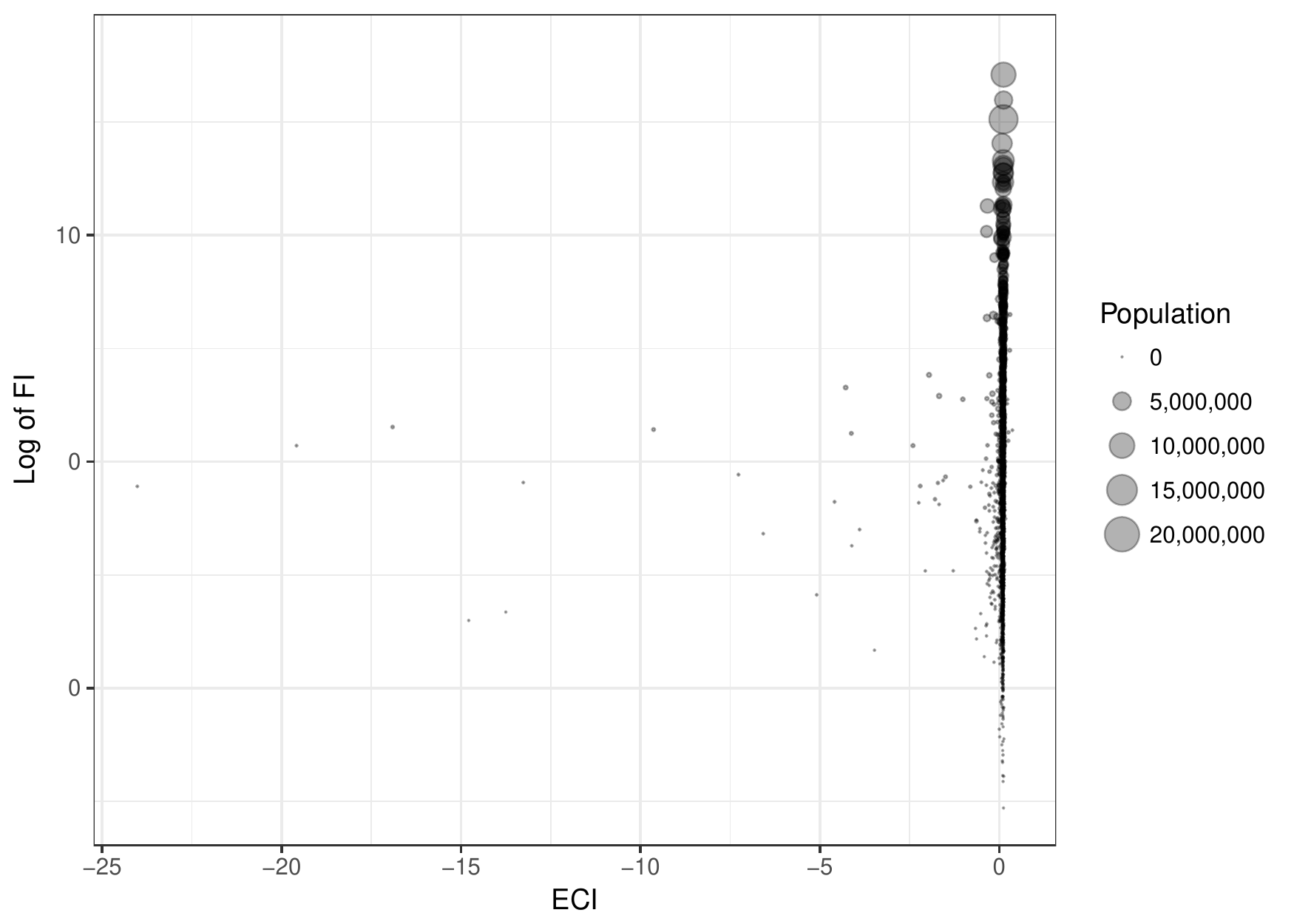}}}	
	\caption{Scatterplots $ECI$ vs. $FI$ ($WM_{r,i}$, NAICS, 2015)}%
	\source{Own calculations and CBP.}
\end{figure}

\FloatBarrier
\subsection{Panel Regressions}

\begin{table}[!htbp] \centering 
  \caption{Panel Regression: Income per Capita (in 1,000 USD) 2015 (NAICS, $BM_{r,i}$, ECI)} 
  \label{} 
\footnotesize 
\begin{tabular}{@{\extracolsep{5pt}}lccccc} 
\\[-1.8ex]\hline 
\hline \\[-1.8ex] 
 & \multicolumn{5}{c}{\textit{Dependent variable:}} \\ 
\cline{2-6} 
\\[-1.8ex] & \multicolumn{5}{c}{Income per Capita (in 1,000 USD) 2015} \\ 
\\[-1.8ex] & (1) & (2) & (3) & (4) & (5)\\ 
\hline \\
\end{tabular} 
\end{table}

\begin{table}[!htbp] \centering 
  \caption{Panel Regression: Population (in 1,000) 2015 (NAICS, $BM_{r,i}$, ECI)} 
  \label{} 
\footnotesize 
\begin{tabular}{@{\extracolsep{5pt}}lccccc} 
\\[-1.8ex]\hline 
\hline \\[-1.8ex] 
 & \multicolumn{5}{c}{\textit{Dependent variable:}} \\ 
\cline{2-6} 
\\[-1.8ex] & \multicolumn{5}{c}{Population (in 1,000) 2015} \\ 
\\[-1.8ex] & (1) & (2) & (3) & (4) & (5)\\ 
\hline \\[-1.8ex] 
 l.ECI & $-$55.299$^{**}$ & $-$32.757$^{*}$ & $-$48.970$^{**}$ & $-$56.020$^{**}$ & $-$32.198$^{*}$ \\ 
  & (24.646) & (19.702) & (24.161) & (24.368) & (19.540) \\ 
  & & & & & \\ 
 Unemployment &  & 0.244 &  &  & 0.114 \\ 
  &  & (0.266) &  &  & (0.260) \\ 
  & & & & & \\ 
 ManuShare &  & 0.651 &  &  & 0.753 \\ 
  &  & (0.724) &  &  & (0.733) \\ 
  & & & & & \\ 
 Patent &  & 0.097$^{***}$ &  &  & 0.097$^{***}$ \\ 
  &  & (0.031) &  &  & (0.031) \\ 
  & & & & & \\ 
 Income per Capita &  & $-$1.769 &  &  & $-$1.743$^{*}$ \\ 
  &  & (1.088) &  &  & (0.996) \\ 
  & & & & & \\ 
 MedianAge &  &  & $-$3.833$^{**}$ &  & $-$0.800 \\ 
  &  &  & (1.636) &  & (0.912) \\ 
  & & & & & \\ 
 Education &  &  & 0.902$^{**}$ &  & 0.327 \\ 
  &  &  & (0.436) &  & (0.287) \\ 
  & & & & & \\ 
 BlackShare &  &  & 1.066 &  & 2.502 \\ 
  &  &  & (3.156) &  & (2.369) \\ 
  & & & & & \\ 
 Foreign &  &  & 2.144 &  & 0.612 \\ 
  &  &  & (2.018) &  & (1.484) \\ 
  & & & & & \\ 
 UnionCoverage &  &  &  & 6.081$^{***}$ & 5.089$^{***}$ \\ 
  &  &  &  & (1.895) & (1.480) \\ 
  & & & & & \\ 
 MinimumWage &  &  &  & 7.059$^{*}$ & 7.805$^{**}$ \\ 
  &  &  &  & (4.135) & (3.518) \\ 
  & & & & & \\ 
 Vote &  &  &  & $-$0.085 & 0.492$^{*}$ \\ 
  &  &  &  & (0.410) & (0.272) \\ 
  & & & & & \\ 
\hline \\[-1.8ex] 
Observations & 2,715 & 2,715 & 2,715 & 2,715 & 2,715 \\ 
R$^{2}$ & 0.011 & 0.338 & 0.019 & 0.027 & 0.353 \\ 
Adjusted R$^{2}$ & $-$0.121 & 0.248 & $-$0.114 & $-$0.105 & 0.263 \\ 
F Statistic & 27.046$^{***}$ & 243.556$^{***}$ & 9.351$^{***}$ & 16.509$^{***}$ & 108.277$^{***}$ \\ 
\hline 
\hline \\[-1.8ex] 
\textit{Note:}  & \multicolumn{5}{r}{$^{*}$p$<$0.1; $^{**}$p$<$0.05; $^{***}$p$<$0.01} \\ 
\end{tabular} 
\end{table} 

\label{appendix:InpMat:PanelReg}

\begin{table}[!htbp] \centering 
  \caption{Panel Regression: Income per Capita (in 1,000 USD) 2015 (NAICS, $Presence_{r,i}$, ECI)} 
  \label{} 
\footnotesize 
\begin{tabular}{@{\extracolsep{5pt}}lccccc} 
\\[-1.8ex]\hline 
\hline \\[-1.8ex] 
 & \multicolumn{5}{c}{\textit{Dependent variable:}} \\ 
\cline{2-6} 
\\[-1.8ex] & \multicolumn{5}{c}{Income per Capita (in 1,000 USD) 2015} \\ 
\\[-1.8ex] & (1) & (2) & (3) & (4) & (5)\\ 
\hline \\[-1.8ex] 
 l.ECI & 0.355 & 1.597 & 0.142 & 0.446 & 0.766 \\ 
  & (1.527) & (1.464) & (1.483) & (1.473) & (1.318) \\ 
  & & & & & \\ 
 Unemployment &  & $-$0.153$^{***}$ &  &  & $-$0.145$^{***}$ \\ 
  &  & (0.023) &  &  & (0.021) \\ 
  & & & & & \\ 
 ManuShare &  & 0.020 &  &  & 0.026 \\ 
  &  & (0.034) &  &  & (0.034) \\ 
  & & & & & \\ 
 Patent &  & 0.001$^{***}$ &  &  & 0.001$^{***}$ \\ 
  &  & (0.0002) &  &  & (0.0002) \\ 
  & & & & & \\ 
 Population &  &  & 0.00000 &  & $-$0.00000$^{**}$ \\ 
  &  &  & (0.00000) &  & (0.00000) \\ 
  & & & & & \\ 
 MedianAge &  &  & $-$0.061 &  & $-$0.030 \\ 
  &  &  & (0.124) &  & (0.121) \\ 
  & & & & & \\ 
 Education &  &  & 0.028 &  & 0.025 \\ 
  &  &  & (0.024) &  & (0.024) \\ 
  & & & & & \\ 
 BlackShare &  &  & $-$0.236$^{***}$ &  & $-$0.198$^{***}$ \\ 
  &  &  & (0.075) &  & (0.071) \\ 
  & & & & & \\ 
 Foreign &  &  & 0.257$^{**}$ &  & 0.246$^{*}$ \\ 
  &  &  & (0.131) &  & (0.131) \\ 
  & & & & & \\ 
 UnionCoverage &  &  &  & 0.037 & 0.064 \\ 
  &  &  &  & (0.056) & (0.055) \\ 
  & & & & & \\ 
 MinimumWage &  &  &  & 0.044 & 0.003 \\ 
  &  &  &  & (0.169) & (0.156) \\ 
  & & & & & \\ 
 Vote &  &  &  & $-$0.055$^{***}$ & $-$0.047$^{***}$ \\ 
  &  &  &  & (0.012) & (0.010) \\ 
  & & & & & \\ 
\hline \\[-1.8ex] 
Observations & 2,715 & 2,715 & 2,715 & 2,715 & 2,715 \\ 
R$^{2}$ & 0.0001 & 0.038 & 0.021 & 0.010 & 0.066 \\ 
Adjusted R$^{2}$ & $-$0.134 & $-$0.092 & $-$0.112 & $-$0.124 & $-$0.063 \\ 
F Statistic & 0.202 & 23.700$^{***}$ & 8.529$^{***}$ & 5.969$^{***}$ & 14.111$^{***}$ \\ 
\hline 
\hline \\[-1.8ex] 
\textit{Note:}  & \multicolumn{5}{r}{$^{*}$p$<$0.1; $^{**}$p$<$0.05; $^{***}$p$<$0.01} \\ 
\end{tabular} 
\end{table}

\begin{table}[!htbp] \centering 
  \caption{Panel Regression: Population (in 1,000) 2015 (NAICS, $Presence_{r,i}$, ECI)} 
  \label{} 
\footnotesize 
\begin{tabular}{@{\extracolsep{5pt}}lccccc} 
\\[-1.8ex]\hline 
\hline \\[-1.8ex] 
 & \multicolumn{5}{c}{\textit{Dependent variable:}} \\ 
\cline{2-6} 
\\[-1.8ex] & \multicolumn{5}{c}{Population (in 1,000) 2015} \\ 
\\[-1.8ex] & (1) & (2) & (3) & (4) & (5)\\ 
\hline \\[-1.8ex] 
 l.ECI & $-$260.761$^{***}$ & $-$132.244$^{**}$ & $-$255.896$^{***}$ & $-$270.461$^{***}$ & $-$140.653$^{**}$ \\ 
  & (78.312) & (56.060) & (77.724) & (78.494) & (56.055) \\ 
  & & & & & \\ 
 Unemployment &  & 0.251 &  &  & 0.117 \\ 
  &  & (0.279) &  &  & (0.272) \\ 
  & & & & & \\ 
 ManuShare &  & 0.793 &  &  & 0.947 \\ 
  &  & (0.712) &  &  & (0.727) \\ 
  & & & & & \\ 
 Patent &  & 0.093$^{***}$ &  &  & 0.092$^{***}$ \\ 
  &  & (0.031) &  &  & (0.030) \\ 
  & & & & & \\ 
 Income per Capita &  & $-$1.475 &  &  & $-$1.482 \\ 
  &  & (1.015) &  &  & (0.928) \\ 
  & & & & & \\ 
 MedianAge &  &  & $-$3.072 &  & $-$0.570 \\ 
  &  &  & (1.928) &  & (1.079) \\ 
  & & & & & \\ 
 Education &  &  & 1.133$^{**}$ &  & 0.488 \\ 
  &  &  & (0.520) &  & (0.319) \\ 
  & & & & & \\ 
 BlackShare &  &  & $-$0.225 &  & 1.813 \\ 
  &  &  & (3.053) &  & (2.235) \\ 
  & & & & & \\ 
 Foreign &  &  & 2.707 &  & 0.933 \\ 
  &  &  & (1.977) &  & (1.487) \\ 
  & & & & & \\ 
 UnionCoverage &  &  &  & 6.853$^{***}$ & 5.533$^{***}$ \\ 
  &  &  &  & (1.961) & (1.553) \\ 
  & & & & & \\ 
 MinimumWage &  &  &  & 7.724$^{*}$ & 8.148$^{**}$ \\ 
  &  &  &  & (4.337) & (3.667) \\ 
  & & & & & \\ 
 Vote &  &  &  & $-$0.026 & 0.492$^{*}$ \\ 
  &  &  &  & (0.383) & (0.269) \\ 
  & & & & & \\ 
\hline \\[-1.8ex] 
Observations & 2,715 & 2,715 & 2,715 & 2,715 & 2,715 \\ 
R$^{2}$ & 0.061 & 0.348 & 0.068 & 0.081 & 0.365 \\ 
Adjusted R$^{2}$ & $-$0.064 & 0.260 & $-$0.058 & $-$0.043 & 0.277 \\ 
F Statistic & 156.054$^{***}$ & 255.691$^{***}$ & 35.005$^{***}$ & 52.616$^{***}$ & 114.388$^{***}$ \\ 
\hline 
\hline \\[-1.8ex] 
\textit{Note:}  & \multicolumn{5}{r}{$^{*}$p$<$0.1; $^{**}$p$<$0.05; $^{***}$p$<$0.01} \\ 
\end{tabular} 
\end{table}

\begin{table}[!htbp] \centering 
  \caption{Panel Regression: Income per Capita (in 1,000 USD) 2015 (NAICS, $RLQ_{r,i}$, ECI)} 
  \label{} 
\footnotesize 
\begin{tabular}{@{\extracolsep{5pt}}lccccc} 
\\[-1.8ex]\hline 
\hline \\[-1.8ex] 
 & \multicolumn{5}{c}{\textit{Dependent variable:}} \\ 
\cline{2-6} 
\\[-1.8ex] & \multicolumn{5}{c}{Income per Capita (in 1,000 USD) 2015} \\ 
\\[-1.8ex] & (1) & (2) & (3) & (4) & (5)\\ 
\hline \\[-1.8ex] 
 l.ECI & $-$0.029 & $-$0.037 & $-$0.022 & $-$0.024 & $-$0.028 \\ 
  & (0.054) & (0.046) & (0.048) & (0.051) & (0.039) \\ 
  & & & & & \\ 
 Unemployment &  & $-$0.152$^{***}$ &  &  & $-$0.145$^{***}$ \\ 
  &  & (0.023) &  &  & (0.021) \\ 
  & & & & & \\ 
 ManuShare &  & 0.030 &  &  & 0.030 \\ 
  &  & (0.037) &  &  & (0.037) \\ 
  & & & & & \\ 
 Patent &  & 0.001$^{***}$ &  &  & 0.001$^{***}$ \\ 
  &  & (0.0002) &  &  & (0.0002) \\ 
  & & & & & \\ 
 Population &  &  & 0.00000 &  & $-$0.00000$^{**}$ \\ 
  &  &  & (0.00000) &  & (0.00000) \\ 
  & & & & & \\ 
 MedianAge &  &  & $-$0.061 &  & $-$0.027 \\ 
  &  &  & (0.122) &  & (0.120) \\ 
  & & & & & \\ 
 Education &  &  & 0.028 &  & 0.025 \\ 
  &  &  & (0.024) &  & (0.025) \\ 
  & & & & & \\ 
 BlackShare &  &  & $-$0.237$^{***}$ &  & $-$0.204$^{***}$ \\ 
  &  &  & (0.078) &  & (0.074) \\ 
  & & & & & \\ 
 Foreign &  &  & 0.257$^{*}$ &  & 0.248$^{*}$ \\ 
  &  &  & (0.134) &  & (0.134) \\ 
  & & & & & \\ 
 UnionCoverage &  &  &  & 0.039 & 0.067 \\ 
  &  &  &  & (0.059) & (0.059) \\ 
  & & & & & \\ 
 MinimumWage &  &  &  & 0.045 & 0.006 \\ 
  &  &  &  & (0.170) & (0.158) \\ 
  & & & & & \\ 
 Vote &  &  &  & $-$0.055$^{***}$ & $-$0.047$^{***}$ \\ 
  &  &  &  & (0.012) & (0.010) \\ 
  & & & & & \\ 
\hline \\[-1.8ex] 
Observations & 2,715 & 2,715 & 2,715 & 2,715 & 2,715 \\ 
R$^{2}$ & 0.00002 & 0.037 & 0.021 & 0.010 & 0.066 \\ 
Adjusted R$^{2}$ & $-$0.134 & $-$0.094 & $-$0.112 & $-$0.124 & $-$0.064 \\ 
F Statistic & 0.056 & 22.704$^{***}$ & 8.530$^{***}$ & 5.898$^{***}$ & 14.037$^{***}$ \\ 
\hline 
\hline \\[-1.8ex] 
\textit{Note:}  & \multicolumn{5}{r}{$^{*}$p$<$0.1; $^{**}$p$<$0.05; $^{***}$p$<$0.01} \\ 
\end{tabular} 
\end{table}

\begin{table}[!htbp] \centering 
  \caption{Panel Regression: Population (in 1,000) 2015 (NAICS, $RLQ_{r,i}$, ECI)} 
  \label{} 
\footnotesize 
\begin{tabular}{@{\extracolsep{5pt}}lccccc} 
\\[-1.8ex]\hline 
\hline \\[-1.8ex] 
 & \multicolumn{5}{c}{\textit{Dependent variable:}} \\ 
\cline{2-6} 
\\[-1.8ex] & \multicolumn{5}{c}{Population (in 1,000) 2015} \\ 
\\[-1.8ex] & (1) & (2) & (3) & (4) & (5)\\ 
\hline \\[-1.8ex] 
 l.ECI & $-$2.666$^{*}$ & $-$0.436 & $-$2.254 & $-$2.778$^{**}$ & $-$0.378 \\ 
  & (1.451) & (1.269) & (1.457) & (1.367) & (1.343) \\ 
  & & & & & \\ 
 Unemployment &  & 0.120 &  &  & 0.009 \\ 
  &  & (0.264) &  &  & (0.255) \\ 
  & & & & & \\ 
 ManuShare &  & 0.048 &  &  & 0.169 \\ 
  &  & (0.567) &  &  & (0.579) \\ 
  & & & & & \\ 
 Patent &  & 0.098$^{***}$ &  &  & 0.097$^{***}$ \\ 
  &  & (0.032) &  &  & (0.032) \\ 
  & & & & & \\ 
 Income per Capita &  & $-$1.611 &  &  & $-$1.589$^{*}$ \\ 
  &  & (1.033) &  &  & (0.937) \\ 
  & & & & & \\ 
 MedianAge &  &  & $-$4.457$^{***}$ &  & $-$1.144 \\ 
  &  &  & (1.697) &  & (0.911) \\ 
  & & & & & \\ 
 Education &  &  & 0.990$^{**}$ &  & 0.381 \\ 
  &  &  & (0.439) &  & (0.285) \\ 
  & & & & & \\ 
 BlackShare &  &  & 1.292 &  & 2.716 \\ 
  &  &  & (3.155) &  & (2.392) \\ 
  & & & & & \\ 
 Foreign &  &  & 2.204 &  & 0.625 \\ 
  &  &  & (2.026) &  & (1.483) \\ 
  & & & & & \\ 
 UnionCoverage &  &  &  & 5.958$^{***}$ & 4.996$^{***}$ \\ 
  &  &  &  & (1.885) & (1.467) \\ 
  & & & & & \\ 
 MinimumWage &  &  &  & 7.400$^{*}$ & 7.881$^{**}$ \\ 
  &  &  &  & (4.078) & (3.473) \\ 
  & & & & & \\ 
 Vote &  &  &  & $-$0.150 & 0.474$^{*}$ \\ 
  &  &  &  & (0.432) & (0.275) \\ 
  & & & & & \\ 
\hline \\[-1.8ex] 
Observations & 2,715 & 2,715 & 2,715 & 2,715 & 2,715 \\ 
R$^{2}$ & 0.0003 & 0.334 & 0.011 & 0.016 & 0.349 \\ 
Adjusted R$^{2}$ & $-$0.133 & 0.244 & $-$0.123 & $-$0.117 & 0.259 \\ 
F Statistic & 0.634 & 239.648$^{***}$ & 5.215$^{***}$ & 9.553$^{***}$ & 106.661$^{***}$ \\ 
\hline 
\hline \\[-1.8ex] 
\textit{Note:}  & \multicolumn{5}{r}{$^{*}$p$<$0.1; $^{**}$p$<$0.05; $^{***}$p$<$0.01} \\ 
\end{tabular} 
\end{table}

\begin{table}[!htbp] \centering 
  \caption{Panel Regression: Income per Capita (in 1,000 USD) 2015 (NAICS, $WM_{r,i}$, ECI)} 
  \label{} 
\footnotesize 
\begin{tabular}{@{\extracolsep{5pt}}lccccc} 
\\[-1.8ex]\hline 
\hline \\[-1.8ex] 
 & \multicolumn{5}{c}{\textit{Dependent variable:}} \\ 
\cline{2-6} 
\\[-1.8ex] & \multicolumn{5}{c}{Income per Capita (in 1,000 USD) 2015} \\ 
\\[-1.8ex] & (1) & (2) & (3) & (4) & (5)\\ 
\hline \\[-1.8ex] 
 l.ECI & 0.011 & 0.002 & 0.013 & 0.009 & 0.001 \\ 
  & (0.013) & (0.015) & (0.014) & (0.013) & (0.014) \\ 
  & & & & & \\ 
 Unemployment &  & $-$0.152$^{***}$ &  &  & $-$0.145$^{***}$ \\ 
  &  & (0.023) &  &  & (0.021) \\ 
  & & & & & \\ 
 ManuShare &  & 0.029 &  &  & 0.030 \\ 
  &  & (0.037) &  &  & (0.037) \\ 
  & & & & & \\ 
 Patent &  & 0.001$^{***}$ &  &  & 0.001$^{***}$ \\ 
  &  & (0.0002) &  &  & (0.0002) \\ 
  & & & & & \\ 
 Population &  &  & 0.00000 &  & $-$0.00000$^{**}$ \\ 
  &  &  & (0.00000) &  & (0.00000) \\ 
  & & & & & \\ 
 MedianAge &  &  & $-$0.061 &  & $-$0.027 \\ 
  &  &  & (0.123) &  & (0.120) \\ 
  & & & & & \\ 
 Education &  &  & 0.028 &  & 0.025 \\ 
  &  &  & (0.024) &  & (0.025) \\ 
  & & & & & \\ 
 BlackShare &  &  & $-$0.236$^{***}$ &  & $-$0.203$^{***}$ \\ 
  &  &  & (0.078) &  & (0.074) \\ 
  & & & & & \\ 
 Foreign &  &  & 0.258$^{*}$ &  & 0.248$^{*}$ \\ 
  &  &  & (0.134) &  & (0.134) \\ 
  & & & & & \\ 
 UnionCoverage &  &  &  & 0.039 & 0.067 \\ 
  &  &  &  & (0.059) & (0.059) \\ 
  & & & & & \\ 
 MinimumWage &  &  &  & 0.045 & 0.006 \\ 
  &  &  &  & (0.170) & (0.158) \\ 
  & & & & & \\ 
 Vote &  &  &  & $-$0.055$^{***}$ & $-$0.047$^{***}$ \\ 
  &  &  &  & (0.012) & (0.010) \\ 
  & & & & & \\ 
\hline \\[-1.8ex] 
Observations & 2,715 & 2,715 & 2,715 & 2,715 & 2,715 \\ 
R$^{2}$ & 0.00002 & 0.037 & 0.021 & 0.010 & 0.066 \\ 
Adjusted R$^{2}$ & $-$0.134 & $-$0.094 & $-$0.112 & $-$0.124 & $-$0.064 \\ 
F Statistic & 0.045 & 22.679$^{***}$ & 8.533$^{***}$ & 5.895$^{***}$ & 14.032$^{***}$ \\ 
\hline 
\hline \\[-1.8ex] 
\textit{Note:}  & \multicolumn{5}{r}{$^{*}$p$<$0.1; $^{**}$p$<$0.05; $^{***}$p$<$0.01} \\ 
\end{tabular} 
\end{table}

\begin{table}[!htbp] \centering 
  \caption{Panel Regression: Population (in 1,000) 2015 (NAICS, $WM_{r,i}$, ECI)} 
  \label{} 
\footnotesize 
\begin{tabular}{@{\extracolsep{5pt}}lccccc} 
\\[-1.8ex]\hline 
\hline \\[-1.8ex] 
 & \multicolumn{5}{c}{\textit{Dependent variable:}} \\ 
\cline{2-6} 
\\[-1.8ex] & \multicolumn{5}{c}{Population (in 1,000) 2015} \\ 
\\[-1.8ex] & (1) & (2) & (3) & (4) & (5)\\ 
\hline \\[-1.8ex] 
 l.ECI & 0.054 & 0.038 & 0.212 & 0.065 & 0.163 \\ 
  & (0.456) & (0.249) & (0.448) & (0.529) & (0.305) \\ 
  & & & & & \\ 
 Unemployment &  & 0.123 &  &  & 0.011 \\ 
  &  & (0.263) &  &  & (0.254) \\ 
  & & & & & \\ 
 ManuShare &  & 0.042 &  &  & 0.156 \\ 
  &  & (0.569) &  &  & (0.580) \\ 
  & & & & & \\ 
 Patent &  & 0.098$^{***}$ &  &  & 0.097$^{***}$ \\ 
  &  & (0.032) &  &  & (0.032) \\ 
  & & & & & \\ 
 Income per Capita &  & $-$1.611 &  &  & $-$1.589$^{*}$ \\ 
  &  & (1.032) &  &  & (0.936) \\ 
  & & & & & \\ 
 MedianAge &  &  & $-$4.456$^{***}$ &  & $-$1.144 \\ 
  &  &  & (1.697) &  & (0.911) \\ 
  & & & & & \\ 
 Education &  &  & 1.003$^{**}$ &  & 0.384 \\ 
  &  &  & (0.441) &  & (0.285) \\ 
  & & & & & \\ 
 BlackShare &  &  & 1.345 &  & 2.729 \\ 
  &  &  & (3.151) &  & (2.392) \\ 
  & & & & & \\ 
 Foreign &  &  & 2.228 &  & 0.631 \\ 
  &  &  & (2.027) &  & (1.484) \\ 
  & & & & & \\ 
 UnionCoverage &  &  &  & 5.959$^{***}$ & 4.995$^{***}$ \\ 
  &  &  &  & (1.886) & (1.467) \\ 
  & & & & & \\ 
 MinimumWage &  &  &  & 7.365$^{*}$ & 7.878$^{**}$ \\ 
  &  &  &  & (4.076) & (3.475) \\ 
  & & & & & \\ 
 Vote &  &  &  & $-$0.153 & 0.474$^{*}$ \\ 
  &  &  &  & (0.432) & (0.275) \\ 
  & & & & & \\ 
\hline \\[-1.8ex] 
Observations & 2,715 & 2,715 & 2,715 & 2,715 & 2,715 \\ 
R$^{2}$ & 0.00000 & 0.334 & 0.011 & 0.015 & 0.349 \\ 
Adjusted R$^{2}$ & $-$0.134 & 0.244 & $-$0.124 & $-$0.118 & 0.259 \\ 
F Statistic & 0.001 & 239.641$^{***}$ & 5.127$^{***}$ & 9.376$^{***}$ & 106.661$^{***}$ \\ 
\hline 
\hline \\[-1.8ex] 
\textit{Note:}  & \multicolumn{5}{r}{$^{*}$p$<$0.1; $^{**}$p$<$0.05; $^{***}$p$<$0.01} \\ 
\end{tabular} 
\end{table}

\FloatBarrier

\section{Fitness Index}
\label{appendix:Fitness:all}
wait until Robert confirms data and figures
\subsection{Top/Bottom Industries}
\begin{table}[ht]
\centering
\caption{ICI: Top/Bottom Complexity Industries ($FI$, $CM_{r,i}$, NAICS, 2015)} 
\begingroup\footnotesize
\begin{tabular}{l|l|p{10cm}|l|r}
  \hline
 & NAICS & Name & ICI & Type \\ 
  \hline
1 & 523210 & Securities and Commodity Exchanges & 357.01 & Traded \\ 
  2 & 522190 & Other Depository Credit Intermediation & 58.97 & Traded \\ 
  3 & 335224 & Household Laundry Equipment Manufacturing & 35.27 & Traded \\ 
  4 & 485112 & Commuter Rail Systems & 30.05 & Local \\ 
  5 & 483112 & Deep Sea Passenger Transportation & 19.39 & Traded \\ 
  6 & 212392 & Phosphate Rock Mining & 12.17 & Traded \\ 
  7 & 521110 & Monetary Authorities - Central Bank & 11.40 & Traded \\
\hline
\end{tabular}
\endgroup
\end{table}

\FloatBarrier

\subsection{Map}
\begin{figure}[!h]
	\centering
	\includegraphics[width=15cm]{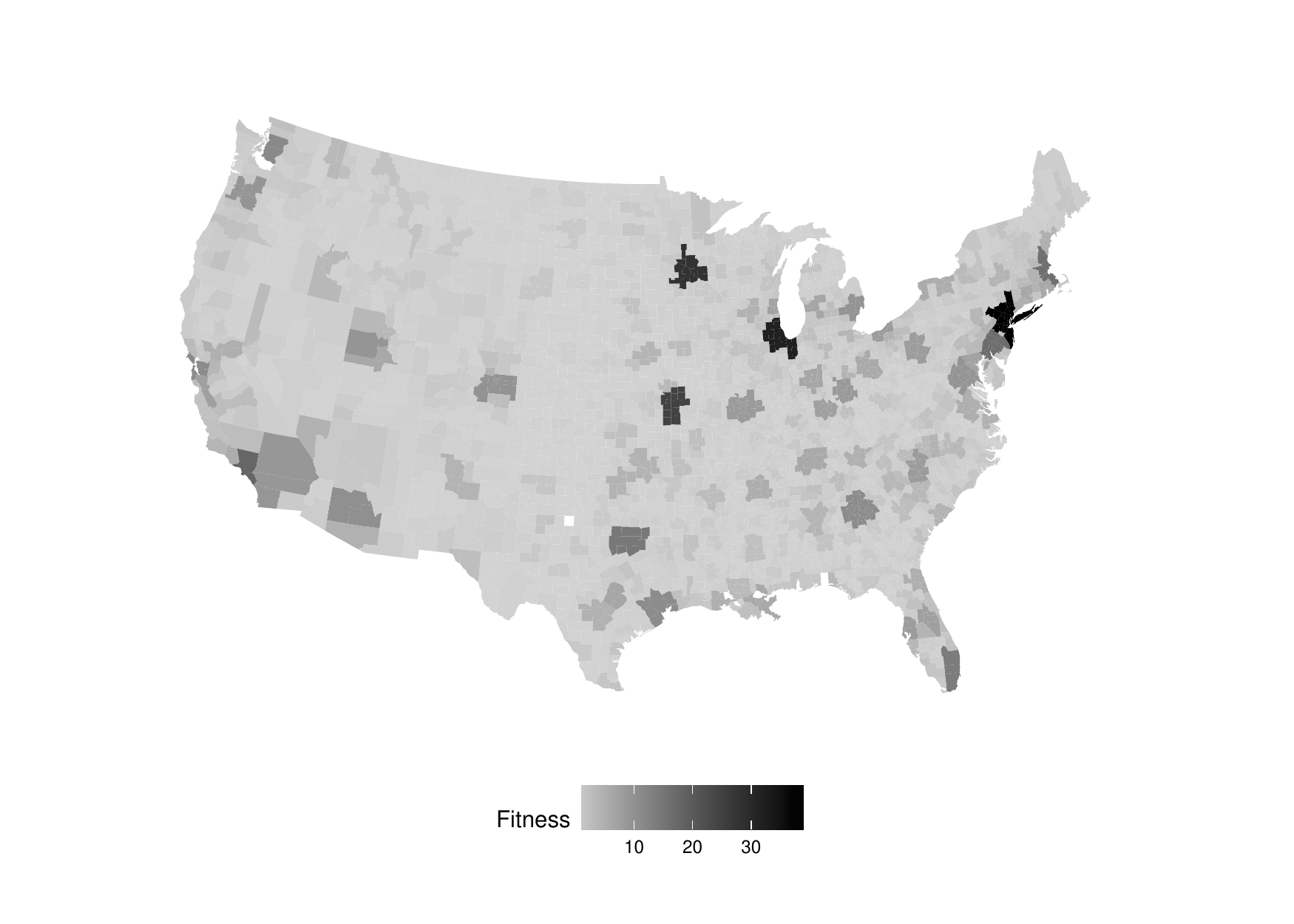}
	\caption{Map: FI ($CM_{r,i}$, NAICS, 2015)}%
	\source{Own calculations and CBP.}
\end{figure}
\FloatBarrier

\subsection{Cross-Sectional Regressions}

\begin{table}[!htbp] \centering 
  \caption{Cross-Section Regression: Income per Capita (in 1,000 USD) 2007 (NAICS, $CM_{r,i}$, FI)} 
  \label{} 
\footnotesize 
\begin{tabular}{@{\extracolsep{5pt}}lccccc} 
\\[-1.8ex]\hline 
\hline \\[-1.8ex] 
 & \multicolumn{5}{c}{\textit{Dependent variable:}} \\ 
\cline{2-6} 
\\[-1.8ex] & \multicolumn{5}{c}{Income per Capita (in 1,000 USD) 2007} \\ 
\\[-1.8ex] & (1) & (2) & (3) & (4) & (5)\\ 
\hline \\
\end{tabular} 
\end{table}

\begin{table}[!htbp] \centering 
  \caption{Cross-Section Regression: Population (in 1,000) 2007 (NAICS, $CM_{r,i}$, FI)} 
  \label{} 
\footnotesize 
\begin{tabular}{@{\extracolsep{5pt}}lccccc} 
\\[-1.8ex]\hline 
\hline \\[-1.8ex] 
 & \multicolumn{5}{c}{\textit{Dependent variable:}} \\ 
\cline{2-6} 
\\[-1.8ex] & \multicolumn{5}{c}{Population (in 1,000) 2007} \\ 
\\[-1.8ex] & (1) & (2) & (3) & (4) & (5)\\ 
\hline \\
\end{tabular} 
\end{table}

\begin{table}[!htbp] \centering 
  \caption{Cross-Section Regression: Income per Capita (in 1,000 USD) 2015 (NAICS, $CM_{r,i}$, FI)} 
  \label{} 
\footnotesize 
\begin{tabular}{@{\extracolsep{5pt}}lccccc} 
\\[-1.8ex]\hline 
\hline \\[-1.8ex] 
 & \multicolumn{5}{c}{\textit{Dependent variable:}} \\ 
\cline{2-6} 
\\[-1.8ex] & \multicolumn{5}{c}{Income per Capita (in 1,000 USD) 2015} \\ 
\\[-1.8ex] & (1) & (2) & (3) & (4) & (5)\\ 
\hline \\
\end{tabular} 
\end{table}

\begin{table}[!htbp] \centering 
  \caption{Cross-Section Regression: Population (in 1,000) 2015 (NAICS, $CM_{r,i}$, FI)} 
  \label{} 
\footnotesize 
\begin{tabular}{@{\extracolsep{5pt}}lccccc} 
\\[-1.8ex]\hline 
\hline \\[-1.8ex] 
 & \multicolumn{5}{c}{\textit{Dependent variable:}} \\ 
\cline{2-6} 
\\[-1.8ex] & \multicolumn{5}{c}{Population (in 1,000) 2015} \\ 
\\[-1.8ex] & (1) & (2) & (3) & (4) & (5)\\ 
\hline \\
\end{tabular} 
\end{table} 

\FloatBarrier

\subsection{Period Regressions}

\begin{table}[!htbp] \centering 
  \caption{Period Regression: Income per Capita (in Pct-Change) between 2007 and 2009 (NAICS, $CM_{r,i}$, FI)} 
  \label{} 
\footnotesize 
\begin{tabular}{@{\extracolsep{5pt}}lccccc} 
\\[-1.8ex]\hline 
\hline \\[-1.8ex] 
 & \multicolumn{5}{c}{\textit{Dependent variable:}} \\ 
\cline{2-6} 
\\[-1.8ex] & \multicolumn{5}{c}{Income per Capita (in Pct-Change) Change between 2007 and 2009} \\ 
\\[-1.8ex] & (1) & (2) & (3) & (4) & (5)\\ 
\hline \\
\end{tabular} 
\end{table}

\begin{table}[!htbp] \centering 
  \caption{Period Regression: Population (in Pct-Change) between 2007 and 2009 (NAICS, $CM_{r,i}$, FI)} 
  \label{} 
\footnotesize 
\begin{tabular}{@{\extracolsep{5pt}}lccccc} 
\\[-1.8ex]\hline 
\hline \\[-1.8ex] 
 & \multicolumn{5}{c}{\textit{Dependent variable:}} \\ 
\cline{2-6} 
\\[-1.8ex] & \multicolumn{5}{c}{Population (in Pct-Change) Change between 2007 and 2009} \\ 
\\[-1.8ex] & (1) & (2) & (3) & (4) & (5)\\ 
\hline \\
\end{tabular} 
\end{table}

\begin{table}[!htbp] \centering 
  \caption{Period Regression: Income per Capita (in Pct-Change) between 2010 and 2015 (NAICS, $CM_{r,i}$, FI)} 
  \label{} 
\footnotesize 
\begin{tabular}{@{\extracolsep{5pt}}lccccc} 
\\[-1.8ex]\hline 
\hline \\[-1.8ex] 
 & \multicolumn{5}{c}{\textit{Dependent variable:}} \\ 
\cline{2-6} 
\\[-1.8ex] & \multicolumn{5}{c}{Income per Capita (in Pct-Change) Change between 2010 and 2015} \\ 
\\[-1.8ex] & (1) & (2) & (3) & (4) & (5)\\ 
\hline \\
\end{tabular} 
\end{table}

\begin{table}[!htbp] \centering 
  \caption{Period Regression: Population (in Pct-Change) between 2010 and 2015 (NAICS, $CM_{r,i}$, FI)} 
  \label{} 
\footnotesize 
\begin{tabular}{@{\extracolsep{5pt}}lccccc} 
\\[-1.8ex]\hline 
\hline \\[-1.8ex] 
 & \multicolumn{5}{c}{\textit{Dependent variable:}} \\ 
\cline{2-6} 
\\[-1.8ex] & \multicolumn{5}{c}{Population (in Pct-Change) Change between 2010 and 2015} \\ 
\\[-1.8ex] & (1) & (2) & (3) & (4) & (5)\\ 
\hline \\
\end{tabular} 
\end{table} 

\FloatBarrier

\subsection{Panel Regression}

\begin{table}[!htbp] \centering 
  \caption{Panel Regression: Income per Capita (in 1,000 USD) 2015 (NAICS, $CM_{r,i}$, FI)} 
  \label{} 
\footnotesize 
\begin{tabular}{@{\extracolsep{5pt}}lccccc} 
\\[-1.8ex]\hline 
\hline \\[-1.8ex] 
 & \multicolumn{5}{c}{\textit{Dependent variable:}} \\ 
\cline{2-6} 
\\[-1.8ex] & \multicolumn{5}{c}{Income per Capita (in 1,000 USD) 2015} \\ 
\\[-1.8ex] & (1) & (2) & (3) & (4) & (5)\\ 
\hline \\
\end{tabular} 
\end{table}

\begin{table}[!htbp] \centering 
  \caption{Panel Regression: Population (in 1,000) 2015 (NAICS, $CM_{r,i}$, FI)} 
  \label{} 
\footnotesize 
\begin{tabular}{@{\extracolsep{5pt}}lccccc} 
\\[-1.8ex]\hline 
\hline \\[-1.8ex] 
 & \multicolumn{5}{c}{\textit{Dependent variable:}} \\ 
\cline{2-6} 
\\[-1.8ex] & \multicolumn{5}{c}{Population (in 1,000) 2015} \\ 
\\[-1.8ex] & (1) & (2) & (3) & (4) & (5)\\ 
\hline \\[-1.8ex] 
 l.FI & $-$0.128$^{***}$ & 0.020 & $-$0.126$^{***}$ & $-$0.128$^{***}$ & 0.018 \\ 
  & (0.045) & (0.059) & (0.044) & (0.045) & (0.059) \\ 
  & & & & & \\ 
 Unemployment &  & 0.122 &  &  & 0.011 \\ 
  &  & (0.263) &  &  & (0.253) \\ 
  & & & & & \\ 
 ManuShare &  & 0.046 &  &  & 0.168 \\ 
  &  & (0.562) &  &  & (0.574) \\ 
  & & & & & \\ 
 Patent &  & 0.099$^{***}$ &  &  & 0.099$^{***}$ \\ 
  &  & (0.035) &  &  & (0.035) \\ 
  & & & & & \\ 
 Income per Capita &  & $-$1.596 &  &  & $-$1.577$^{*}$ \\ 
  &  & (1.050) &  &  & (0.957) \\ 
  & & & & & \\ 
 MedianAge &  &  & $-$4.192$^{***}$ &  & $-$1.138 \\ 
  &  &  & (1.584) &  & (0.922) \\ 
  & & & & & \\ 
 Education &  &  & 0.955$^{**}$ &  & 0.382 \\ 
  &  &  & (0.434) &  & (0.283) \\ 
  & & & & & \\ 
 BlackShare &  &  & 1.860 &  & 2.675 \\ 
  &  &  & (3.031) &  & (2.346) \\ 
  & & & & & \\ 
 Foreign &  &  & 2.130 &  & 0.614 \\ 
  &  &  & (2.011) &  & (1.483) \\ 
  & & & & & \\ 
 UnionCoverage &  &  &  & 5.880$^{***}$ & 4.993$^{***}$ \\ 
  &  &  &  & (1.877) & (1.463) \\ 
  & & & & & \\ 
 MinimumWage &  &  &  & 7.765$^{*}$ & 7.828$^{**}$ \\ 
  &  &  &  & (4.049) & (3.423) \\ 
  & & & & & \\ 
 Vote &  &  &  & $-$0.013 & 0.465$^{*}$ \\ 
  &  &  &  & (0.399) & (0.272) \\ 
  & & & & & \\ 
\hline \\[-1.8ex] 
Observations & 2,715 & 2,715 & 2,715 & 2,715 & 2,715 \\ 
R$^{2}$ & 0.029 & 0.335 & 0.038 & 0.044 & 0.350 \\ 
Adjusted R$^{2}$ & $-$0.101 & 0.244 & $-$0.092 & $-$0.085 & 0.260 \\ 
F Statistic & 70.578$^{***}$ & 240.328$^{***}$ & 19.115$^{***}$ & 27.585$^{***}$ & 106.887$^{***}$ \\ 
\hline 
\hline \\[-1.8ex] 
\textit{Note:}  & \multicolumn{5}{r}{$^{*}$p$<$0.1; $^{**}$p$<$0.05; $^{***}$p$<$0.01} \\ 
\end{tabular} 
\end{table} 

\FloatBarrier

\section{BCD}

\subsection{Input Matrices: Ubiquity vs. Diversity}
\begin{figure}[!h]
	\centering
	\subfloat[$CM_{r,i}$, local+traded]{{\includegraphics[width=6cm]{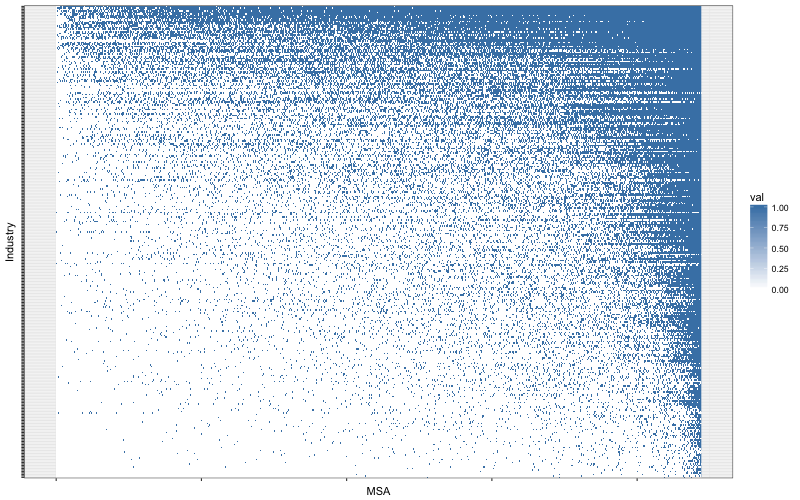}}}%
	\qquad
	\subfloat[$CM_{r,i}$, local+traded]{{\includegraphics[width=6cm]{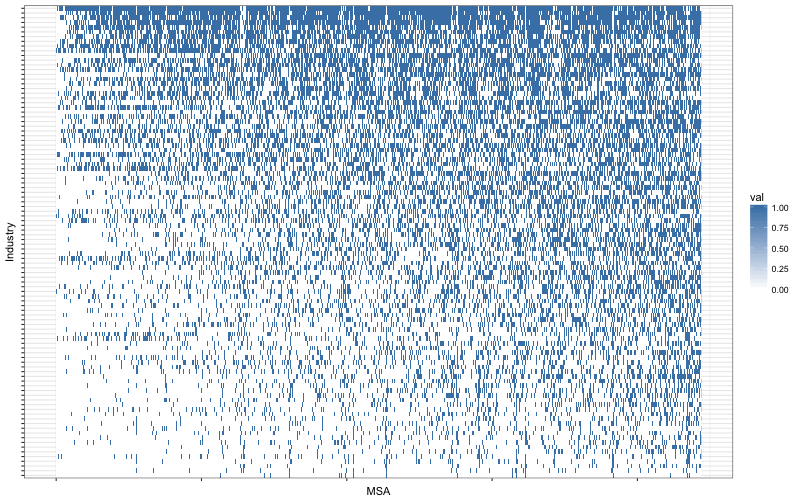}}}%
	\\
	\subfloat[$CM_{r,i}$, local+traded]{{\includegraphics[width=6cm]{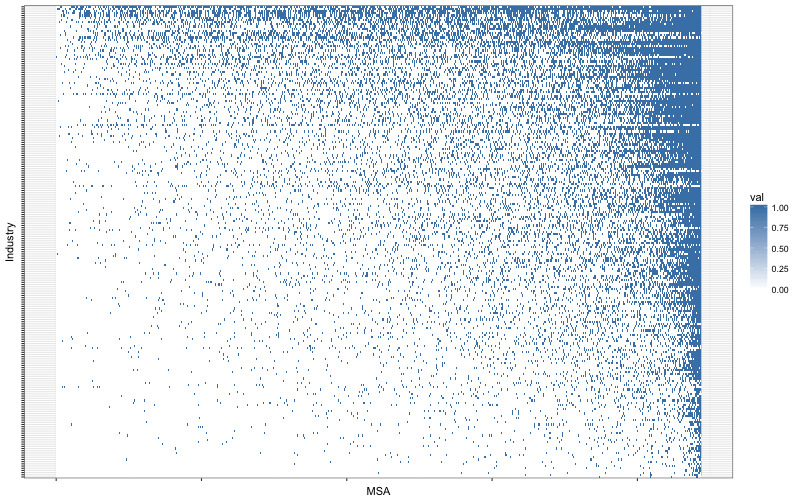}}}%
	\caption{Input matrices: Ubiquity vs. Diversity (BCD, 2015)}%
	\source{Own calculations and CBP.}
\end{figure}
\FloatBarrier

\subsection{Scatter-plots ECI VS. FI}
\begin{figure}[!h]
	\centering
	\subfloat[ECI vs FI]{{\includegraphics[width=6cm]{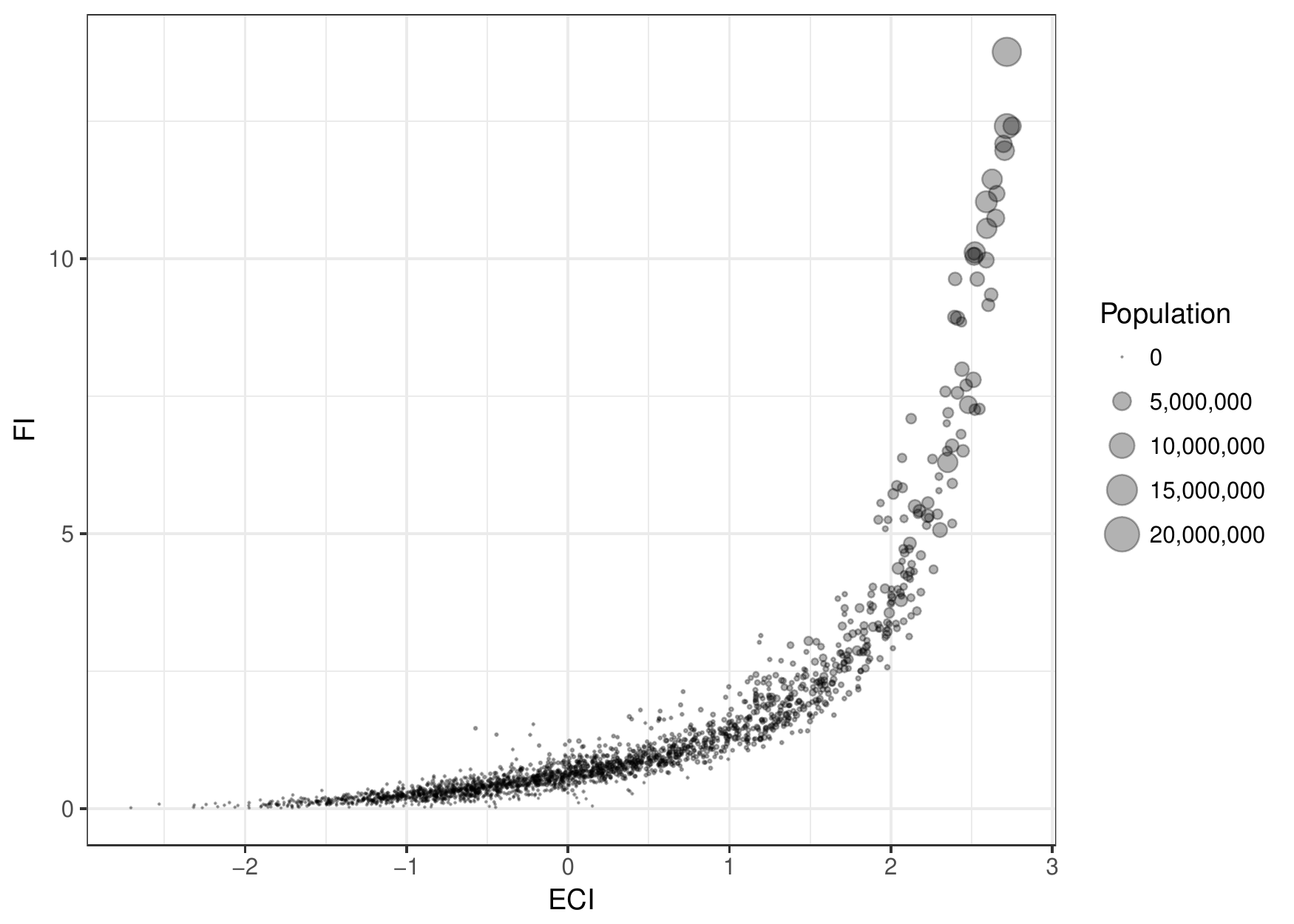}}}%
	\qquad
	\subfloat[ECI vs logged FI]{{\includegraphics[width=6cm]{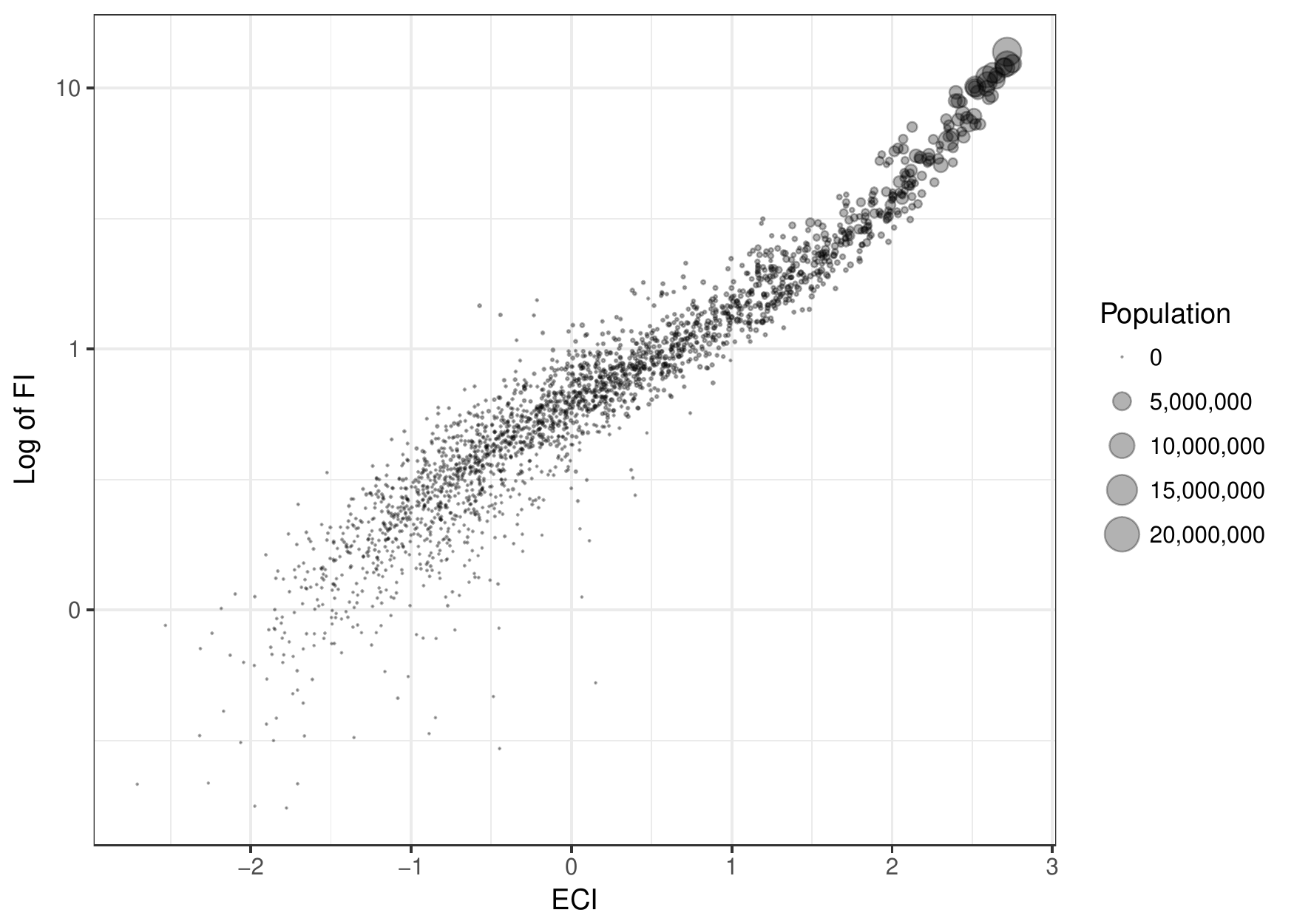}}}%
	\caption{Scatterplots ECI vs FI ($CM_{r,i}$, BCD, 2015)}%
	\source{Own calculations and CBP.}
\end{figure}
\FloatBarrier

\subsection{Top/Bottom Industries}
\begin{table}[ht]
\centering
\caption{ICI: Top/Bottom Complexity Industries ($ECI$, $CM_{r,i}$, BCD, 2015)} 
\begingroup\footnotesize
\begin{tabular}{l|l|p{10cm}|l|r}
  \hline
 & Cluster & Name & ICI & Type \\ 
  \hline
1 & 16\_4 & Monetary Authorities - Central Bank & 3.41 & Traded \\ 
  2 & 24\_3 & Reinsurance Carriers & 2.61 & Traded \\ 
  3 & 5\_3 & Diagnostic Substances & 2.49 & Traded \\ 
  4 & 8\_3 & Communications Equipment Components & 2.29 & Traded \\ 
  5 & 1\_2 & Missiles and Space Vehicles & 2.27 & Traded \\ 
  6 & 23\_7 & Medical Apparatus & 2.14 & Traded \\ 
  7 & 39\_1 & Printing Inputs & 2.10 & Traded \\ 
  8 & 4\_5 & Military Vehicles and Tanks & 2.07 & Traded \\ 
  9 & 5\_2 & Biological Products & 2.02 & Traded \\ 
  10 & 23\_5 & Software Reproducing & 1.99 & Traded \\ 
  11 & 108\_6 & Parking Services & 1.98 & Local \\ 
  12 & 23\_4 & Software Publishers & 1.92 & Traded \\ 
  13 & 27\_4 & Storage Batteries & 1.88 & Traded \\ 
  14 & 23\_3 & Semiconductors & 1.87 & Traded \\ 
  15 & 39\_4 & Greeting Card Printing and Publishing & 1.81 & Traded \\ 
  16 & 1\_3 & Search and Navigation Equipment & 1.78 & Traded \\ 
  17 & 30\_1 & Optical Instruments and Ophthalmic Goods & 1.75 & Traded \\ 
  18 & 107\_2 & Pension, Health, and Welfare Funds & 1.72 & Local \\ 
  19 & 16\_3 & Credit Bureaus & 1.67 & Traded \\ 
  20 & 26\_2 & Women's Handbags and Purses & 1.58 & Traded \\ 
  21 &  & ... &  & ... \\ 
  297 & 103\_4 & Home and Residential Care & -1.58 & Local \\ 
  298 & 108\_9 & Automotive Parts Retailing & -1.58 & Local \\ 
  299 & 111\_1 & Hospitality Establishments & -1.59 & Local \\ 
  300 & 104\_4 & Heating Oil and Other Fuel Dealers & -1.60 & Local \\ 
  301 & 10\_9 & Wholesale of Farm Products and Supplies & -1.61 & Traded \\ 
  302 & 10\_17 & Wholesale of Farm and Garden Machinery and Equipment & -1.64 & Traded \\ 
  303 & 101\_3 & Retail Food Stores & -1.66 & Local \\ 
  304 & 107\_1 & Deposit-taking Institutions & -1.67 & Local \\ 
  305 & 20\_1 & Forestry & -1.68 & Traded \\ 
  306 & 106\_6 & Gardening Products and Supplies Retailing & -1.68 & Local \\ 
  307 & 103\_5 & Funeral Service and Crematories & -1.69 & Local \\ 
  308 & 7\_1 & Coal Mining & -1.79 & Traded \\ 
  309 & 35\_6 & Pipeline Transportation & -1.79 & Traded \\ 
  310 & 108\_5 & Gasoline Stations & -1.81 & Local \\ 
  311 & 2\_1 & Agricultural Services & -1.86 & Traded \\ 
  312 & 114\_3 & Business Associations & -2.00 & Local \\ 
  313 & 18\_14 & Farm Wholesalers & -2.08 & Traded \\ 
  314 & 35\_3 & Drilling Wells & -2.19 & Traded \\ 
  315 & 35\_2 & Support Activities for Oil and Gas Operations & -2.20 & Traded \\ 
  316 & 35\_4 & Oil and Gas Extraction & -2.39 & Traded \\ 
   \hline
\end{tabular}
\endgroup
\end{table}

\FloatBarrier

\subsection{Map}

\begin{figure}[!h]
	\centering
	\includegraphics[width=15cm]{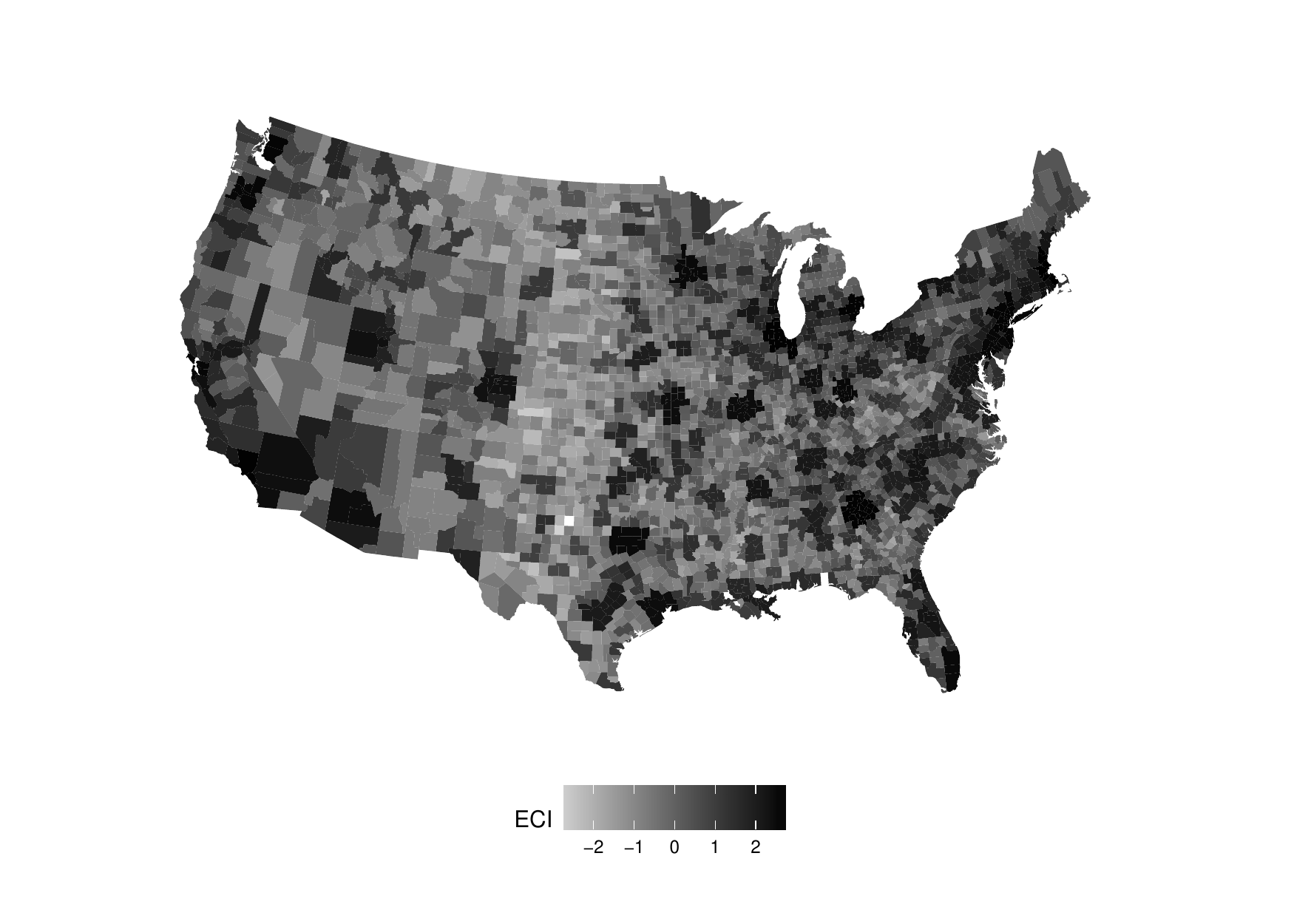}
	\caption{Map: FI ($CM_{r,i}$, BCD, 2015)}%
	\source{Own calculations and CBP.}
\end{figure}

\subsection{Cross-Sectional Regressions}

\begin{table}[!htbp] \centering 
  \caption{Cross-Section Regression: Income per Capita (in 1,000 USD) 2007 (BCD, $CM_{r,i}$, ECI)} 
  \label{} 
\footnotesize 
\begin{tabular}{@{\extracolsep{5pt}}lccccc} 
\\[-1.8ex]\hline 
\hline \\[-1.8ex] 
 & \multicolumn{5}{c}{\textit{Dependent variable:}} \\ 
\cline{2-6} 
\\[-1.8ex] & \multicolumn{5}{c}{Income per Capita (in 1,000 USD) 2007} \\ 
\\[-1.8ex] & (1) & (2) & (3) & (4) & (5)\\ 
\hline \\
\end{tabular} 
\end{table}

\begin{table}[!htbp] \centering 
  \caption{Cross-Section Regression: Population (in 1,000) 2007 (BCD, $CM_{r,i}$, ECI)} 
  \label{} 
\footnotesize 
\begin{tabular}{@{\extracolsep{5pt}}lccccc} 
\\[-1.8ex]\hline 
\hline \\[-1.8ex] 
 & \multicolumn{5}{c}{\textit{Dependent variable:}} \\ 
\cline{2-6} 
\\[-1.8ex] & \multicolumn{5}{c}{Population (in 1,000) 2007} \\ 
\\[-1.8ex] & (1) & (2) & (3) & (4) & (5)\\ 
\hline \\
\end{tabular} 
\end{table}

\begin{table}[!htbp] \centering 
  \caption{Cross-Section Regression: Income per Capita (in 1,000 USD) 2015 (BCD, $CM_{r,i}$, ECI)} 
  \label{} 
\footnotesize 
\begin{tabular}{@{\extracolsep{5pt}}lccccc} 
\\[-1.8ex]\hline 
\hline \\[-1.8ex] 
 & \multicolumn{5}{c}{\textit{Dependent variable:}} \\ 
\cline{2-6} 
\\[-1.8ex] & \multicolumn{5}{c}{Income per Capita (in 1,000 USD) 2015} \\ 
\\[-1.8ex] & (1) & (2) & (3) & (4) & (5)\\ 
\hline \\
\end{tabular} 
\end{table}

\begin{table}[!htbp] \centering 
  \caption{Cross-Section Regression: Population (in 1,000) 2015 (BCD, $CM_{r,i}$, ECI)} 
  \label{} 
\footnotesize 
\begin{tabular}{@{\extracolsep{5pt}}lccccc} 
\\[-1.8ex]\hline 
\hline \\[-1.8ex] 
 & \multicolumn{5}{c}{\textit{Dependent variable:}} \\ 
\cline{2-6} 
\\[-1.8ex] & \multicolumn{5}{c}{Population (in 1,000) 2015} \\ 
\\[-1.8ex] & (1) & (2) & (3) & (4) & (5)\\ 
\hline \\
\end{tabular} 
\end{table} 

\FloatBarrier

\subsection{Period Regressions}

\begin{table}[!htbp] \centering 
  \caption{Period Regression: Income per Capita (in Pct-Change) between 2007 and 2009 (BCD, $CM_{r,i}$, ECI)} 
  \label{} 
\footnotesize 
\begin{tabular}{@{\extracolsep{5pt}}lccccc} 
\\[-1.8ex]\hline 
\hline \\[-1.8ex] 
 & \multicolumn{5}{c}{\textit{Dependent variable:}} \\ 
\cline{2-6} 
\\[-1.8ex] & \multicolumn{5}{c}{Income per Capita (in Pct-Change) Change between 2007 and 2009} \\ 
\\[-1.8ex] & (1) & (2) & (3) & (4) & (5)\\ 
\hline \\
\end{tabular} 
\end{table}

\begin{table}[!htbp] \centering 
  \caption{Period Regression: Population (in Pct-Change) between 2007 and 2009 (BCD, $CM_{r,i}$, ECI)} 
  \label{} 
\footnotesize 
\begin{tabular}{@{\extracolsep{5pt}}lccccc} 
\\[-1.8ex]\hline 
\hline \\[-1.8ex] 
 & \multicolumn{5}{c}{\textit{Dependent variable:}} \\ 
\cline{2-6} 
\\[-1.8ex] & \multicolumn{5}{c}{Population (in Pct-Change) Change between 2007 and 2009} \\ 
\\[-1.8ex] & (1) & (2) & (3) & (4) & (5)\\ 
\hline \\
\end{tabular} 
\end{table}

\begin{table}[!htbp] \centering 
  \caption{Period Regression: Income per Capita (in Pct-Change) between 2010 and 2015 (BCD, $CM_{r,i}$, ECI)} 
  \label{} 
\footnotesize 
\begin{tabular}{@{\extracolsep{5pt}}lccccc} 
\\[-1.8ex]\hline 
\hline \\[-1.8ex] 
 & \multicolumn{5}{c}{\textit{Dependent variable:}} \\ 
\cline{2-6} 
\\[-1.8ex] & \multicolumn{5}{c}{Income per Capita (in Pct-Change) Change between 2010 and 2015} \\ 
\\[-1.8ex] & (1) & (2) & (3) & (4) & (5)\\ 
\hline \\
\end{tabular} 
\end{table}

\begin{table}[!htbp] \centering 
  \caption{Period Regression: Population (in Pct-Change) between 2010 and 2015 (BCD, $CM_{r,i}$, ECI)} 
  \label{} 
\footnotesize 
\begin{tabular}{@{\extracolsep{5pt}}lccccc} 
\\[-1.8ex]\hline 
\hline \\[-1.8ex] 
 & \multicolumn{5}{c}{\textit{Dependent variable:}} \\ 
\cline{2-6} 
\\[-1.8ex] & \multicolumn{5}{c}{Population (in Pct-Change) Change between 2010 and 2015} \\ 
\\[-1.8ex] & (1) & (2) & (3) & (4) & (5)\\ 
\hline \\
\end{tabular} 
\end{table}

\FloatBarrier

\subsection{Panel Regression}

\begin{table}[!htbp] \centering 
  \caption{Panel Regression: Income per Capita (in 1,000 USD) 2015 (BCD, $CM_{r,i}$, ECI)} 
  \label{} 
\footnotesize 
\begin{tabular}{@{\extracolsep{5pt}}lccccc} 
\\[-1.8ex]\hline 
\hline \\[-1.8ex] 
 & \multicolumn{5}{c}{\textit{Dependent variable:}} \\ 
\cline{2-6} 
\\[-1.8ex] & \multicolumn{5}{c}{Income per Capita (in 1,000 USD) 2015} \\ 
\\[-1.8ex] & (1) & (2) & (3) & (4) & (5)\\ 
\hline \\[-1.8ex] 
 l.ECI & 0.044 & $-$0.080 & 0.065 & 0.073 & $-$0.057 \\ 
  & (0.485) & (0.472) & (0.411) & (0.487) & (0.386) \\ 
  & & & & & \\ 
 Unemployment &  & $-$0.152$^{***}$ &  &  & $-$0.145$^{***}$ \\ 
  &  & (0.023) &  &  & (0.021) \\ 
  & & & & & \\ 
 ManuShare &  & 0.030 &  &  & 0.030 \\ 
  &  & (0.037) &  &  & (0.037) \\ 
  & & & & & \\ 
 Patent &  & 0.001$^{***}$ &  &  & 0.001$^{***}$ \\ 
  &  & (0.0002) &  &  & (0.0002) \\ 
  & & & & & \\ 
 Population &  &  & 0.00000 &  & $-$0.00000$^{**}$ \\ 
  &  &  & (0.00000) &  & (0.00000) \\ 
  & & & & & \\ 
 MedianAge &  &  & $-$0.061 &  & $-$0.027 \\ 
  &  &  & (0.121) &  & (0.119) \\ 
  & & & & & \\ 
 Education &  &  & 0.028 &  & 0.025 \\ 
  &  &  & (0.024) &  & (0.025) \\ 
  & & & & & \\ 
 BlackShare &  &  & $-$0.236$^{***}$ &  & $-$0.203$^{***}$ \\ 
  &  &  & (0.078) &  & (0.074) \\ 
  & & & & & \\ 
 Foreign &  &  & 0.258$^{*}$ &  & 0.248$^{*}$ \\ 
  &  &  & (0.134) &  & (0.134) \\ 
  & & & & & \\ 
 UnionCoverage &  &  &  & 0.039 & 0.067 \\ 
  &  &  &  & (0.059) & (0.059) \\ 
  & & & & & \\ 
 MinimumWage &  &  &  & 0.045 & 0.006 \\ 
  &  &  &  & (0.170) & (0.158) \\ 
  & & & & & \\ 
 Vote &  &  &  & $-$0.055$^{***}$ & $-$0.047$^{***}$ \\ 
  &  &  &  & (0.012) & (0.010) \\ 
  & & & & & \\ 
\hline \\[-1.8ex] 
Observations & 2,715 & 2,715 & 2,715 & 2,715 & 2,715 \\ 
R$^{2}$ & 0.00001 & 0.037 & 0.021 & 0.010 & 0.066 \\ 
Adjusted R$^{2}$ & $-$0.134 & $-$0.094 & $-$0.112 & $-$0.124 & $-$0.064 \\ 
F Statistic & 0.012 & 22.690$^{***}$ & 8.528$^{***}$ & 5.897$^{***}$ & 14.034$^{***}$ \\ 
\hline 
\hline \\[-1.8ex] 
\textit{Note:}  & \multicolumn{5}{r}{$^{*}$p$<$0.1; $^{**}$p$<$0.05; $^{***}$p$<$0.01} \\ 
\end{tabular} 
\end{table}

\begin{table}[!htbp] \centering 
  \caption{Panel Regression: Population (in 1,000) 2015 (BCD, $CM_{r,i}$, ECI)} 
  \label{} 
\footnotesize 
\begin{tabular}{@{\extracolsep{5pt}}lccccc} 
\\[-1.8ex]\hline 
\hline \\[-1.8ex] 
 & \multicolumn{5}{c}{\textit{Dependent variable:}} \\ 
\cline{2-6} 
\\[-1.8ex] & \multicolumn{5}{c}{Population (in 1,000) 2015} \\ 
\\[-1.8ex] & (1) & (2) & (3) & (4) & (5)\\ 
\hline \\[-1.8ex] 
 l.ECI & 10.011 & $-$2.728 & 16.884$^{*}$ & 11.133 & 0.394 \\ 
  & (8.588) & (6.766) & (9.025) & (8.666) & (6.753) \\ 
  & & & & & \\ 
 Unemployment &  & 0.123 &  &  & 0.011 \\ 
  &  & (0.262) &  &  & (0.254) \\ 
  & & & & & \\ 
 ManuShare &  & 0.062 &  &  & 0.164 \\ 
  &  & (0.567) &  &  & (0.577) \\ 
  & & & & & \\ 
 Patent &  & 0.098$^{***}$ &  &  & 0.097$^{***}$ \\ 
  &  & (0.032) &  &  & (0.032) \\ 
  & & & & & \\ 
 Income per Capita &  & $-$1.612 &  &  & $-$1.589$^{*}$ \\ 
  &  & (1.032) &  &  & (0.936) \\ 
  & & & & & \\ 
 MedianAge &  &  & $-$4.654$^{***}$ &  & $-$1.149 \\ 
  &  &  & (1.679) &  & (0.911) \\ 
  & & & & & \\ 
 Education &  &  & 1.012$^{**}$ &  & 0.384 \\ 
  &  &  & (0.438) &  & (0.284) \\ 
  & & & & & \\ 
 BlackShare &  &  & 1.447 &  & 2.726 \\ 
  &  &  & (3.145) &  & (2.389) \\ 
  & & & & & \\ 
 Foreign &  &  & 2.228 &  & 0.628 \\ 
  &  &  & (2.025) &  & (1.485) \\ 
  & & & & & \\ 
 UnionCoverage &  &  &  & 5.984$^{***}$ & 4.996$^{***}$ \\ 
  &  &  &  & (1.892) & (1.470) \\ 
  & & & & & \\ 
 MinimumWage &  &  &  & 7.346$^{*}$ & 7.875$^{**}$ \\ 
  &  &  &  & (4.075) & (3.473) \\ 
  & & & & & \\ 
 Vote &  &  &  & $-$0.158 & 0.474$^{*}$ \\ 
  &  &  &  & (0.432) & (0.275) \\ 
  & & & & & \\ 
\hline \\[-1.8ex] 
Observations & 2,715 & 2,715 & 2,715 & 2,715 & 2,715 \\ 
R$^{2}$ & 0.0004 & 0.334 & 0.012 & 0.016 & 0.349 \\ 
Adjusted R$^{2}$ & $-$0.133 & 0.244 & $-$0.122 & $-$0.117 & 0.259 \\ 
F Statistic & 0.842 & 239.668$^{***}$ & 5.603$^{***}$ & 9.644$^{***}$ & 106.659$^{***}$ \\ 
\hline 
\hline \\[-1.8ex] 
\textit{Note:}  & \multicolumn{5}{r}{$^{*}$p$<$0.1; $^{**}$p$<$0.05; $^{***}$p$<$0.01} \\ 
\end{tabular} 
\end{table}

\end{document}